\documentstyle[preprint,float,aps,psfig]{revtex} 
\tightenlines
\begin{document}

\title{Analysis of Pion Photoproduction Data}

\author{R.~A.~Arndt\thanks{arndt@reo.ntelos.net},
        W.~J.~Briscoe\thanks{briscoe@gwu.edu},
        I.~I.~Strakovsky\thanks{igor@gwu.edu},
        R. L.~Workman\thanks{rworkman@gwu.edu}}
\address{Center for Nuclear Studies, Department of
        Physics, \\
        The George Washington University, Washington, 
        D.C. 20052}

\draft
\date{\today}
\maketitle

\begin{abstract}

A partial-wave analysis of single-pion photoproduction
data has been completed.  This study extends from threshold
to 2 GeV in the laboratory photon energy, focusing mainly
on the influence of new measurements and model-dependence
in the choice of parameterization employed above the 
two-pion threshold.  Results are used to evaluate sum rules 
and estimate resonance photo-decay amplitudes. These are 
compared to values obtained in the MAID analysis.

\end{abstract}


\narrowtext
\section{Introduction}
\label{sec:intro}

Meson-nucleon scattering, meson photoproduction, and meson 
electroproduction have been extensively studied within a 
comprehensive program exploring the spectroscopy of 
$N^{\ast}$ and $\Delta ^{\ast}$ resonances.  An objective 
of this program is the determination of all relevant 
characteristics of these resonances, \textit{i.e.} pole 
positions, widths, principal decay channels, and branching 
ratios.  In order to compare directly with QCD-inspired 
models, there has also been a considerable effort to find 
``hidden" or ``missing" resonances, predicted by quark 
models (see, for example, the states predicted by Capstick 
and Roberts~\cite{caro}) but not yet confirmed.

Here we will give detailed results from an ongoing 
analysis of pion photoproduction data.  This work 
complements our studies of pion-nucleon elastic 
scattering~\cite{pin}, both reactions having the same 
resonance content, and provides the real-photon limit for 
our pion-electroproduction fits~\cite{electro}.
Resonance characteristics are determined through a 
two-step procedure.  The full database is initially 
fitted to determine the underlying multipole 
contributions.  Those multipoles having resonant 
contributions are then fitted to a form containing 
both resonance and background terms. 
 
The availability of multipole amplitudes greatly
simplifies certain numerical aspects of coupled-channel 
analysis.  The number of fitted multipole amplitudes, 
associated with a dataset, may be smaller (than the 
count of individual data) by one or two orders of 
magnitude, and can account for issues associated with 
statistical/systematic errors, data rejection, and 
incomplete sets of observables.  In general, our 
partial-wave analyses (PWA) have been as 
model-independent as possible, so as to avoid bias when 
used in resonance extraction or coupled-channel analysis. 
However, in the absense of complete experimental 
information, all multipole analyses above the two-pion 
production threshold are model-dependent to some degree. 
This issue will be discussed in Section~\ref{sec:ampl}.

The amplitudes from these analyses can be utilized 
in evaluating contributions to the Gerasimov-Drell-Hearn 
(GDH) sum rule~\cite{gdh} and sum rules related to the 
nucleon polarizability.  We display our results for 
these and compare with recent Mainz experimental data 
and predictions.

In the next section, we summarize changes in the 
database since our last published analysis~\cite{ar96}.  
Results of our multipole analyses, as well as the
photo-decay amplitudes for resonances within our energy 
region, are given in Section~\ref{sec:ampl}.  In 
Section~\ref{sec:conc}, we summarize our findings and 
consider what improvements can be expected in the future.

\section{Database}
\label{sec:expt}

Our three previous pion photoproduction analyses~
\cite{ar96,ar90,li93} extended to 1.0, 1.8, and 2.0~GeV, 
respectively.  The present database~\cite{said} is 
considerably larger, due mainly to the addition of new 
data at low to intermediate (below 800~MeV) energies.  

In 1994, bremsstrahlung data comprised over 85\% of the 
existing measurements.  These data often suffered from 
significant uncertainties in normalization that were not 
completely understood or not quoted.  The available 
tagged-photon data were generally measured in 
low-statistics experiments, and hence were presented 
with large energy and angular binning.  Much of the 
remaining dataset was comprised of excitation cross 
sections with no extensive angular range.  
Inconsistencies were obvious almost everywhere that 
comparisons could be made.  

\begin{figure}[ht]
\centerline{
\psfig{file=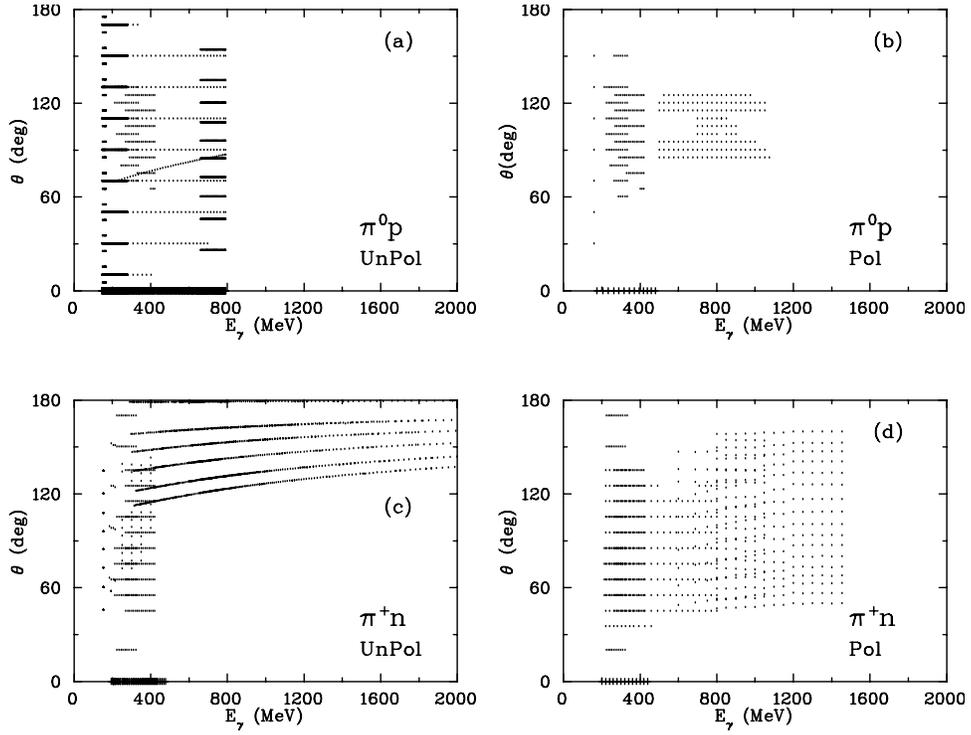,width=5in,clip=,silent=,angle=90}}
\vspace{3mm}
\caption[fig1]{\label{g1}
      Energy-angle distribution of recent (post-1995)
      data: (a) unpolarized $\pi^0p$, (b) polarized
      $\pi^0p$, (c) unpolarized $\pi^+n$, (d)
      polarized $\pi^+n$.  (a,c) total cross sections
      and (b,d) ($\sigma_{1/2} - \sigma_{3/2}$) are
      plotted at zero degrees.}
\end{figure}

The full database has increased by 30\% since the 
publication of Ref.~\cite{ar96}, and is about 40\% 
larger than the set available for the analysis of Ref.~
\cite{li93}.  The majority of these new data have come 
from tagged-photon facilities at MAMI (Mainz), 
GRAAL (Grenoble), and LEGS (Brookhaven.)  [The total 
database has doubled over the last two decades (see 
Table~\ref{tbl1}.)]  The distribution of recent 
(post-1995) $\pi^0p$ and $\pi^+n$ data is given in 
Fig.~\ref{g1} (there are no new $\pi^-p$ and $\pi^0n$ 
data.) 

As the full database contains conflicting results, some 
of these have been excluded from our fits.  We have, 
however, retained all available data sets (and labeled 
these excluded data as ``flagged") so that comparisons 
can be made through our on-line facility~\cite{said}.  
Data taken before 1960 were not analyzed, nor were 
those single-angle and single-energy points measured 
prior to 1970.  Some individual data points were also 
removed from the analysis in order to resolve conflicts 
or upon authors' requests (since our previous analysis~
\cite{ar96}, we have flagged 950 $\pi^0p$, 1060 $\pi^+n$, 
and 150 $\pi^-p$ bremsstrahlung data taken prior to 
1983.)  Some of the data, listed as new, were available 
in unpublished form at the time of our previous 
analysis~\cite{ar96}.  A complete description of the 
database and those data not included in our fits is 
available from the authors.

Since 1995, 87\% (21\%) of all new $\pi^0p$ ($\pi^+n$) 
data have been produced at Mainz using the MAMI facility~
\cite{fu96,ma96,be97,ha96,sc97,kr99,ah00,ah01,be00,br00,sc01}.
These measurements of total and differential cross 
sections, $\Sigma$ beam asymmetry, and 
the GDH-related quantity ($\sigma_{1/2} - 
\sigma_{3/2}$) have increased the database by a factor 
of about two over the energy range from the threshold 
to 800~MeV.  The angular range of cross sections extends 
from 10$^{\circ}$ to 170$^{\circ}$, and thus increases 
the sensitivity to contributions from higher partial 
waves.

Results from other laboratories include low-energy 
unpolarized $\pi^0p$ total (47 data) and differential 
(198 data) cross sections measured at SAL~\cite{be96,bg97}, 
and $\pi^+n$ threshold data (45 points), covering a range 
of 2~MeV in $E_{\gamma}$, produced by the TRIUMF--SAL 
Collaboration~\cite{ko99}.  At energies spanning the 
$\Delta$ resonance, $\Sigma$ (169 data) and differential 
cross section (157 data) for both $\pi^0p$ and $\pi^+n$ 
channels have been measured by the LEGS group at BNL
~\cite{bl01}.

In the medium-energy range, $\pi^+n$ $\Sigma$ 
beam-asymmetry data between 600 and 1500~MeV (329 data) 
have been measured at GRAAL~\cite{aj00,ku01} and 
$\pi^0p$ $\Sigma$ data between 500 and 1100~MeV (158 data) 
have been measured at the 4.5~GeV Yerevan Synchrotron~
\cite{ad01}.  Target asymmetry $T$ measurements 
between 220 and 800~MeV for both $\pi^0p$ (52 data) and 
$\pi^+n$ (210 data) have come from ELSA at Bonn~
\cite{du96,bo98}.  Excitation $\pi^+n$ differential 
cross sections for backward scattering between 290 and 
2110~MeV, also from the Bonn facility, have been replaced 
by finalized data~\cite{da01}, with final versions 
of other $\pi ^+n$ and $\pi ^0p$ differential cross 
sections expected~\cite{du80,sc81,he88,ze88}.

Further experimental efforts will provide data in the 
intermediate energy region.  Above 400~MeV, a large 
amount of new data is expected from CLAS at Hall~B of 
Jefferson Lab~\cite{he01}.  Differential cross 
sections associated with the Mainz GDH experiment~
\cite{pr01} (related to the double-polarization 
quantity $E$~\cite{bds}) should also have an impact on the 
analysis when combined with 4$^{\circ}$ to 177$^{\circ}$ 
cross section (200 to 790~MeV) and 10$^{\circ}$ to 
160$^{\circ}$ $\Sigma$ beam asymmetry (250 to 440~MeV) 
measurements at MAMI~\cite{le01}.  Beam asymmetry data 
for $\pi^0$ photoproduction below 1100~MeV will also be 
available from GRAAL~\cite{re99}.  Of particular interest 
are the polarized $\pi^0$ photoproduction experiments 
(including the polarization transfers $C_{x'}$ and 
$C_{z'}$ from circularly polarized photons to recoil 
protons) above 800~MeV  carried out in Hall~A of JLab~
\cite{gi00}.  From Brookhaven, we expect final LEGS 
$\Sigma$ beam asymmetries around the $\Delta$ resonance~
\cite{sa95} and new radiative capture cross sections 
which have been taken at BNL--AGS using the Crystal Ball 
Spectrometer (E913/914) at $p_{\pi}$ from 400 to 
750~MeV/c (E$_{\gamma}$ = 430 -- 780~MeV)~\cite{aziz}.

\section{Multipole and Photo-decay Amplitudes}
\label{sec:ampl}

\subsection{Analysis}
\label{sec:anal}

Fits to the expanded database were first attempted
within the formalism we have used and described 
previously~\cite{ar90,li93,ar96}. Multipoles were
parameterized using the form
\begin{eqnarray}
M = \left( Born + A\right) \left( 1 + i T_{\pi N}
\right) + B T_{\pi N}
\label{1}\end{eqnarray}
with $T_{\pi N}$ being the associated elastic 
pion-nucleon $T-$matrix, and the terms $A$ and $B$ 
being purely phenomenological polynomials with 
the correct threshold properties.  As in our most 
recent analysis~\cite{ar96}, some multipoles were 
allowed an additional overall phase $e^{i\Phi}$, 
where the angle $\Phi$ was proportional to (${\rm Im} 
T_{\pi N} - T_{\pi N}^2$).  This form satisfied 
Watson's theorem for elastic $\pi N$ amplitudes~
\cite{wa54} while exploiting the undetermined phase 
for $\pi N$ inelastic amplitudes.  For $T_{\pi N}$,
we utilized our most recent fit (SM02) to elastic 
scattering data~\cite{pin2}. 

New and precise $\Sigma$ measurements proved difficult
to describe, using this choice of phenomenology.
Searching for a more successful form, we found an 
improved description was possible if the dependence on 
(${\rm Im} T_{\pi N} - T_{\pi N}^2$) was additive 
rather than multiplicative. As a result, we re-fitted 
the full database, removing the overall phase and 
instead added a term of the form
\begin{eqnarray}
(C + i D) \left( {\rm Im} T_{\pi N} - T_{\pi N}^2 \right)
\label{2}\end{eqnarray}
with $C$ and $D$ again being energy-dependent 
polynomials.

The resulting energy-dependent solution (SM02) had a 
$\chi^2$ of 35296 for 17571 ($\pi^0p$, $\pi^+n$, 
$\pi^-p$, and $\pi^0n$) data to 2~GeV.  The overall 
$\chi^2$/data was significantly lower than that 
found in our previously published result 
($\chi^2$/data = 2.4)~\cite{ar96}.  This change is 
partly a reflection of the the database changes 
discussed in Section~\ref{sec:expt}.  Our present 
and previous energy-dependent solutions are compared 
in Table~\ref{tbl1}.  As in previous analyses, we 
used the systematic uncertainty as an overall 
normalization factor for angular distributions
\footnote{For total cross sections and excitation 
          data, we combined statistical and 
          systematic uncertainties in quadrature.}.  
This renormalization freedom provided a significant 
improvement for our best fit results as shown in 
Table~\ref{tbl2}.

In order to see if the inclusion of the full 
existing database had resulted in a bias towards 
older and possibly outdated measurements, we compared
the fit quality versus measurement date.  Generally, 
we found no problem of bias, as illustrated in Table~
\ref{tbl3}, where we have displayed our fit quality 
over the region covering the N(1535) resonance.  
Except for the above mentioned $\Sigma$ measurements, 
over this energy region, our overall fit to the most 
recent measurements (1985$-$Present) is charactorized 
by a $\chi^2$/data of about 1.3.  For all data to 
2~GeV, the date restriction yields a $\chi^2$/data 
near 1.5.  Unfortunately, the modern measurements do 
not completely overlap older data.  Much of the older 
data is required in an analysis extending over the 
full resonance region. 

The very low energy region is complicated by different 
thresholds for $\pi^0p$ and $\pi^+n$ final states.  
While we have obtained a reasonable fit to the 
available $\pi^0p$ differential and total cross 
sections (we fit rather well the new threshold 
TRIUMF--SAL $\pi^+n$ cross sections~\cite{ko99}, and 
Mainz $\pi^0p$ $\Sigma$ at 160~MeV~\cite{sc01}), the 
multipole amplitudes have no cusp built into the 
$\pi^+n$ threshold region.  

Both energy-dependent and single-energy solutions 
(SES) were obtained from fits to the combined $\pi^0p$, 
$\pi^+n$, $\pi^-p$, and $\pi^0n$ databases to 2~GeV.  
In Table~\ref{tbl4}, we compare the energy-dependent 
and single-energy results over the energy bins used in 
these single-energy analyses.  Also listed are the 
number of parameters varied in each single-energy 
solution.  A total of 148 parameters were varied in 
the energy-dependent analysis SM02.  The extended 
database allowed an increase in the number of SES 
versus our previous result~\cite{ar96} over the same 
energy range to 2~GeV.  

\begin{figure}[ht]
\centerline{
\psfig{file=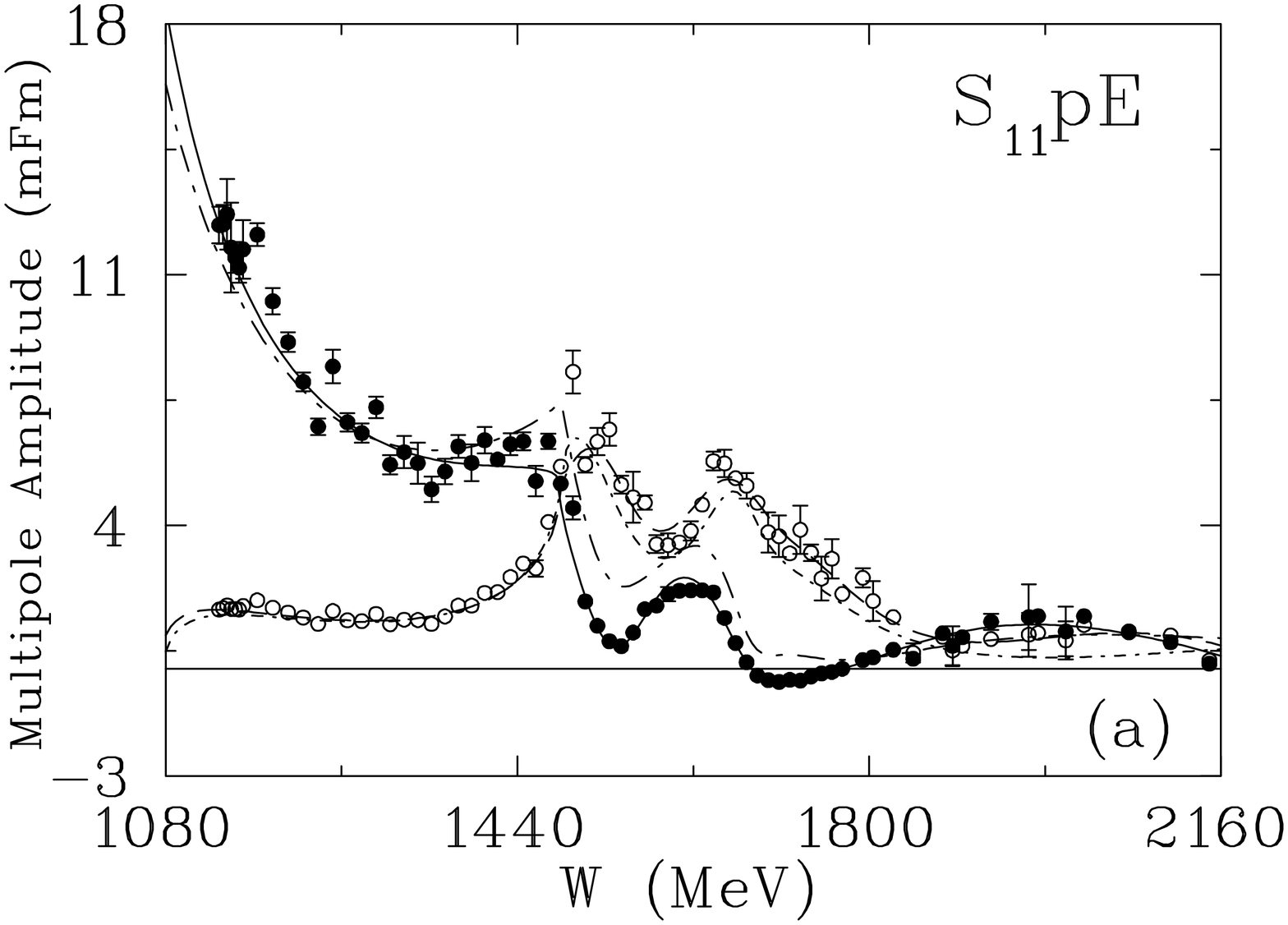,width=3in,clip=,silent=,angle=90}\hfill
\psfig{file=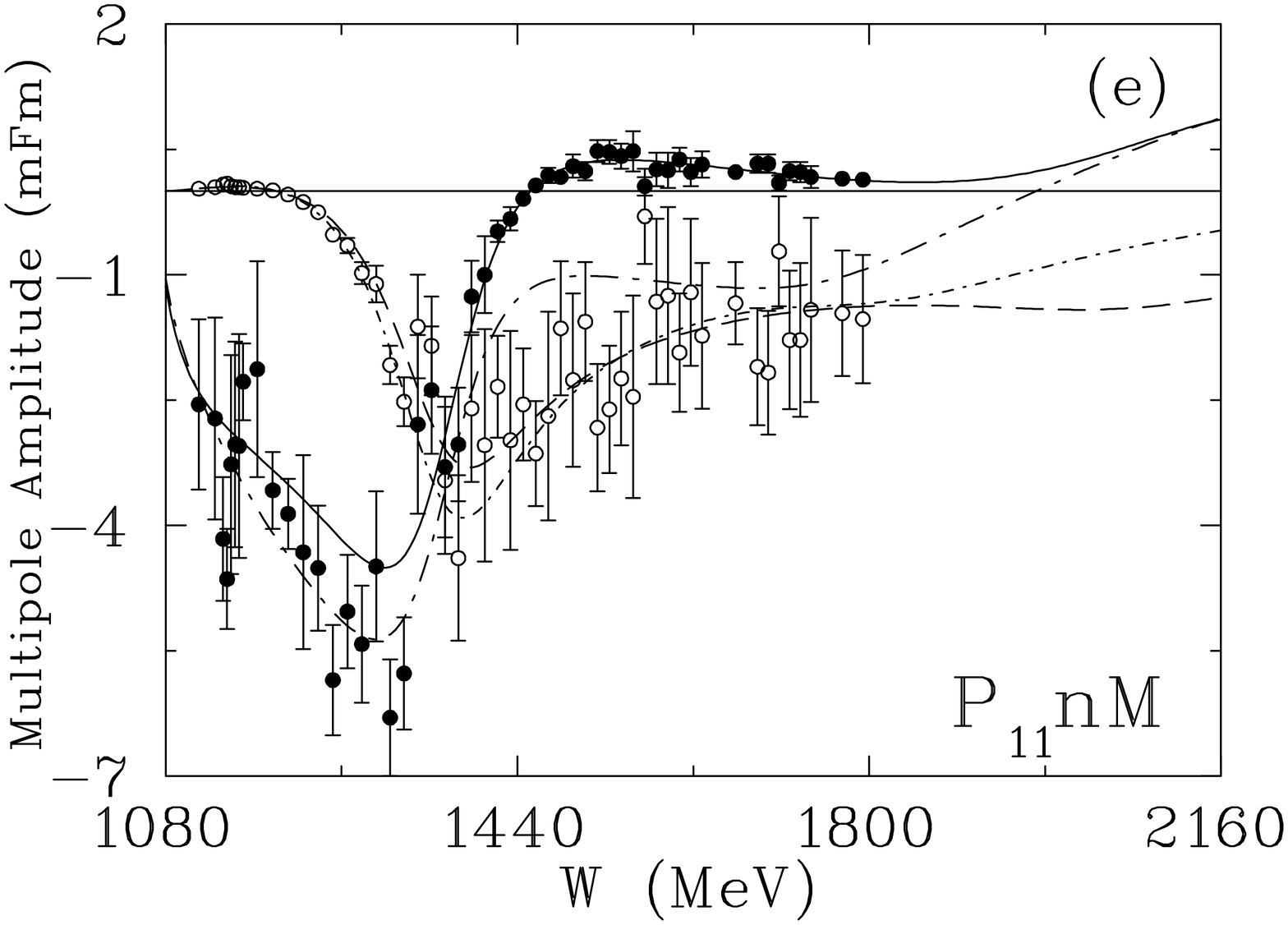,width=3in,clip=,silent=,angle=90}}
\centerline{
\psfig{file=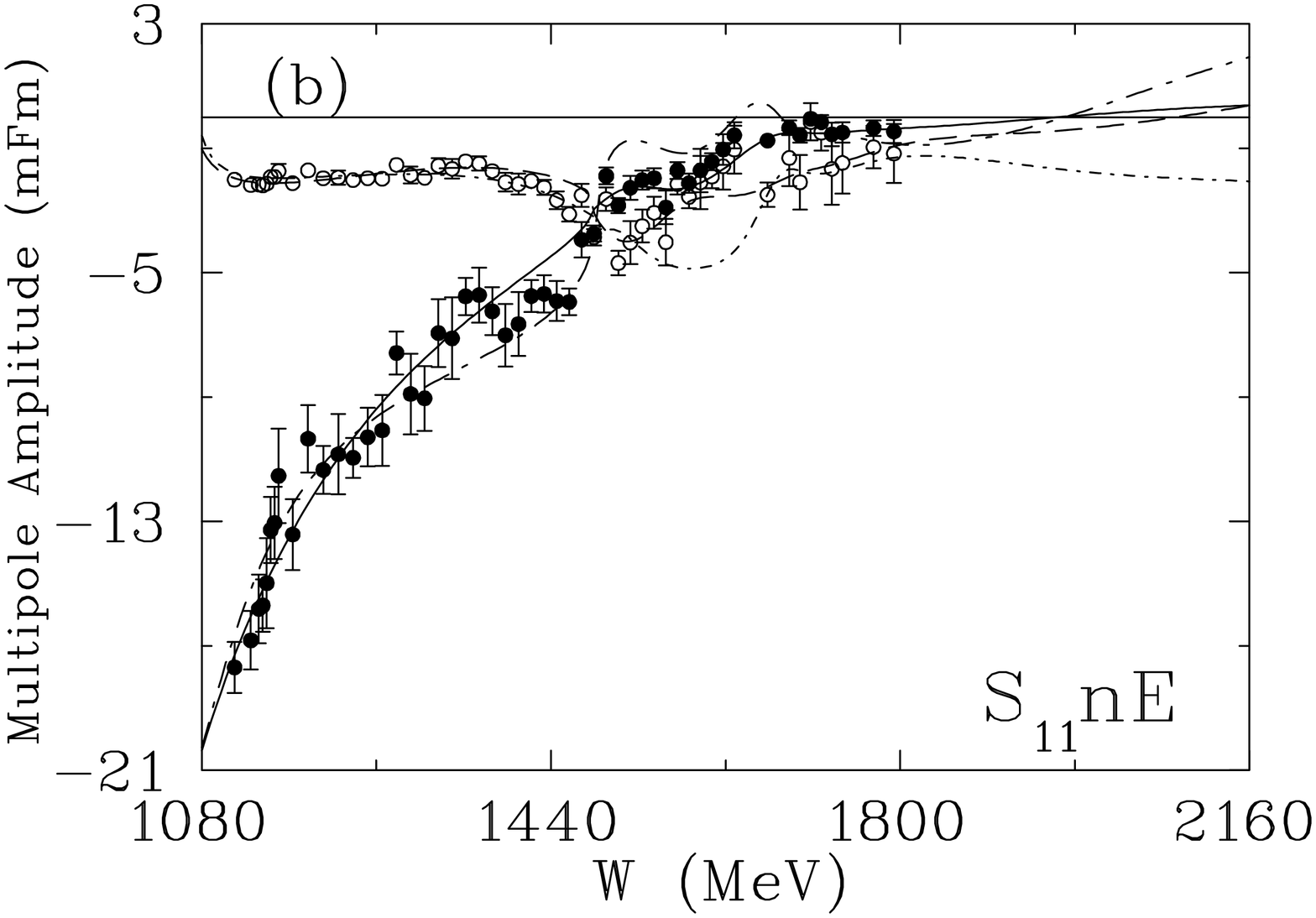,width=3in,clip=,silent=,angle=90}\hfill
\psfig{file=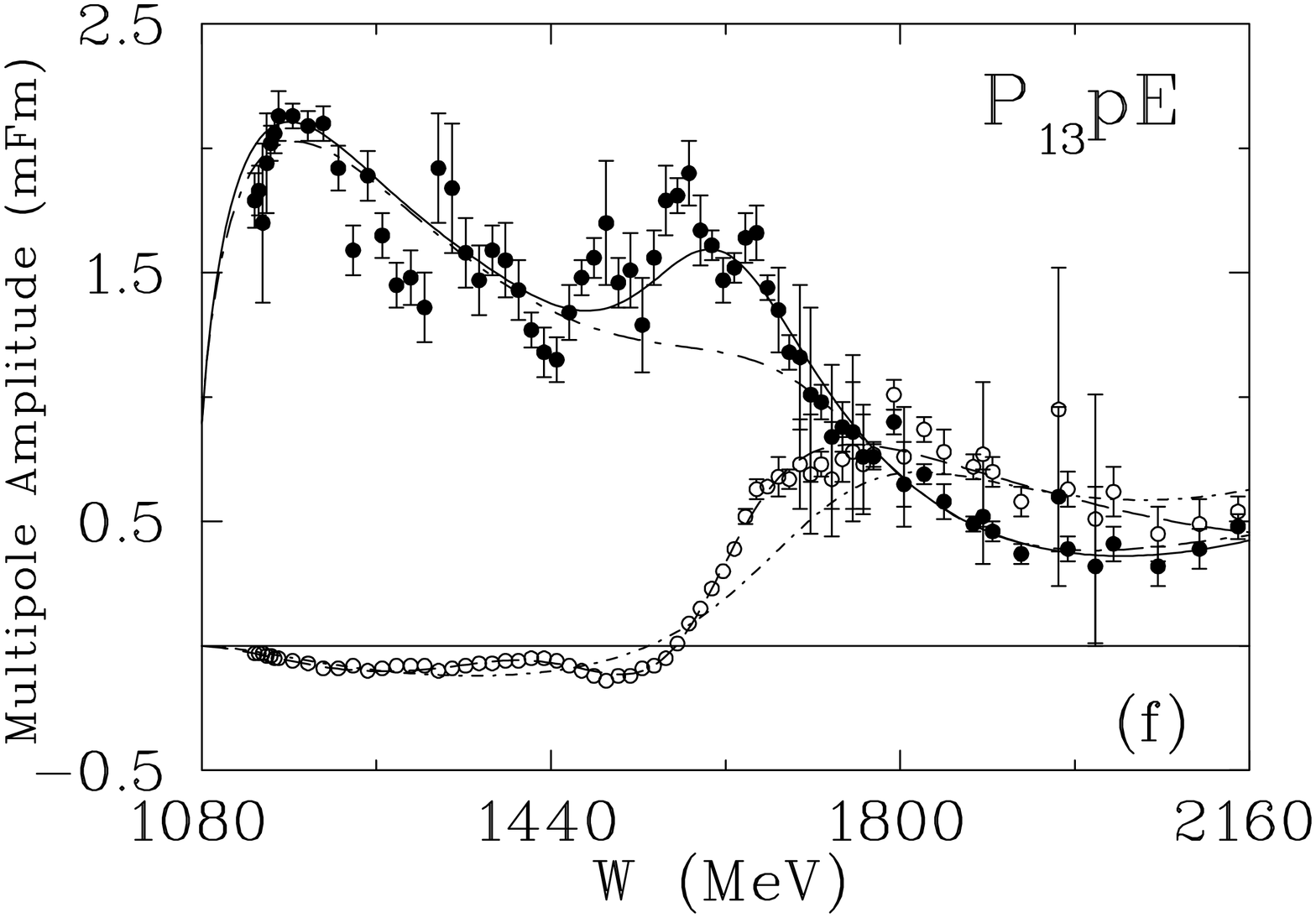,width=3in,clip=,silent=,angle=90}}
\centerline{
\psfig{file=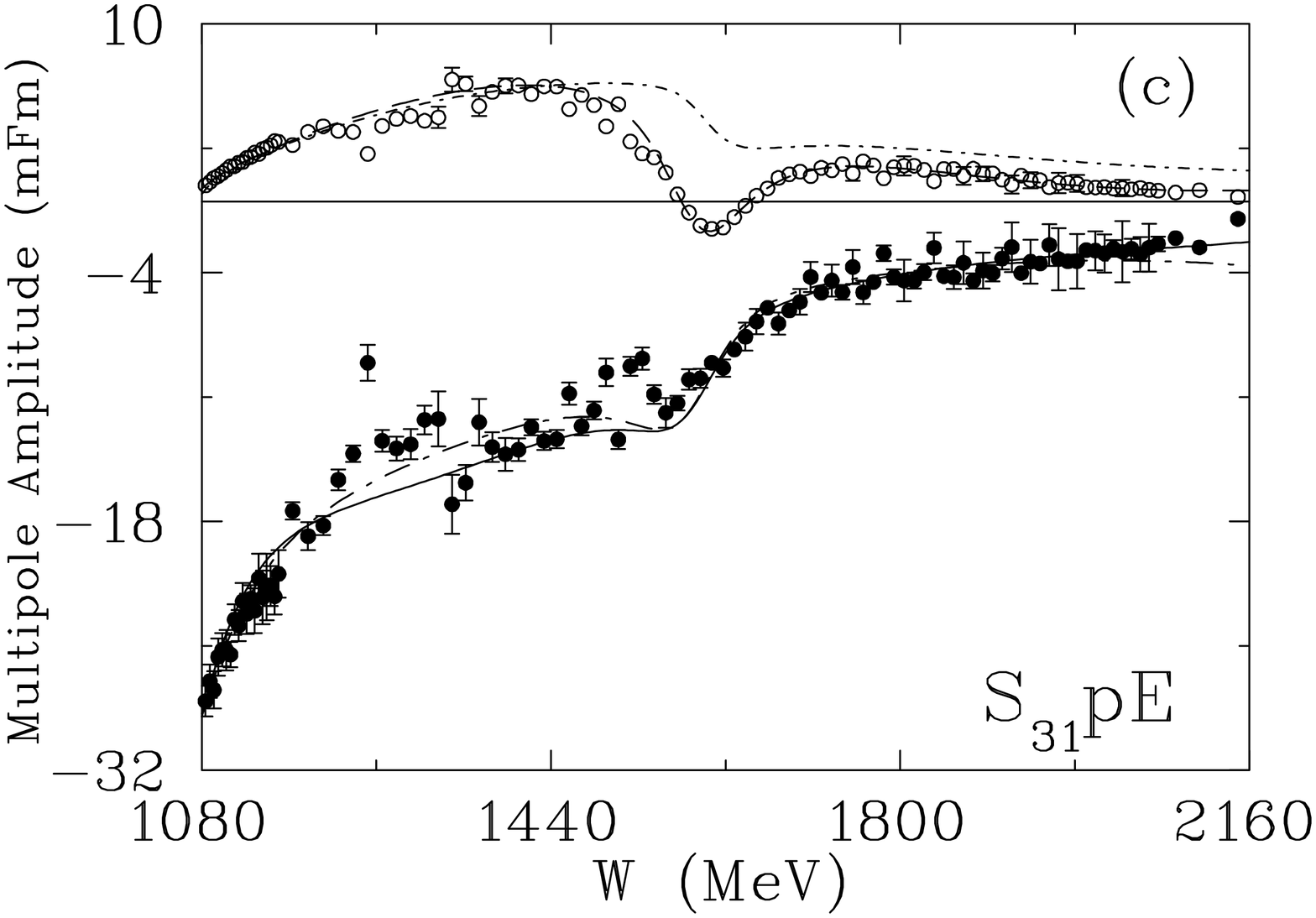,width=3in,clip=,silent=,angle=90}\hfill
\psfig{file=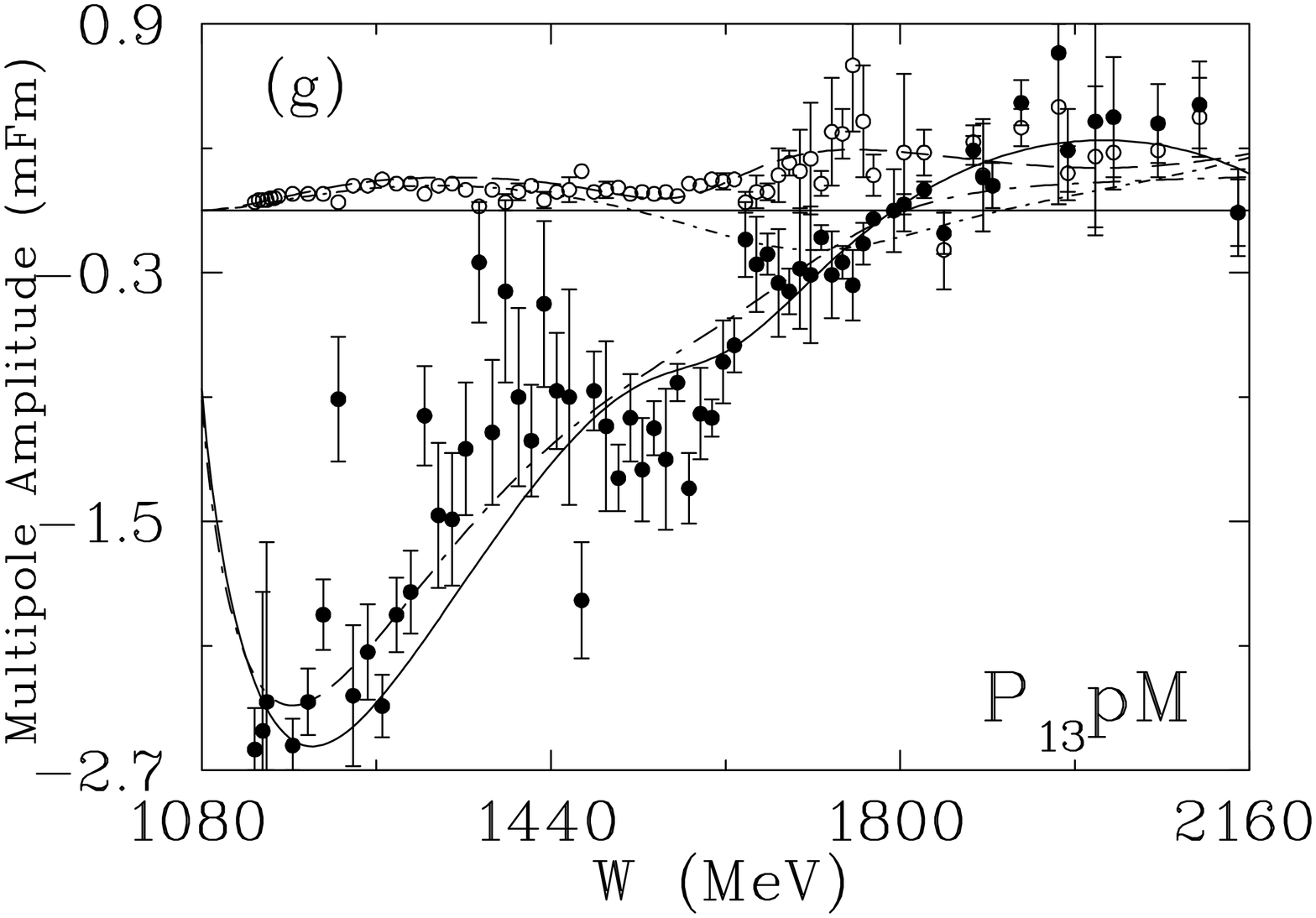,width=3in,clip=,silent=,angle=90}}
\centerline{
\psfig{file=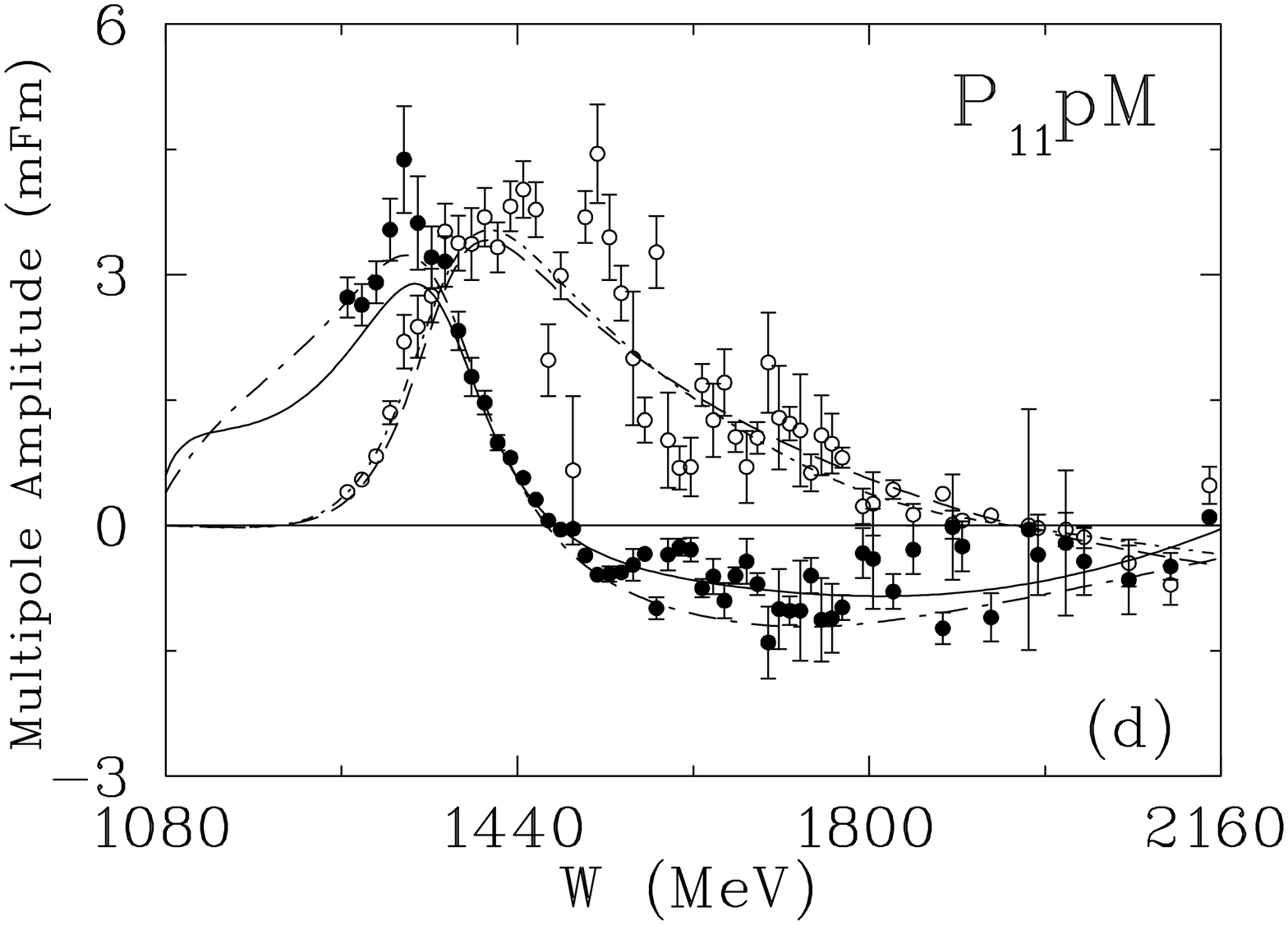,width=3in,clip=,silent=,angle=90}\hfill
\psfig{file=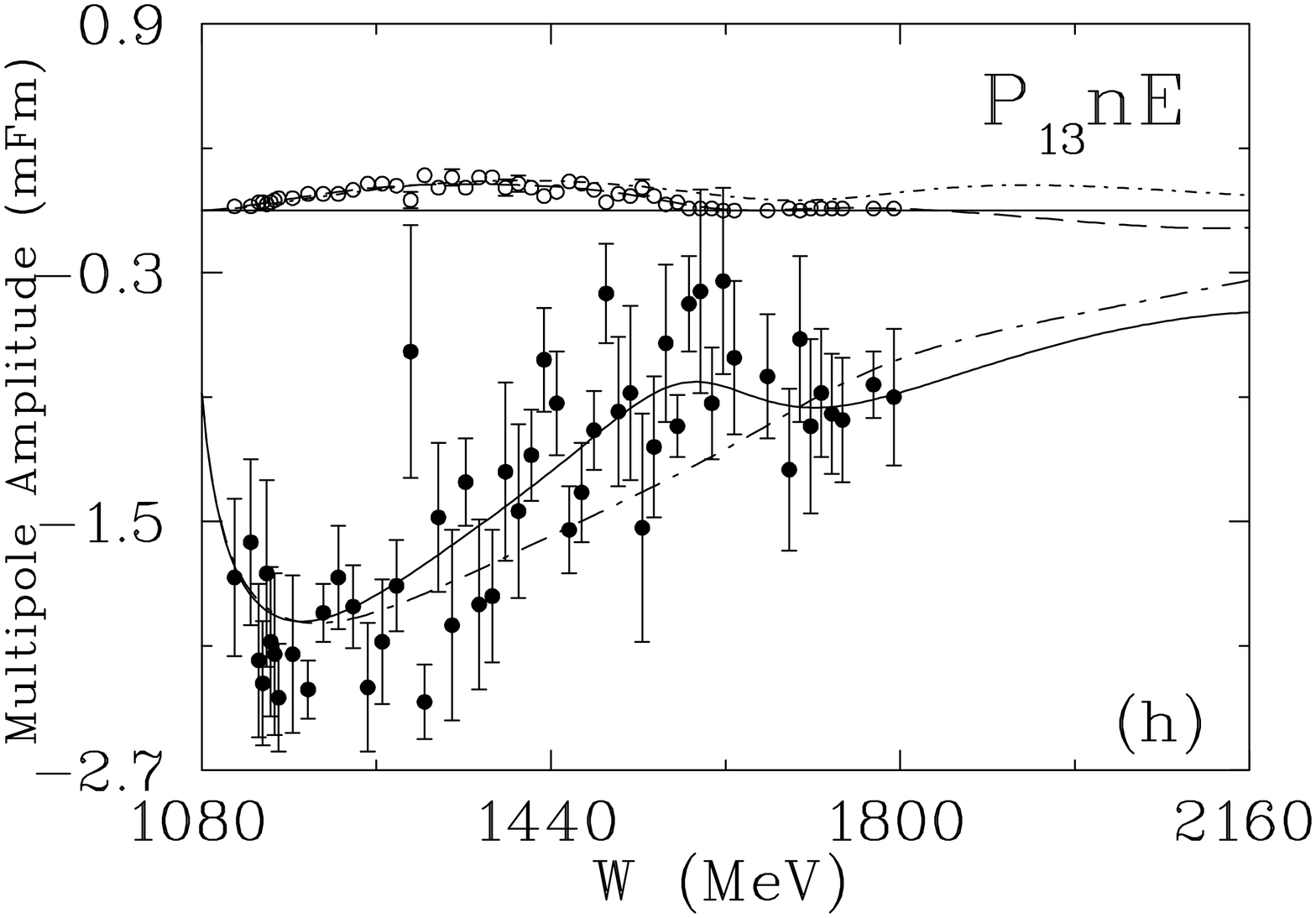,width=3in,clip=,silent=,angle=90}}
\vspace{3mm}
\centerline{
\psfig{file=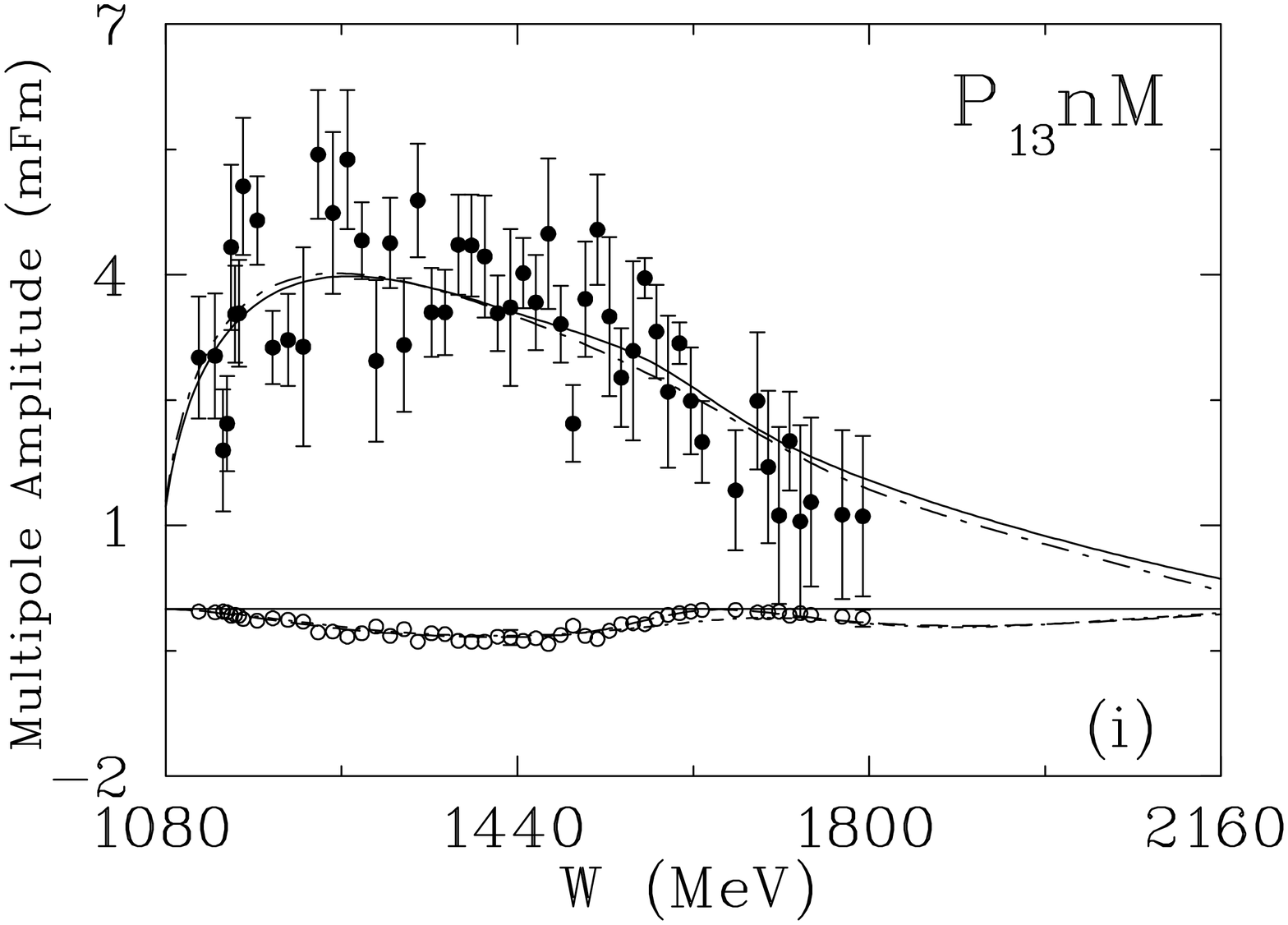,width=3in,clip=,silent=,angle=90}\hfill
\psfig{file=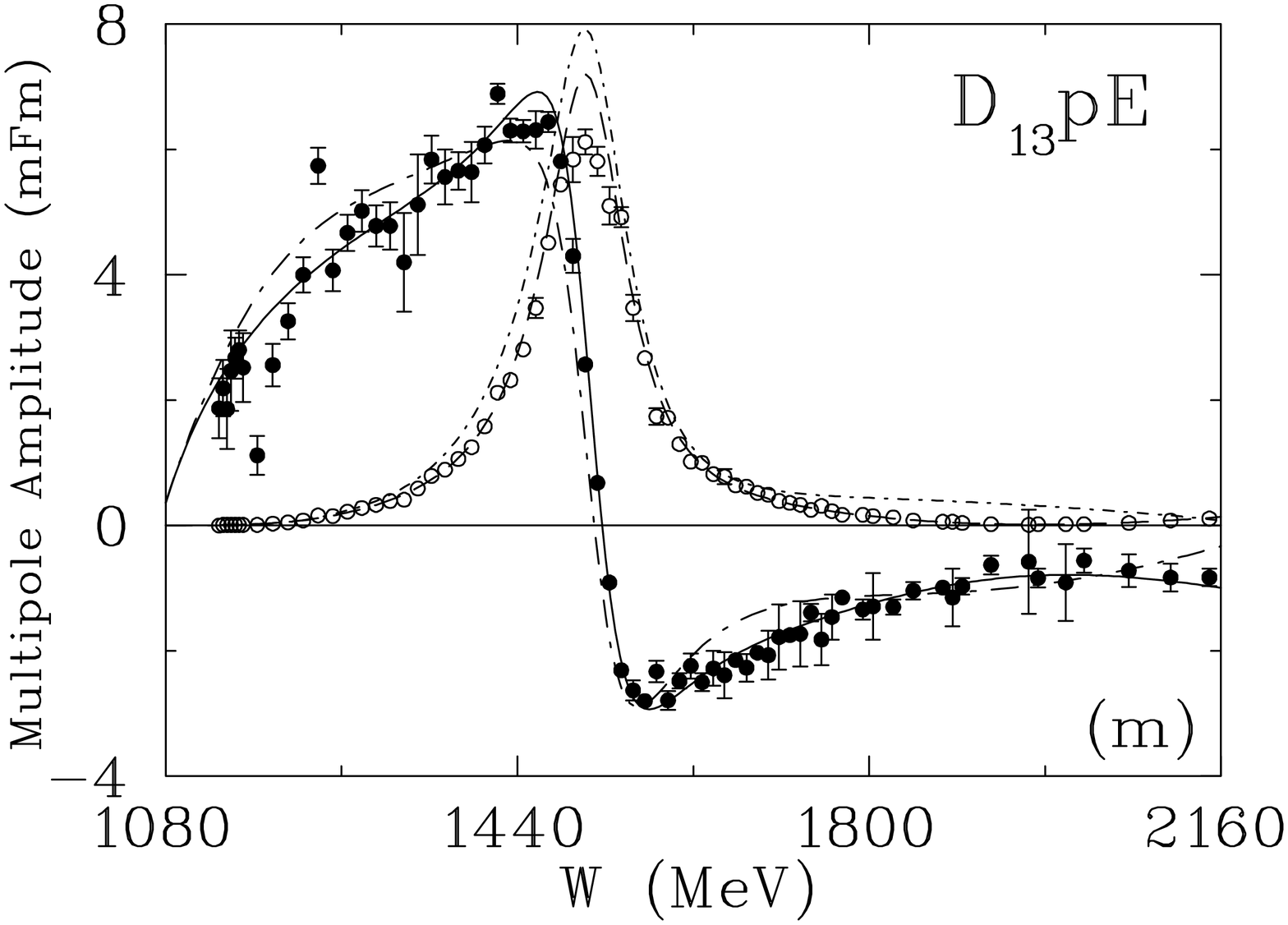,width=3in,clip=,silent=,angle=90}}
\centerline{
\psfig{file=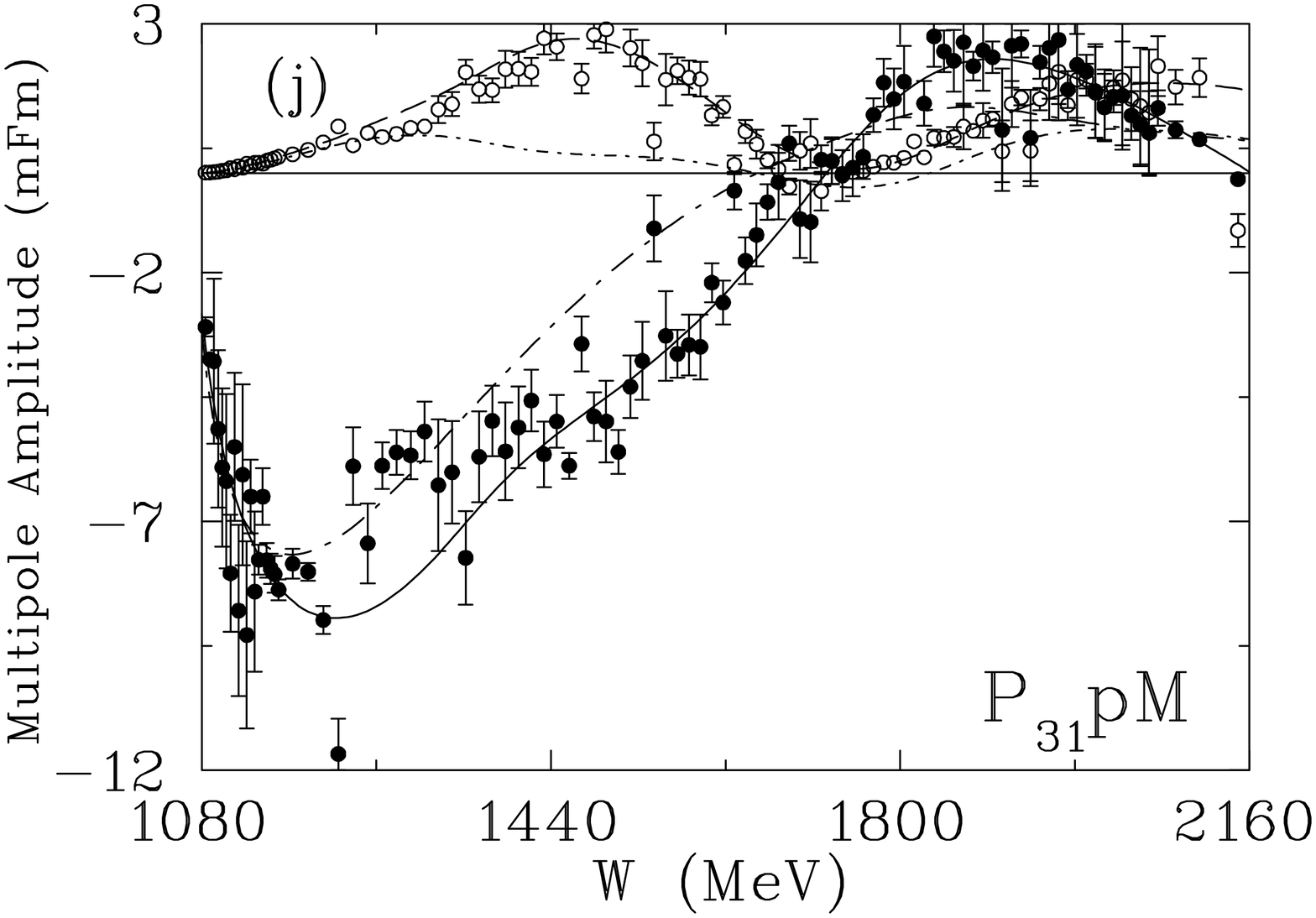,width=3in,clip=,silent=,angle=90}\hfill
\psfig{file=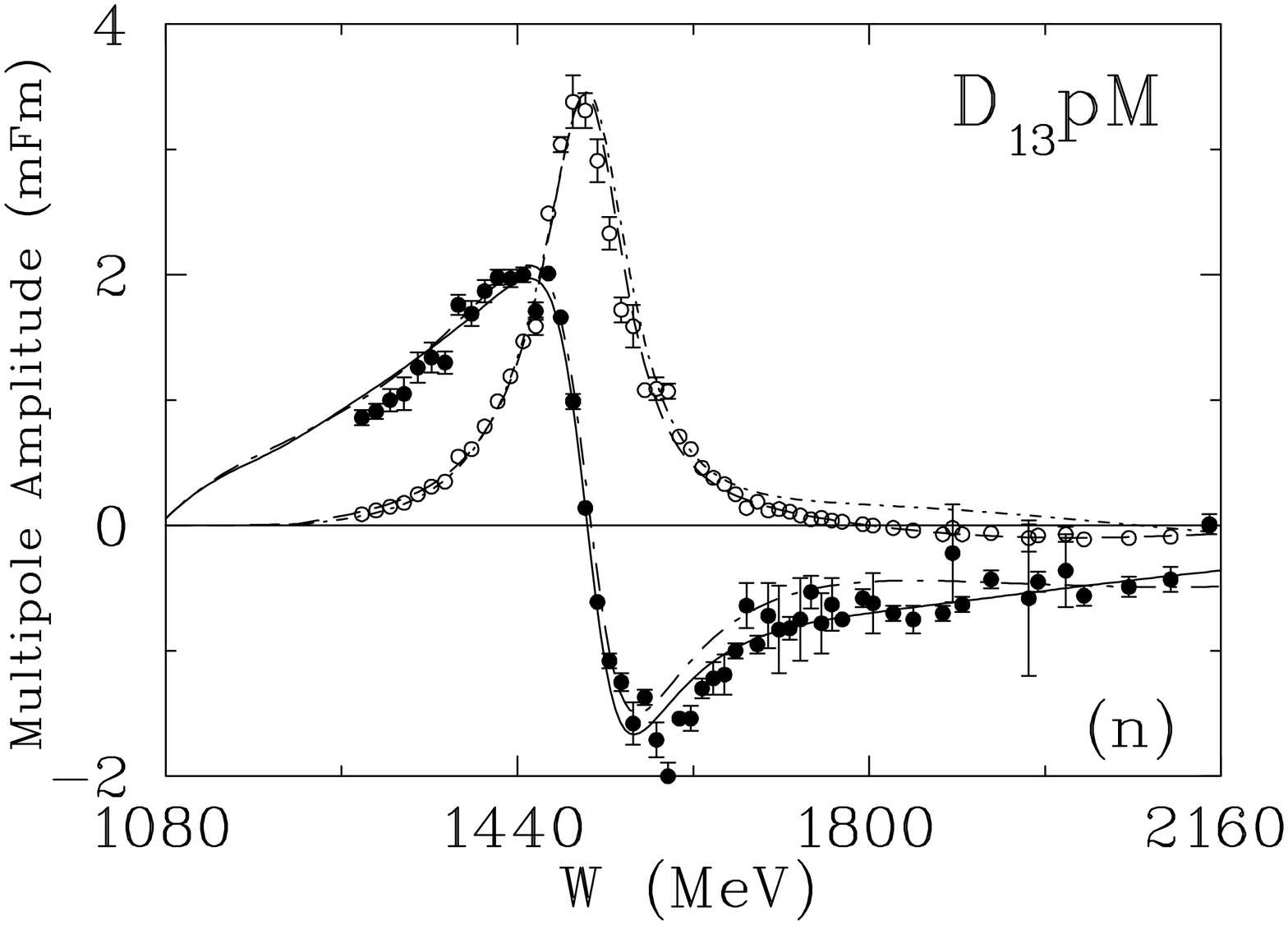,width=3in,clip=,silent=,angle=90}}
\centerline{
\psfig{file=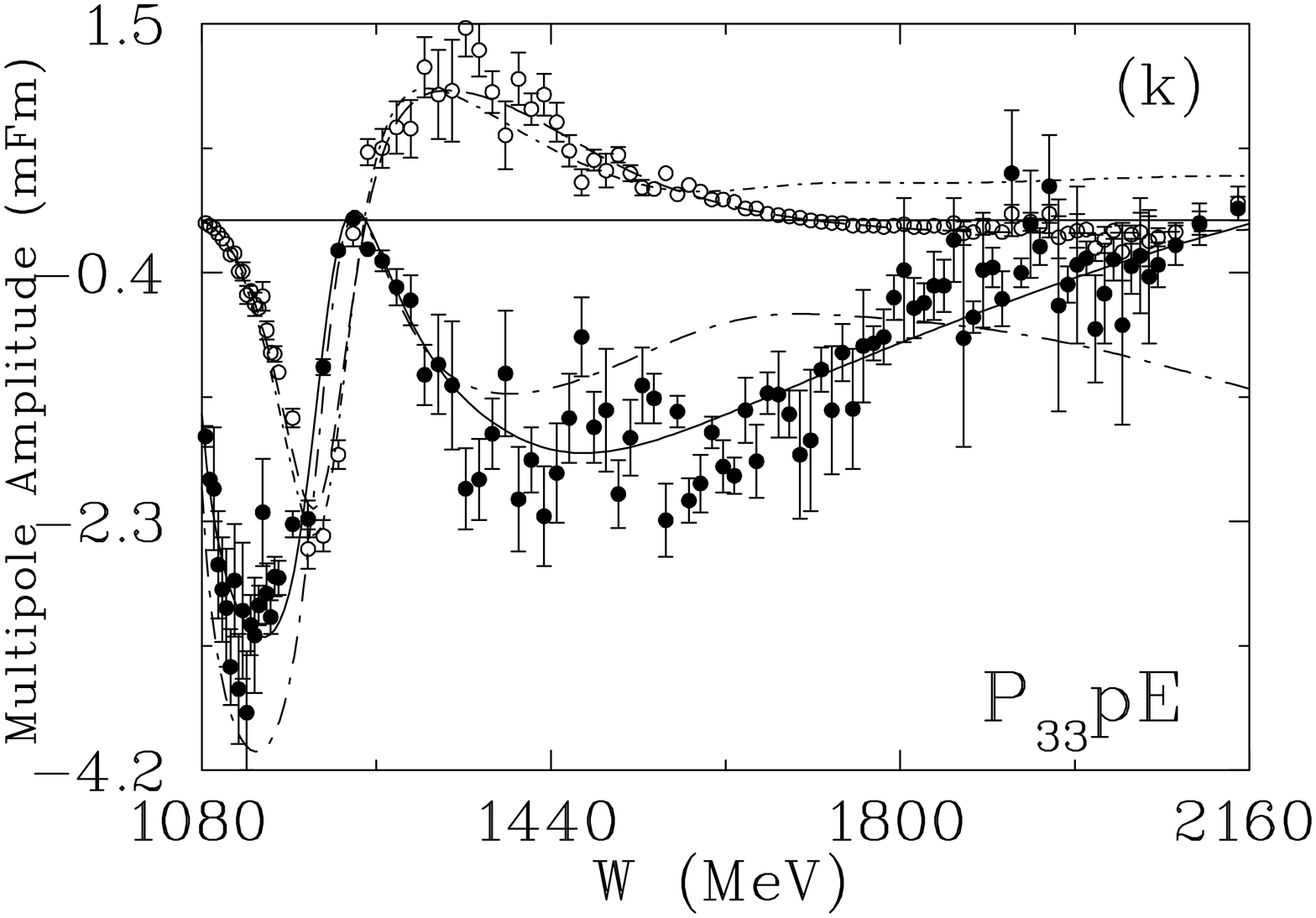,width=3in,clip=,silent=,angle=90}\hfill
\psfig{file=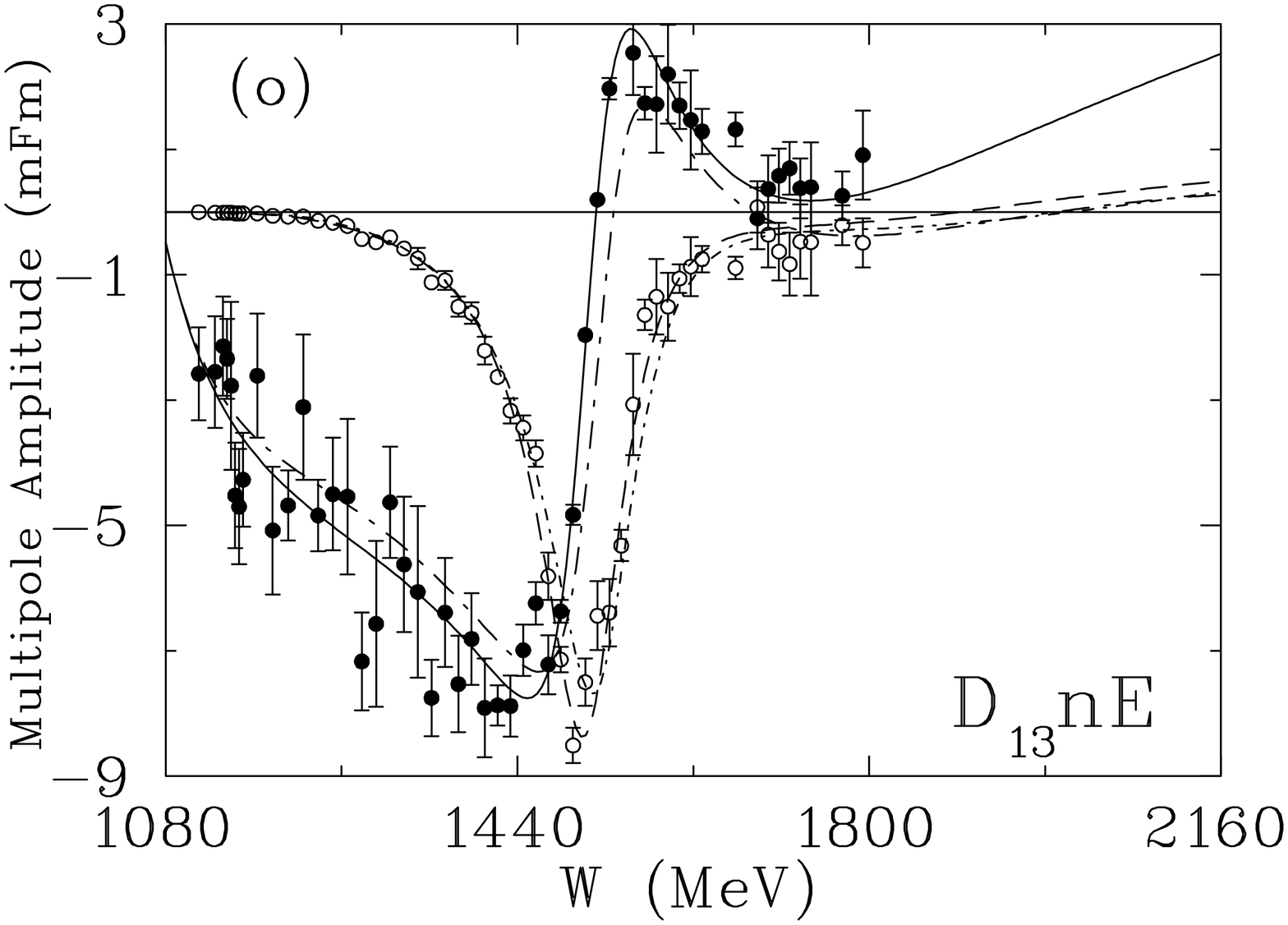,width=3in,clip=,silent=,angle=90}}
\centerline{
\psfig{file=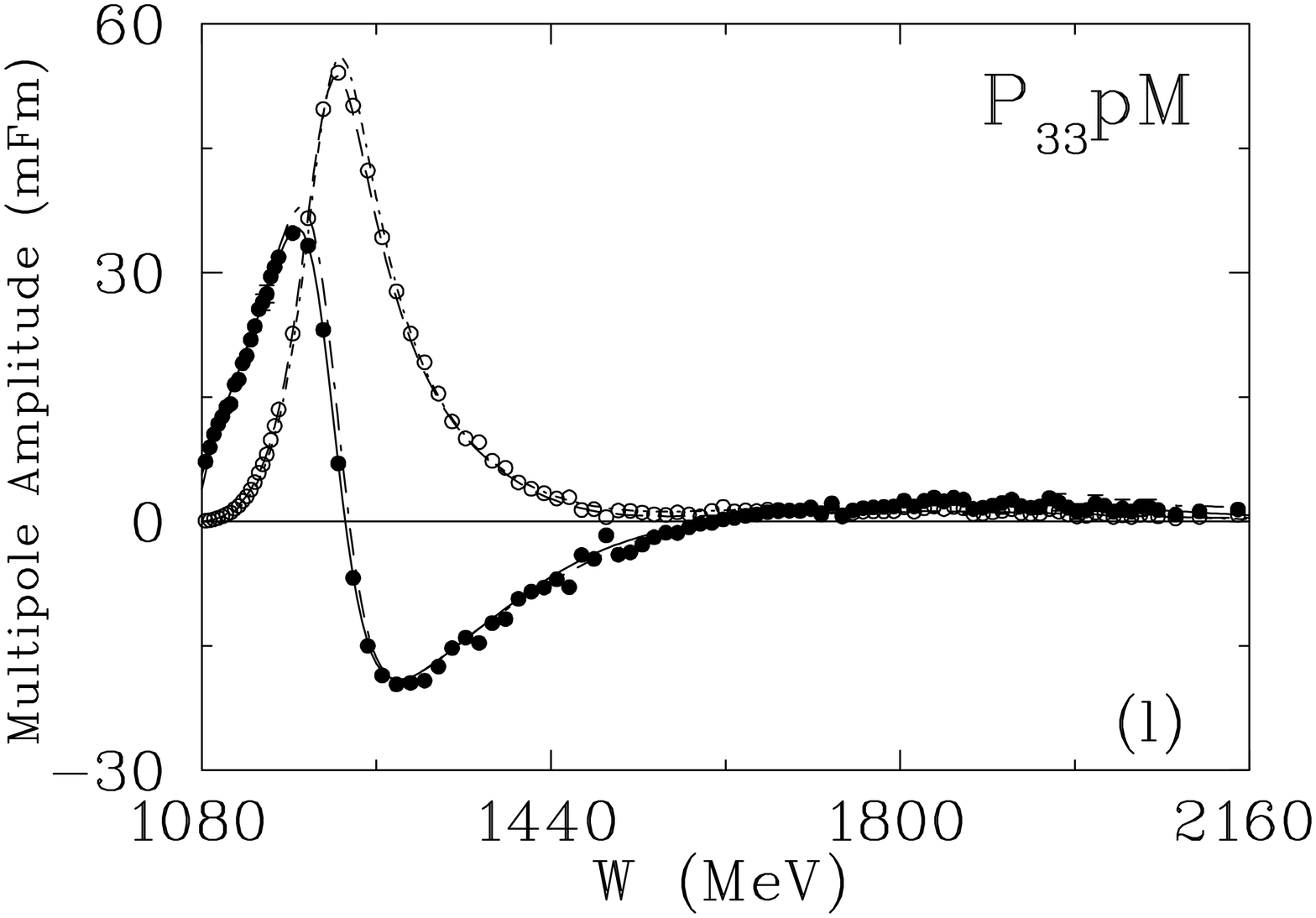,width=3in,clip=,silent=,angle=90}\hfill
\psfig{file=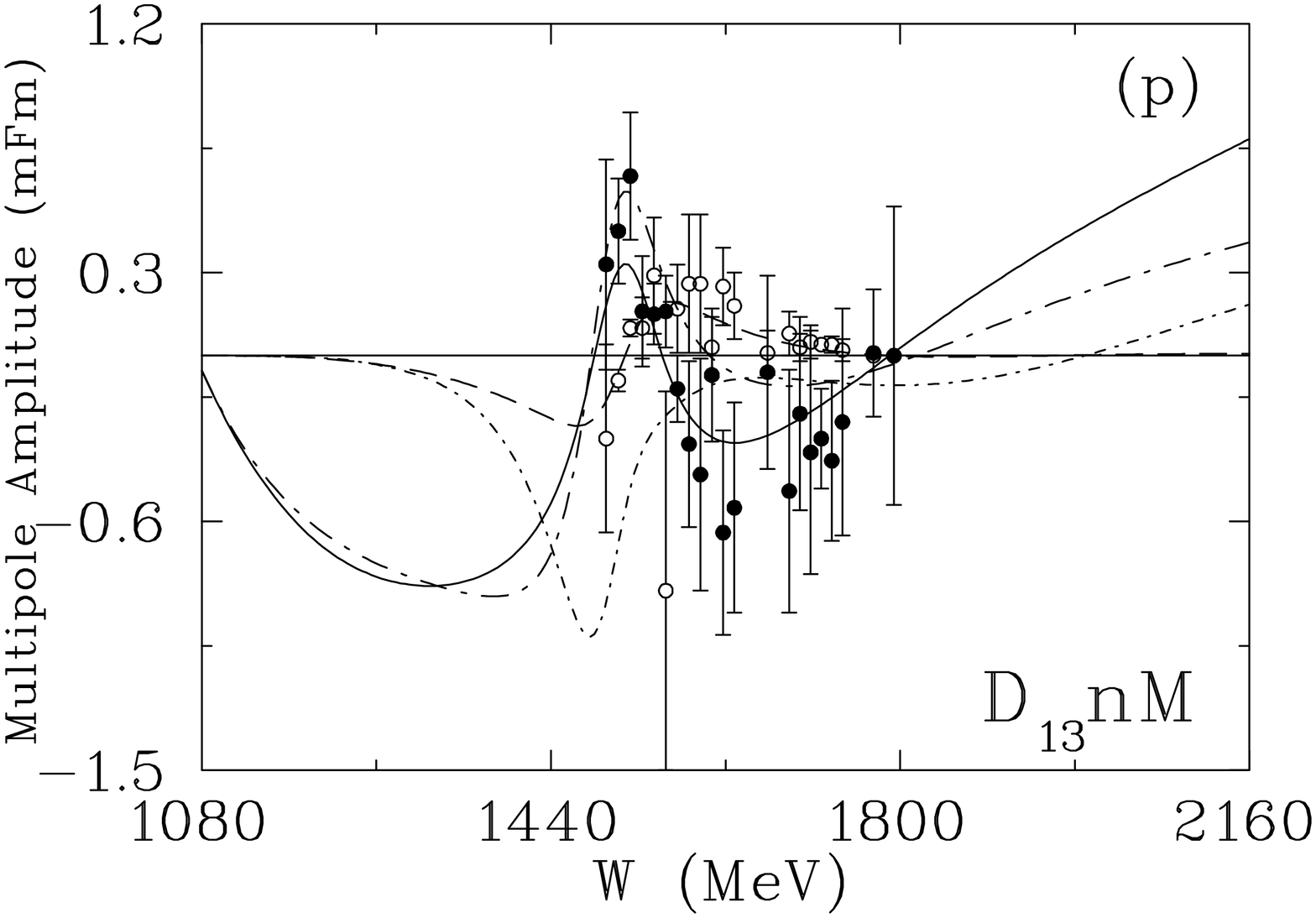,width=3in,clip=,silent=,angle=90}}
\vspace{3mm}
\centerline{
\psfig{file=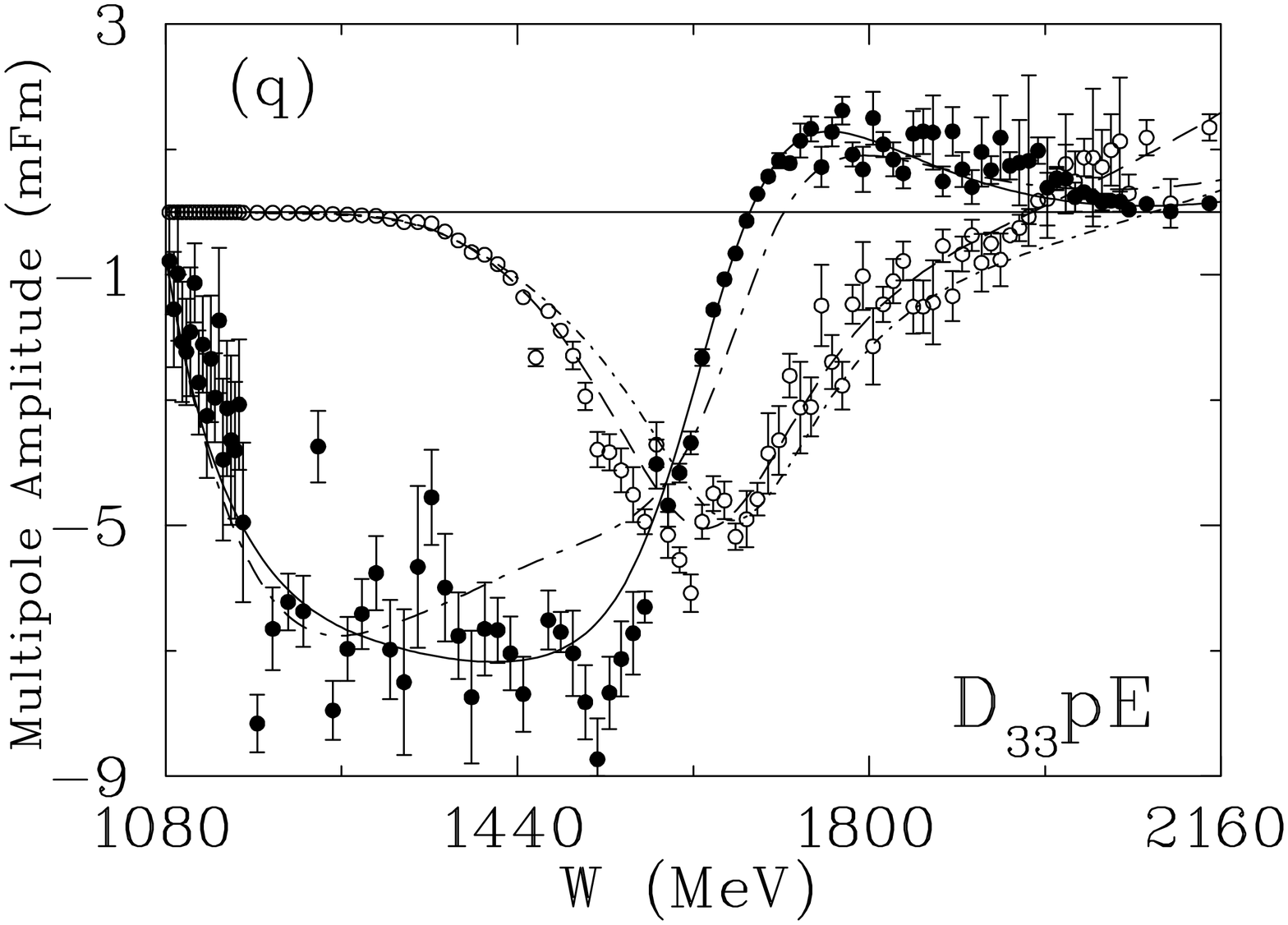,width=3in,clip=,silent=,angle=90}\hfill
\psfig{file=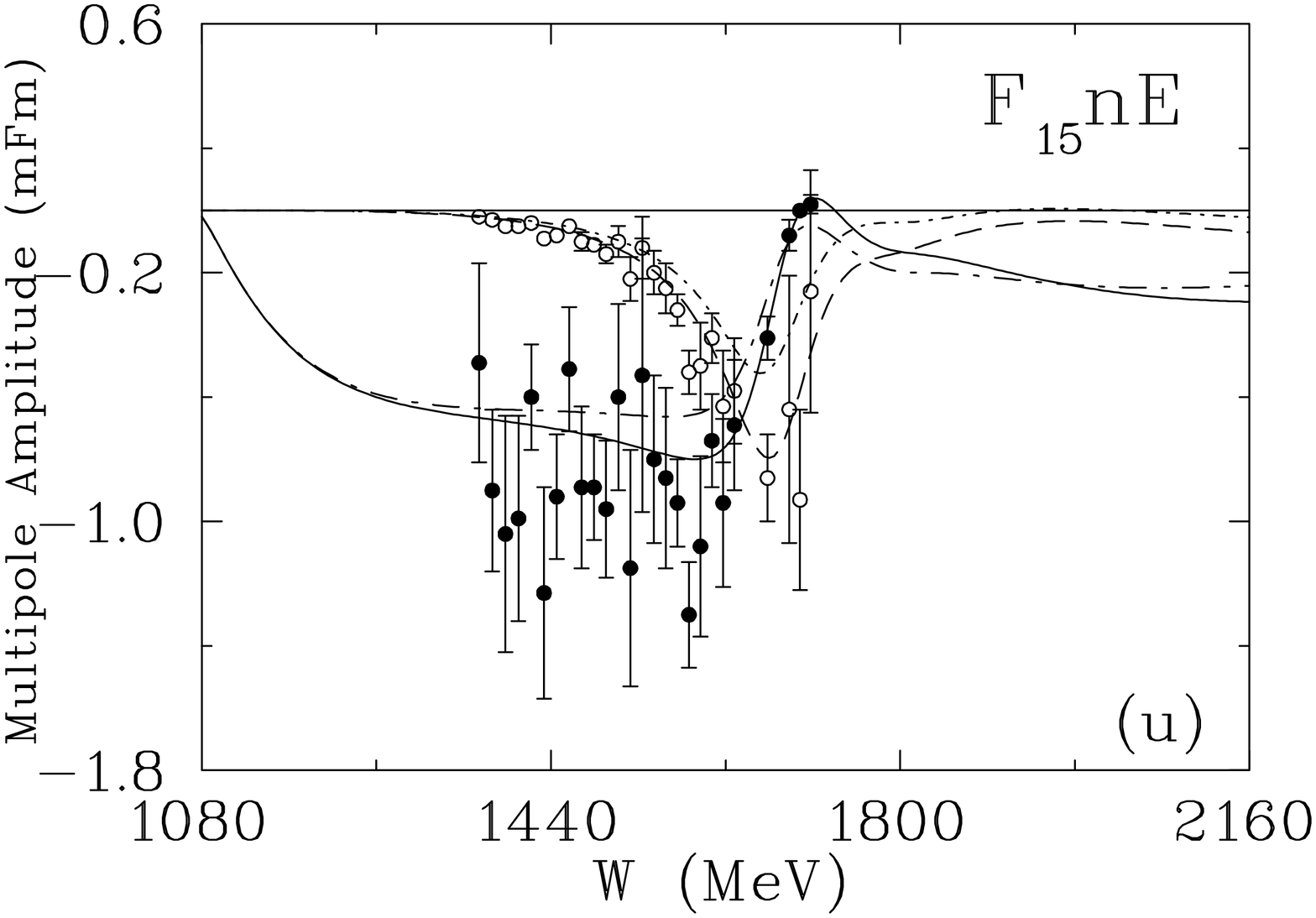,width=3in,clip=,silent=,angle=90}}
\centerline{
\psfig{file=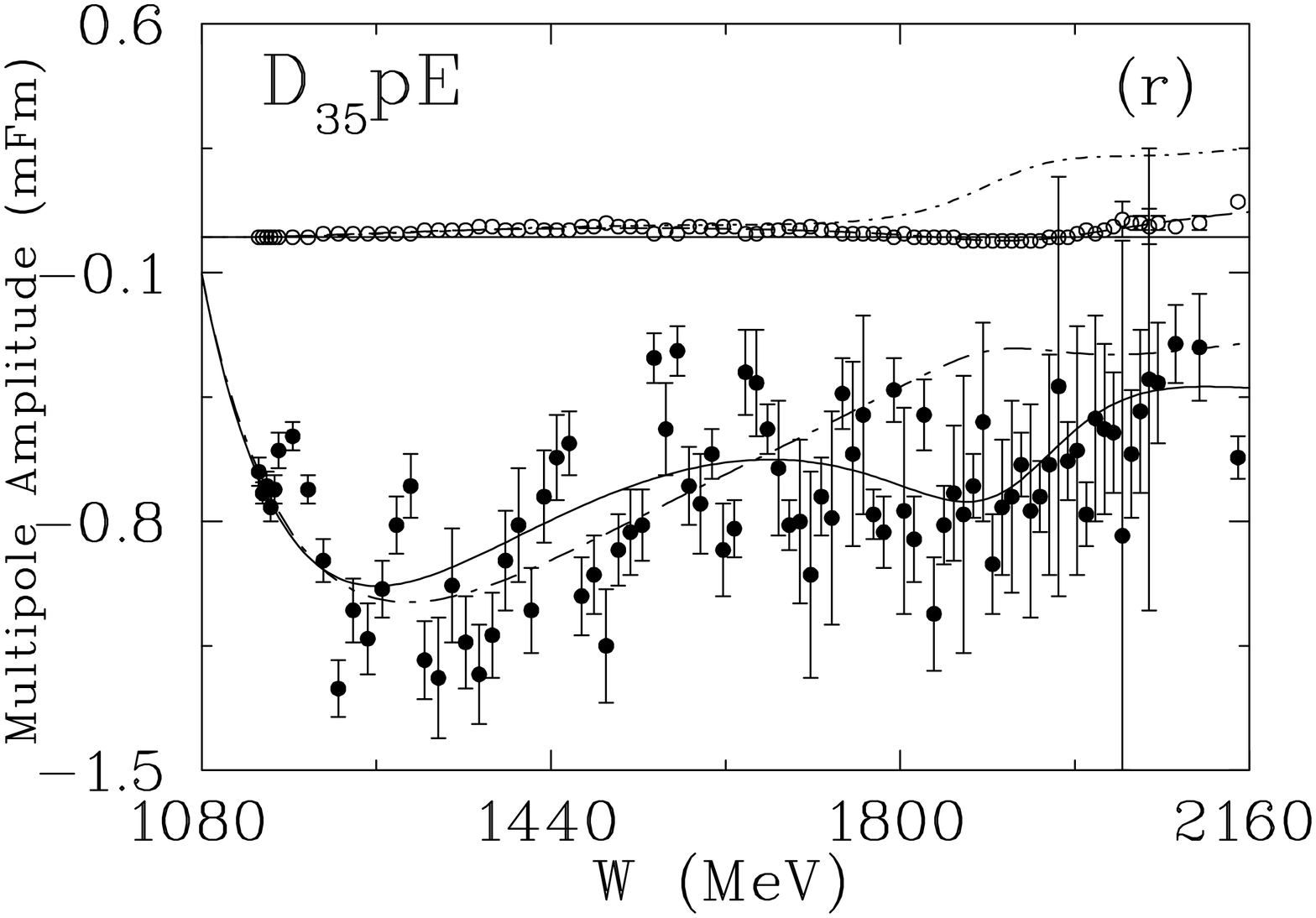,width=3in,clip=,silent=,angle=90}\hfill
\psfig{file=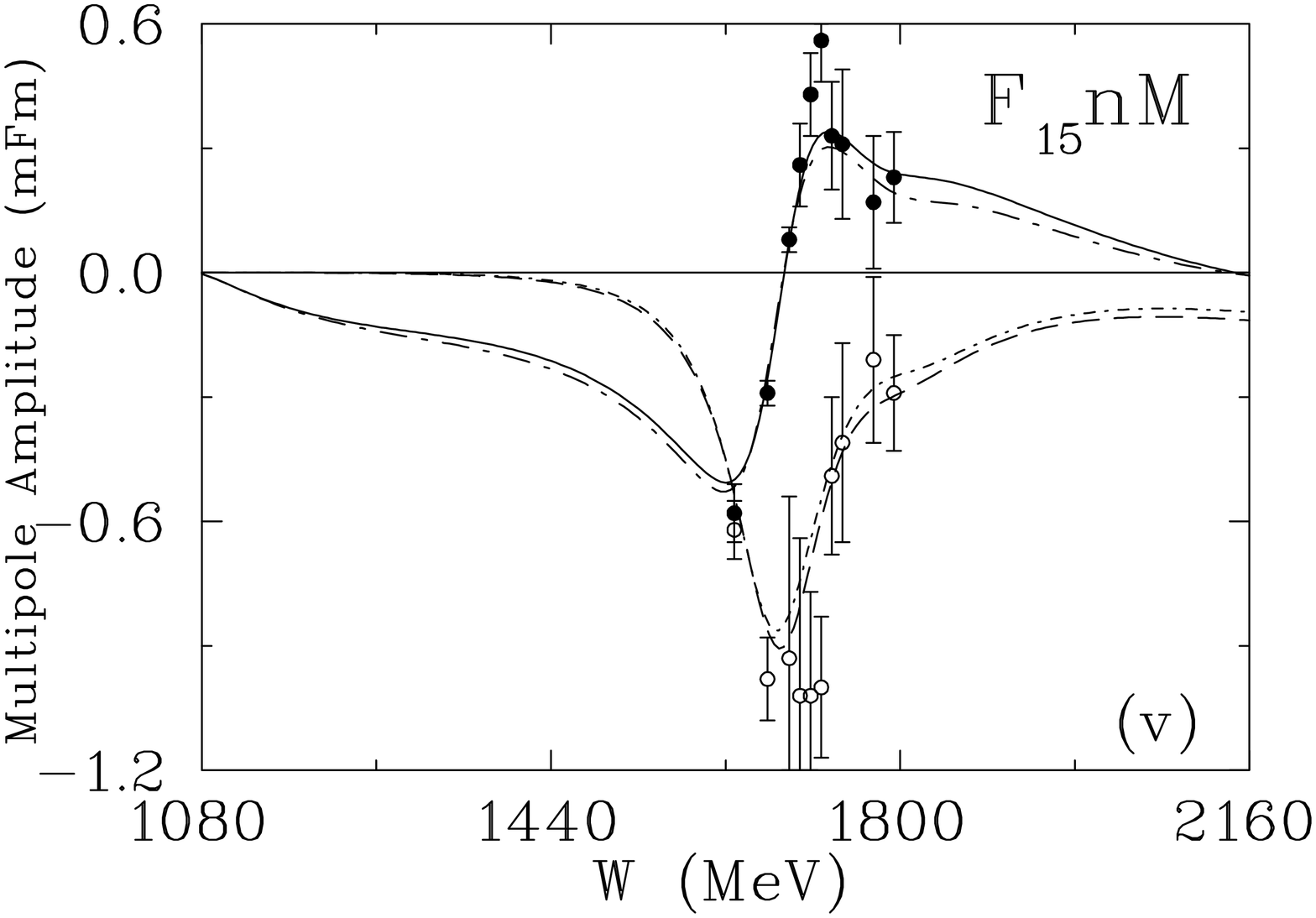,width=3in,clip=,silent=,angle=90}}
\centerline{
\psfig{file=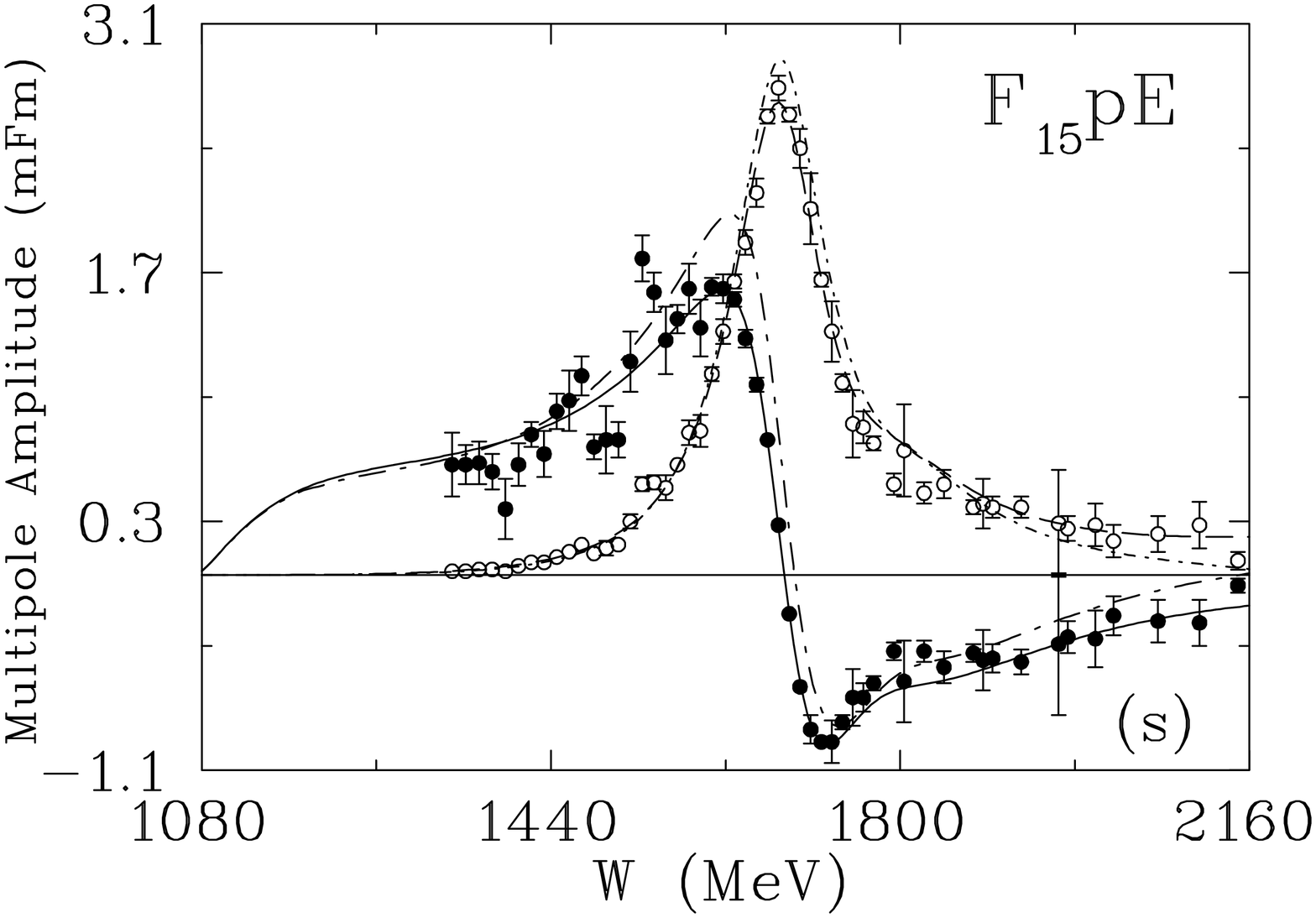,width=3in,clip=,silent=,angle=90}\hfill
\psfig{file=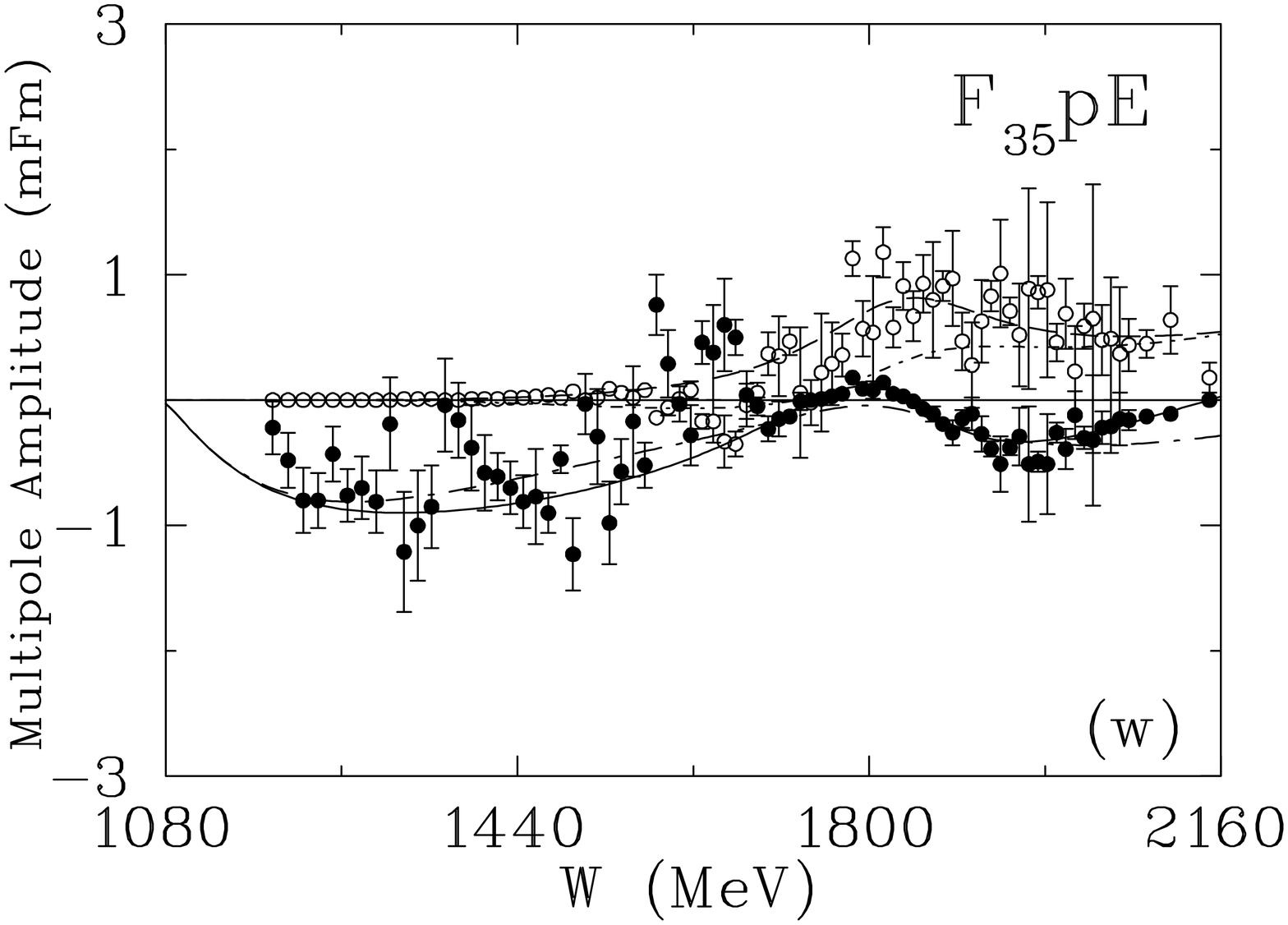,width=3in,clip=,silent=,angle=90}}
\centerline{
\psfig{file=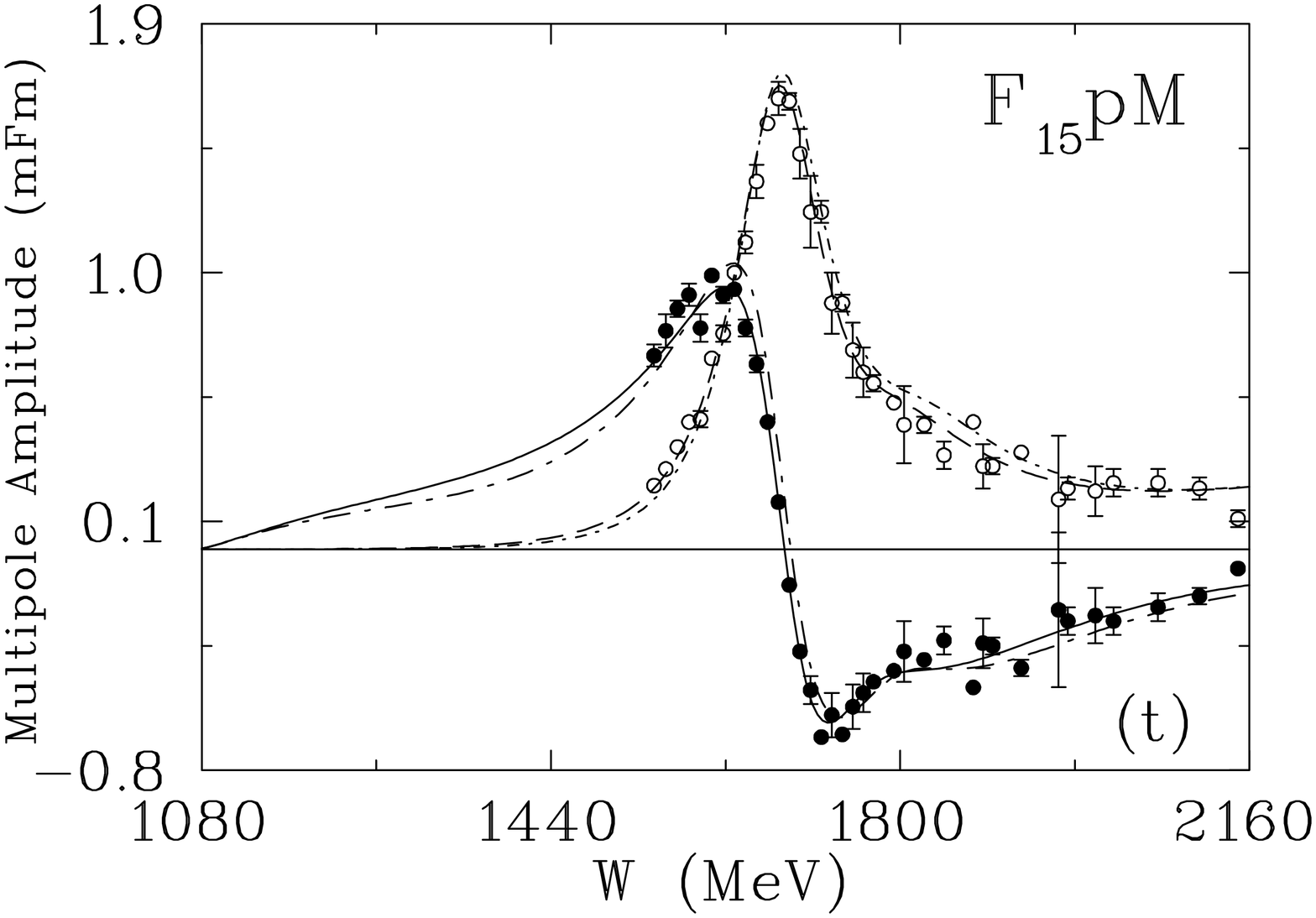,width=3in,clip=,silent=,angle=90}\hfill
\psfig{file=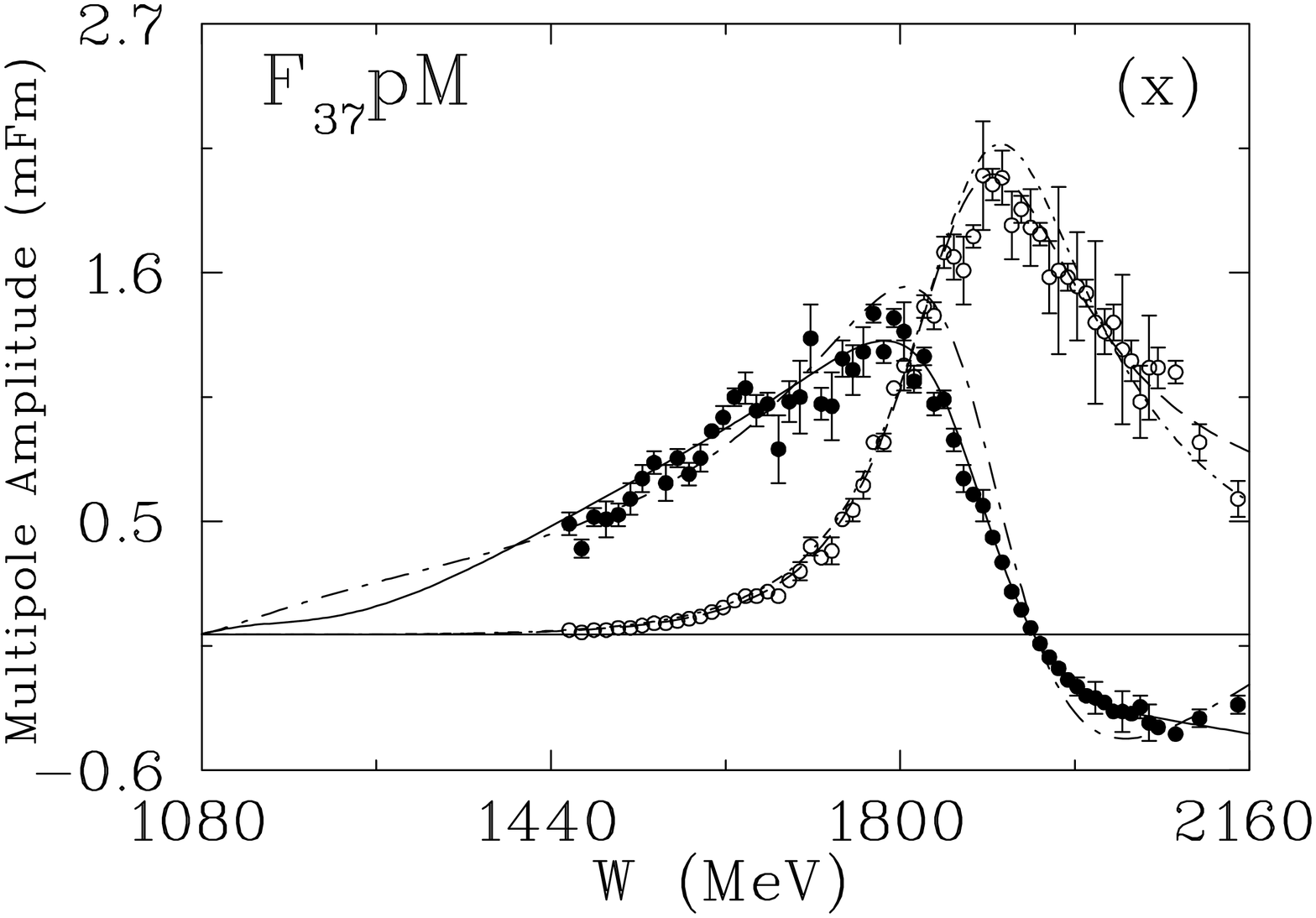,width=3in,clip=,silent=,angle=90}}
\vspace{3mm}
\caption[fig2]{\label{g2}
      Partial-wave amplitudes (L$_{2I, 2J}$) from 
      threshold to $E_{\gamma}$ = 2~GeV.  Solid (dashed) 
      curves give the real (imaginary) parts of amplitudes 
      corresponding to the SM02 solution.  The real 
      (imaginary) parts of single-energy solutions are 
      plotted as filled (open) circles.  The previous 
      SM95 solution \protect\cite{ar96} is plotted with 
      long dash-dotted (real part) and short dash-dotted 
      (imaginary part) lines.  Plotted are the multipole 
      amplitudes
      (a) $\rm _pE_{0+}^{1/2}$, (b) $\rm _nE_{0+}^{1/2}$, 
      (c) $\rm _pE_{0+}^{3/2}$, (d) $\rm _pM_{1-}^{1/2}$,
      (e) $\rm _nM_{1-}^{1/2}$, (f) $\rm _pE_{1+}^{1/2}$, 
      (g) $\rm _pM_{1+}^{1/2}$, (h) $\rm _nE_{1+}^{1/2}$, 
      (i) $\rm _nM_{1+}^{1/2}$, (j) $\rm _pM_{1-}^{3/2}$,
      (k) $\rm _pE_{1+}^{3/2}$, (l) $\rm _pM_{1+}^{3/2}$, 
      (m) $\rm _pE_{2-}^{1/2}$, (n) $\rm _pM_{2-}^{1/2}$, 
      (o) $\rm _nE_{2-}^{1/2}$, (p) $\rm _nM_{2-}^{1/2}$,
      (q) $\rm _pE_{2-}^{3/2}$, (r) $\rm _pE_{2+}^{3/2}$, 
      (s) $\rm _pE_{3-}^{1/2}$, (t) $\rm _pM_{3-}^{1/2}$, 
      (u) $\rm _nE_{3-}^{1/2}$, (v) $\rm _nM_{3-}^{1/2}$,
      (w) $\rm _pE_{3-}^{3/2}$, and (x) $\rm _pM_{3+}^{3/2}$.  
      The subscript p (n) denotes a proton (neutron) target.}
\end{figure}

Fig.~\ref{g2} is a plot of the energy dependent fits
SM02 and SM95 over the full energy region.  The SES 
are also shown with uncertainties coming from the 
error matrix.  In the SES fits, initial values for 
the partial-wave amplitudes and their (fixed) energy 
derivatives were obtained from the energy-dependent 
solution.  A comparison of global and single-energy 
solutions then serves as a check for structures that 
could have been ``smoothed over'' in the 
energy-dependent analysis.  Partial waves with $J < 4$ 
are displayed, whereas the analysis fitted waves up to 
$J = 5$.  Significant deviations from SM95 are visible 
in multipoles connected to the $\pi N$ S- and P-waves, 
as well as $D_{35}$, $F_{35}$, and $D_{13}$ (for the 
neutron.)

\begin{figure}[ht]
\centerline{
\psfig{file=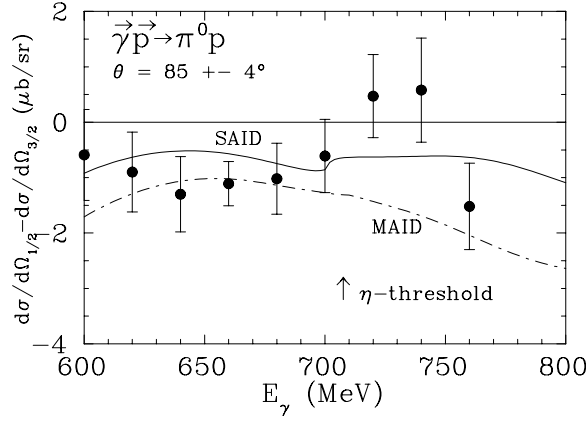,width=3in,clip=,silent=,angle=90}}
\vspace{3mm}
\caption[fig3]{\label{g3}   
       Differential cross section ($d \sigma /d \Omega
       _{1/2}-d \sigma /d \Omega _{3/2}$) for $\vec{\gamma}
       \vec{p}\to\pi^0p$ at $\theta = 85\pm 4^{\circ}$.  
       The solid (dash-dotted) line plots the SM02 
       (MAID2001~\protect\cite{maid}) solution.  
       Experimental data are from Mainz~
       \protect\cite{dx13}.}
\end{figure}

As mentioned above, the parametrization used in our 
previous analysis SM95~\cite{ar96} did not allow a 
good to fit recent Yerevan $\pi^0p$~\cite{ad01} and 
GRAAL $\pi^+n$~\cite{aj00,ku01} $\Sigma$ data (the 
critical range extends from 700 to 800~MeV.)  For 
comparison purposes, in Table~\ref{tbl5}, we compare 
SM95, SM02, and SX99.  The test fit SX99 retains the 
form used in previous fits (as in SM95) and is applied 
to the present full database.  One can see that recent 
Mainz $\pi^+n$ ($\sigma_{1/2}-\sigma_{3/2}$)~
\cite{ah00} data are also problematic for the SX99 fit, 
with the most difficult region again covering the 700 
to 800~MeV range.  In Fig.~\ref{g3}, we display the 
energy dependence of a differential cross section 
measurement~\cite{dx13}, at a fixed angle, displaying 
rapid variation over the region in question.  These 
Mainz data from 560 to 780~MeV are reasonably well 
reproduced ($\chi^2 = 106/76$), though not included 
in our analysis.

\subsection{Comparing SAID to MAID}
\label{sec:SAID-MAID}

While the SAID and MAID analyses are qualitatively 
similar from threshold to 1 GeV, some significant 
differences exist and these have been mentioned in a 
recent multi-analysis study which fitted to
a benchmark dataset~\cite{benchmark}.  While some 
multipoles show significant differences, the 
photo-decay amplitudes from MAID and SM95 are quite 
similar, with larger differences between the MAID 
and SM02 solutions.  The data fit quality shows 
greater variability and a few cases are given below.

\begin{figure}[ht]
\centerline{
\psfig{file=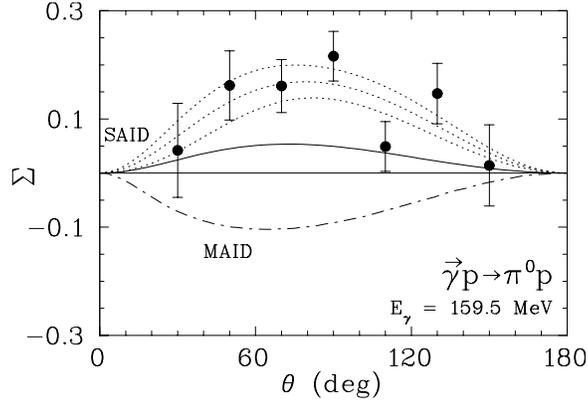,width=3in,clip=,silent=,angle=90}}
\vspace{3mm}
\caption[fig4]{\label{g4}
       Photon asymmetry for $\pi^0$ photoproduction on
       the proton at 159.5~MeV.  Data are from Mainz
       (solid circles)~\protect\cite{sc01}.  Plotted
       are the SM02 (solid line), the 162~MeV-SES (158 $-$
       165~MeV) fit associated with SM02 (dotted 
       lines represent uncertainties of the SES fit)
       and the MAID2000 results (dash-dotted)~
       \protect\cite{maid}.}
\end{figure}

The measurement of $\Sigma$ for $\pi^0p$ at 
threshold (160~MeV) has been discussed in~
\cite{sabit,hans} and is particularly sensitive (Fig.~
\ref{g4}.)  The plotted curves differ mainly in the 
P-wave multipoles.  Our SES at 162~MeV, which covers 
158 $-$ 165~MeV (see Table~\ref{tbl4},) fits the new 
Mainz data~\cite{sc01} rather well, while the 
energy-dependent MAID2000 and SM02 solutions are less 
successful.

\begin{figure}[ht]
\centerline{
\psfig{file=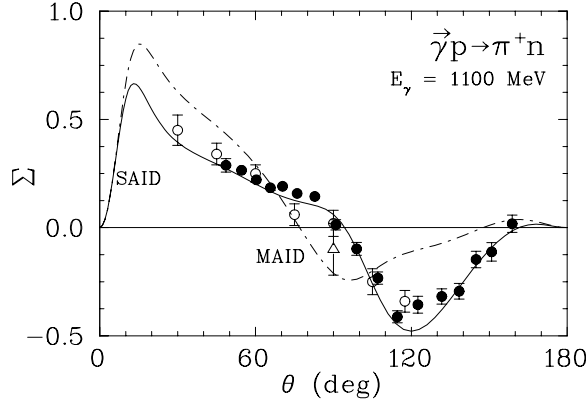,width=3in,clip=,silent=,angle=90}} 
\vspace{3mm}
\caption[fig5]{\label{g5}                                                                                                                     
       $\Sigma$ beam asymmetry for $\pi^+n$ at 1100~MeV.
       Black circles show GRAAL results~
       \protect\cite{ku01}, open circles
       indicate the results of the Daresbury group~
       \protect\cite{bs79}, open triangles indicate
       the results from Saclay~\protect\cite{as72}.
       The solid (dash-dotted) line represents the
       SM02 (MAID2001~\protect\cite{maid}) solution.}  
\end{figure}

At higher energies, close to the upper limit of MAID,
new $\pi^+n$ $\Sigma$ measurements, shown in Fig.~
\ref{g5}, are clearly problematic.  This has been used 
to suggest a change in the N(1650) photo-decay 
amplitude~\cite{aj00}.

\begin{figure}[ht]
\centerline{
\psfig{file=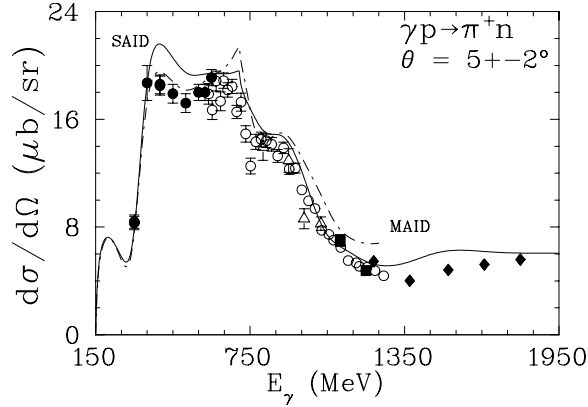,width=3in,clip=,silent=,angle=90}}
\vspace{3mm}
\caption[fig6]{\label{g6}                                                                                                                     
       Forward ($5^{\circ}$) differential cross section
       for $\gamma p\to\pi^+n$ as a function of energy.
       Experimental data for the range of $5\pm
       2^{\circ}$ are from Orsay~\protect\cite{bt68}
       (black circles), SLAC~\protect\cite{ec67} (open 
       circles), \protect\cite{by61} (open triangles), 
       \protect\cite{ki62} (black square), and 
       DESY~\protect\cite{bu67} (black diamonds.)
       The solid (dash-dotted) line represents the 
       SM02 (MAID2001~\protect\cite{maid}) solution.}  
\end{figure}

For the forward peaking in $\pi^+n$ unpolarized 
differential cross sections, displayed in Fig.~
\ref{g6}, the disagreement between SAID and MAID 
results reaches as much as 30\%.  In the SAID fits, 
some of these problems are resolved once systematic 
uncertainties are taken in account.

\begin{figure}[ht]
\centerline{
\psfig{file=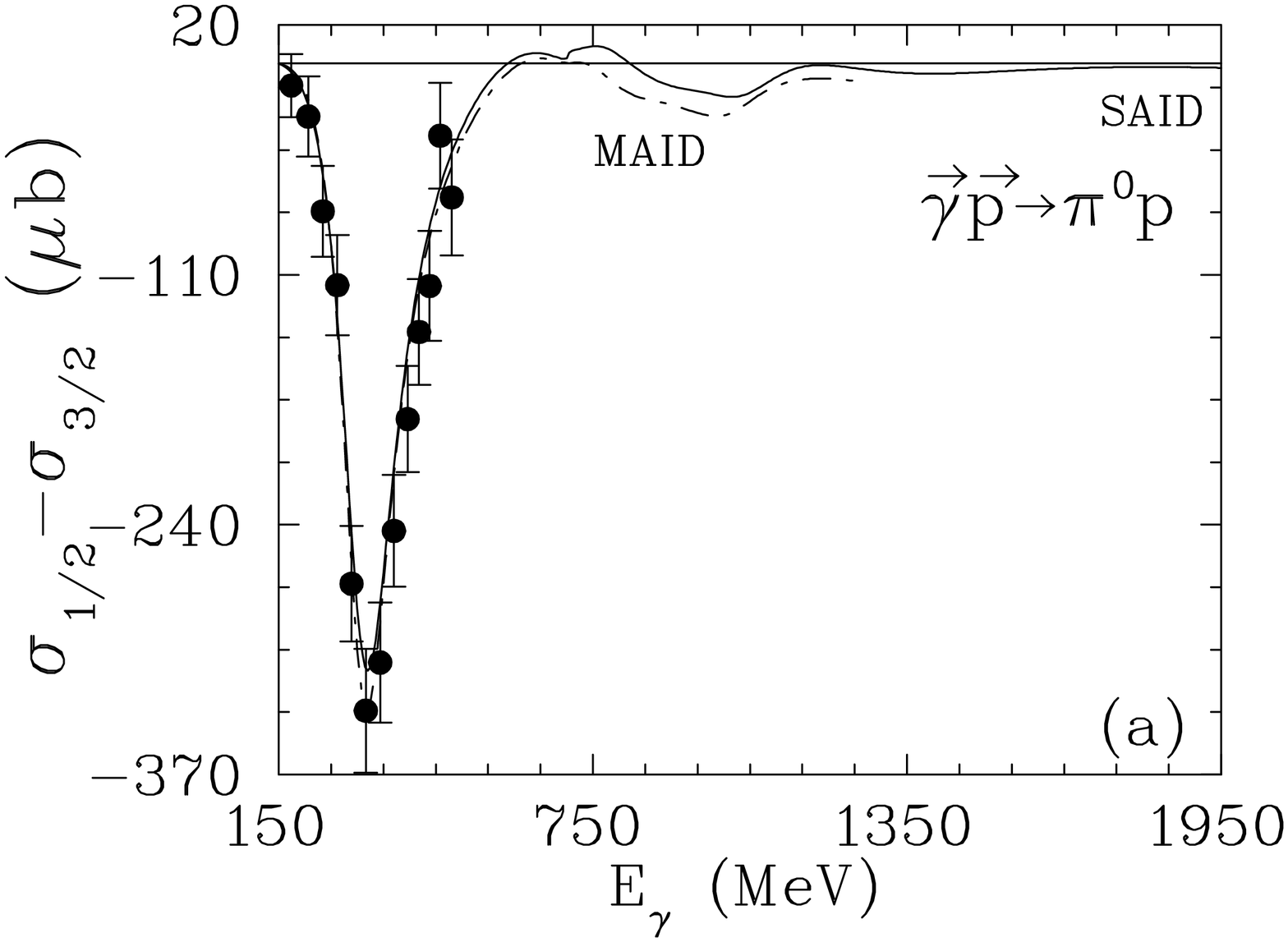,width=3in,clip=,silent=,angle=90}\hfill
\psfig{file=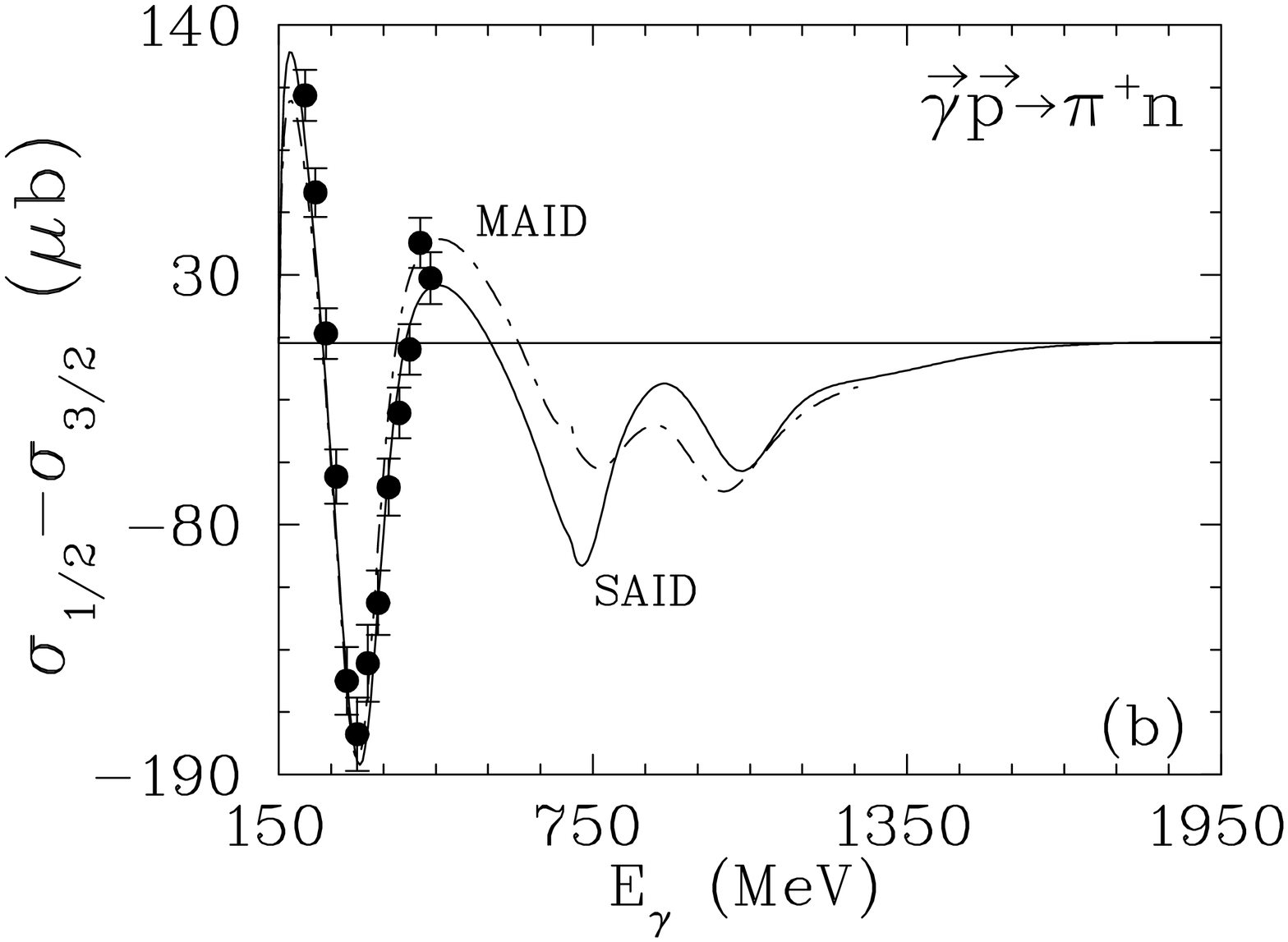,width=3in,clip=,silent=,angle=90}}
\vspace{3mm}
\caption[fig7]{\label{g7}                                                                                                                     
       Difference of the total cross sections for the 
       helicity states 1/2 and 3/2.
       (a) $\vec{\gamma}\vec{p}\to\pi^0p$ and
       (b) $\vec{\gamma}\vec{p}\to\pi^+n$.
       The solid (dash-dotted) line represents the 
       SM02 (MAID2001~\protect\cite{maid}) solution.  
       Experimental data are from Mainz~
       \protect\cite{ah00}.}
\end{figure}

The double-polarization quantity 
$\vec{\gamma}\vec{p}\to\pi^0p$, measured in
the A2 Collaboration GDH experiment~\cite{ah00} and 
displayed in Fig.~\ref{g7}, is well described by
both MAID2001 and SAID solution SM02.  At higher 
energies, deviations become more apparent (as was 
also shown in Fig.~\ref{g3}.)  CLAS at JLab~
\cite{sober}, ELSA at Bonn~\cite{elsa}, SPring--8 at 
Hyogo~\cite{iwata}, and LEGS at Brookhaven~\cite{andy}
have further programs underway to study this process.

\begin{figure}[ht]
\centerline{
\psfig{file=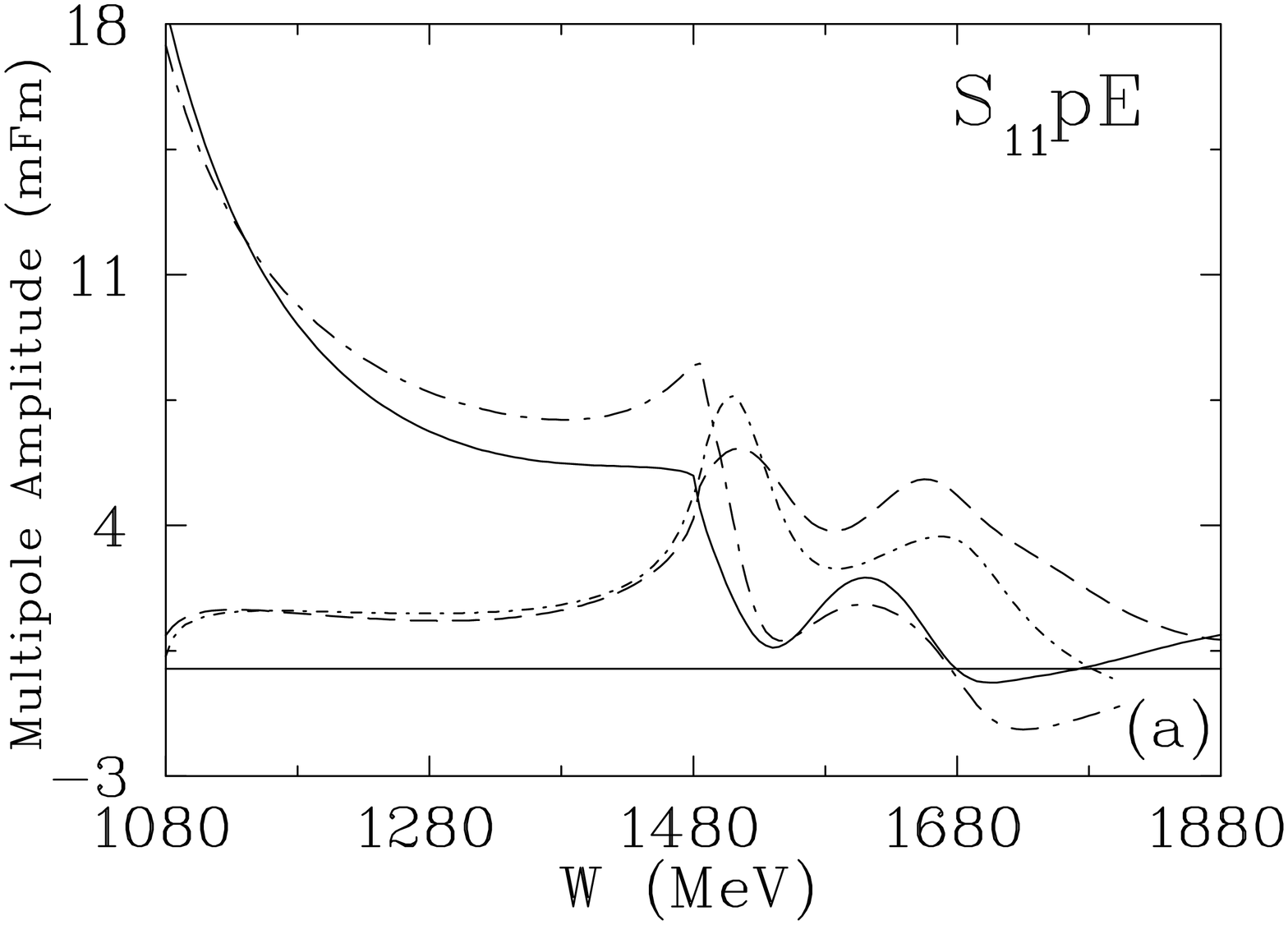,width=3in,clip=,silent=,angle=90}\hfill
\psfig{file=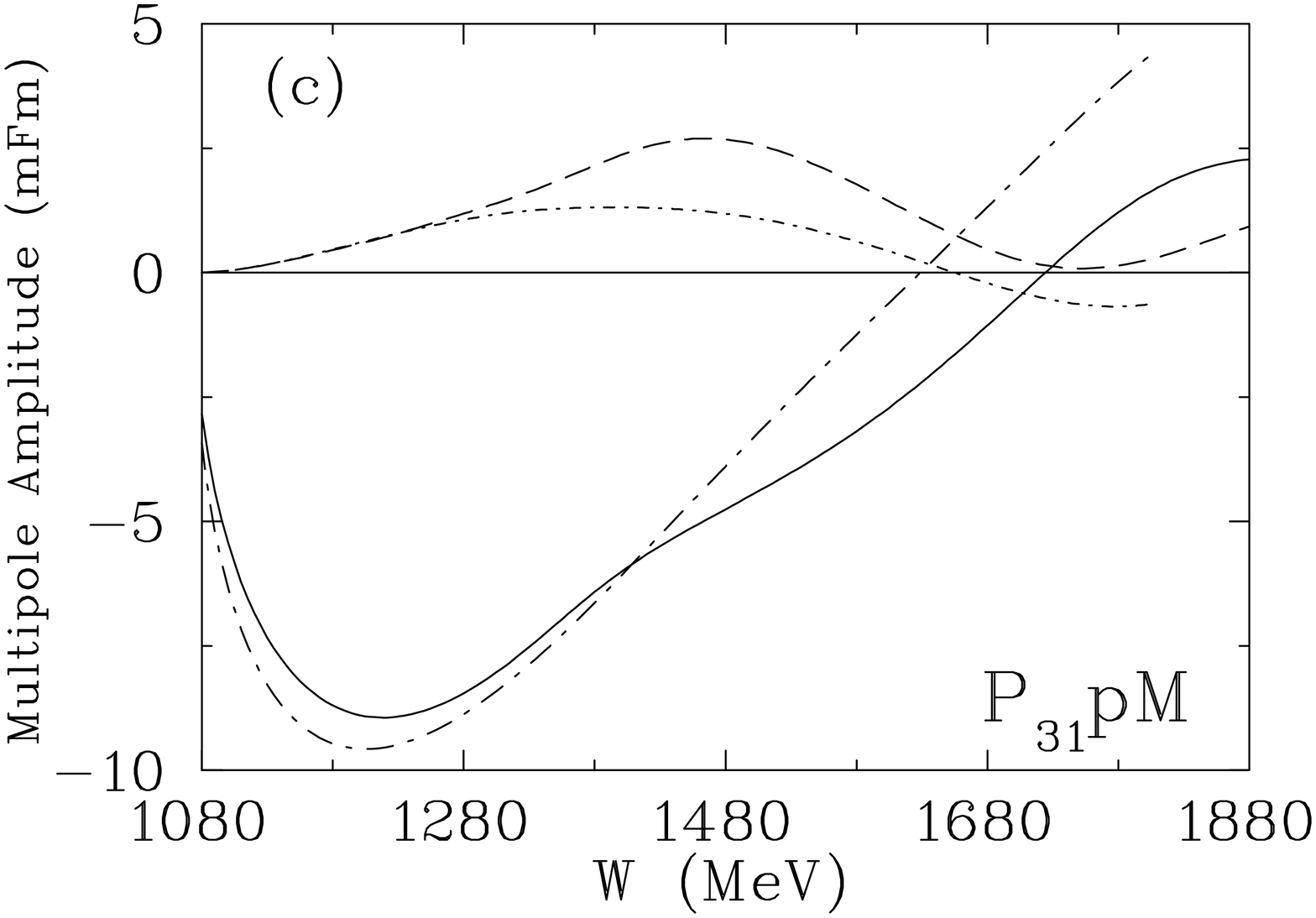,width=3in,clip=,silent=,angle=90}
}
\centerline{
\psfig{file=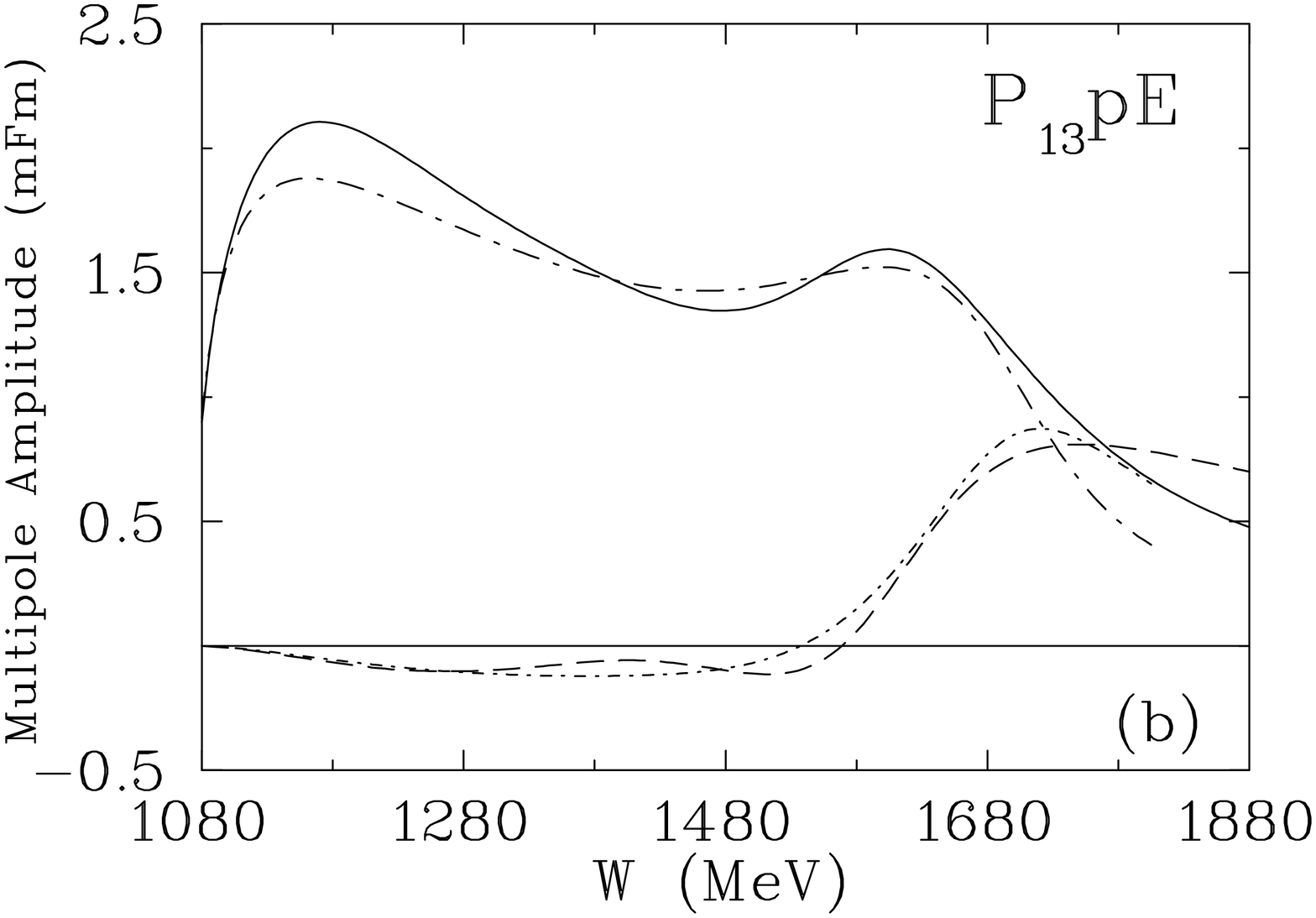,width=3in,clip=,silent=,angle=90}\hfill
\psfig{file=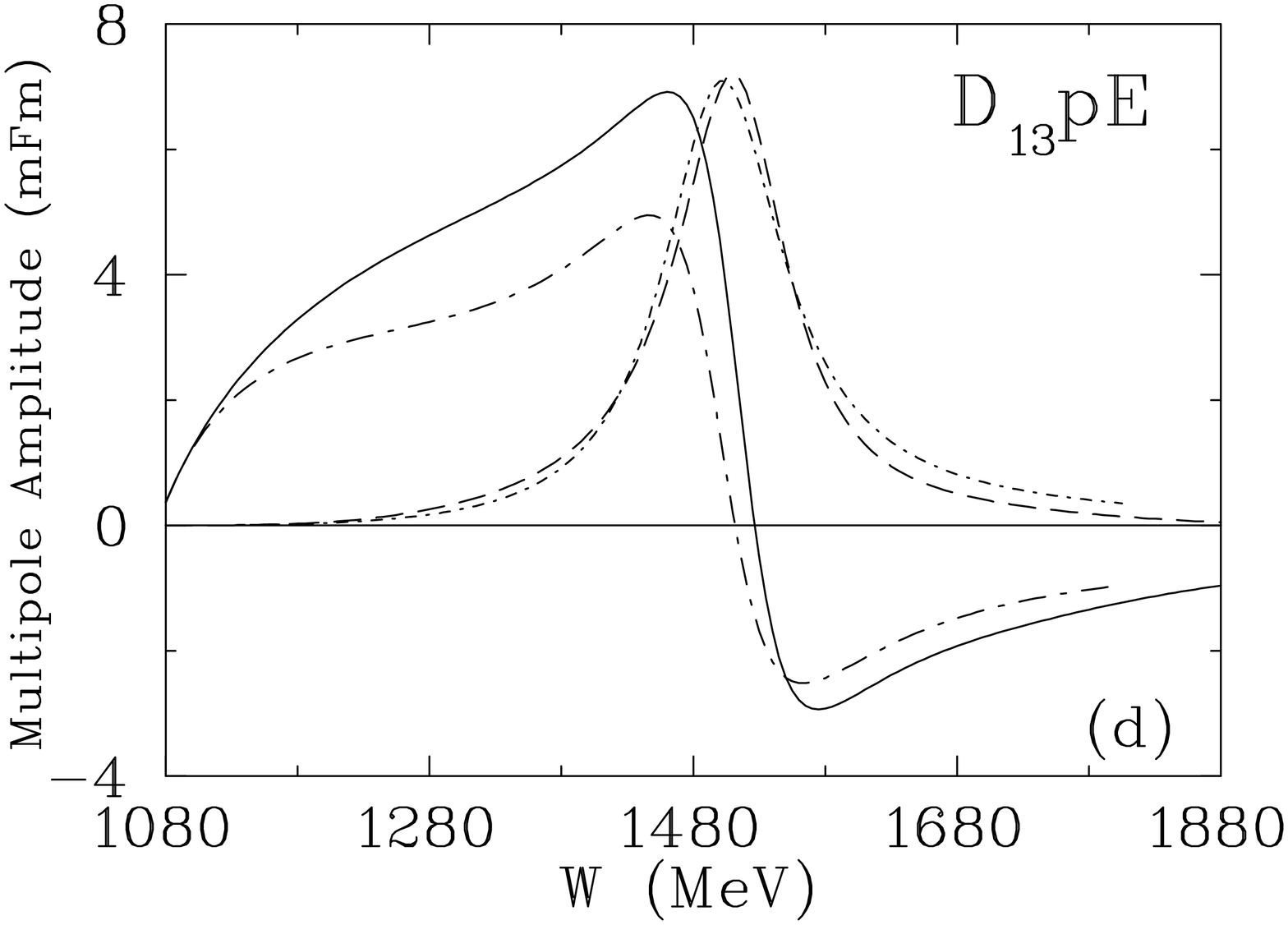,width=3in,clip=,silent=,angle=90}
}
\vspace{3mm}
\caption[fig8]{\label{g8}
      Selected partial-wave amplitudes to $E_{\gamma}$ =
      1250~MeV.  Solid (dashed) curves give the real 
      (imaginary) parts of amplitudes corresponding to 
      the SM02 solution.  The recent MAID2001 solution~
      \protect\cite{maid} is plotted with long 
      dash-dotted (real part) and short dash-dotted
      (imaginary part) lines.  Plotted are the 
      multipole amplitudes (a) $S_{11}pE$ 
      [$\rm _pE_{0+}^{1/2}$], (b) $P_{13}pE$ 
      [$\rm _pE_{1+}^{1/2}$], (c) $P_{31}pM$
      [$\rm _pM_{1-}^{3/2}$], and (d) $D_{13}pE$
      [$\rm _pE_{2-}^{1/2}$].  The subscript p (n) 
      denotes a proton (neutron) target.}
\end{figure}

In Fig.~\ref{g8}, we compare our results with the 
MAID analysis~\cite{maid}
\footnote{This MAID solution is valid to W = 1800~MeV
          (E$_{\gamma}$ = 1250~MeV)~\cite{lot}.}
where more substantial differences are seen for the
$S_{11}pE$, $P_{13}pE$, $P_{31}pM$, and $D_{13}pE$ 
multipoles.

We have fitted our multipoles using a simple   
Breit-Wigner plus background function, as described
in Ref.~\cite{ar90}.  We have employed both
single-energy and energy-dependent solutions
over a variety of energy ranges in order to estimate 
uncertainties.  A listing of our resonance couplings
is given in Table~\ref{tbl6}.  Here, values for the 
resonance mass ($W_R$), full width ($\Gamma$), and 
the decay width to $\pi N$ final states ($\Gamma 
_{\pi} /\Gamma$) were taken from our elastic $\pi N$ 
analysis~\cite{pin2} and were not varied in the fits 
and error estimates.

\begin{figure}[ht]
\centerline{
\psfig{file=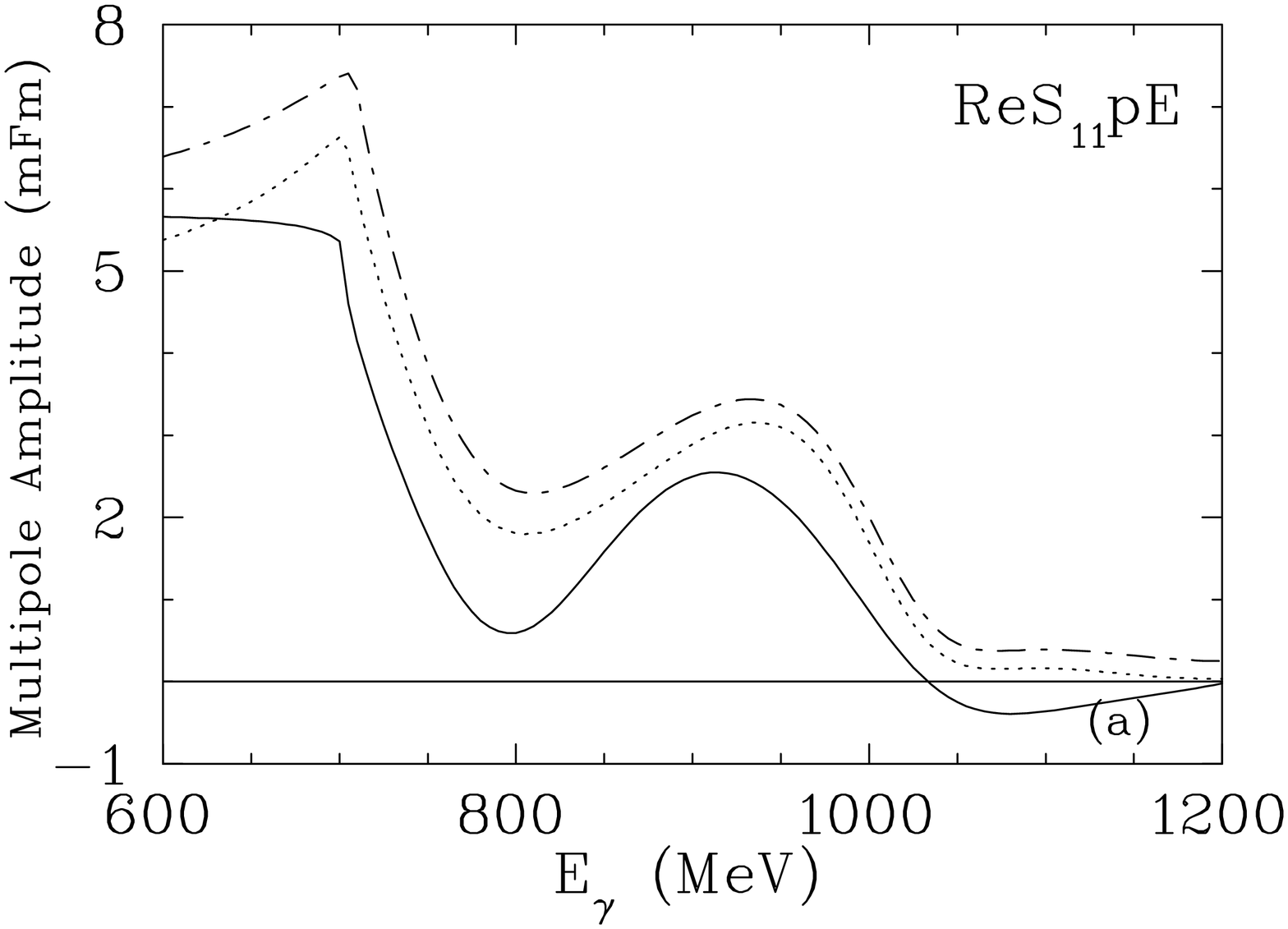,width=3in,clip=,silent=,angle=90}\hfill
\psfig{file=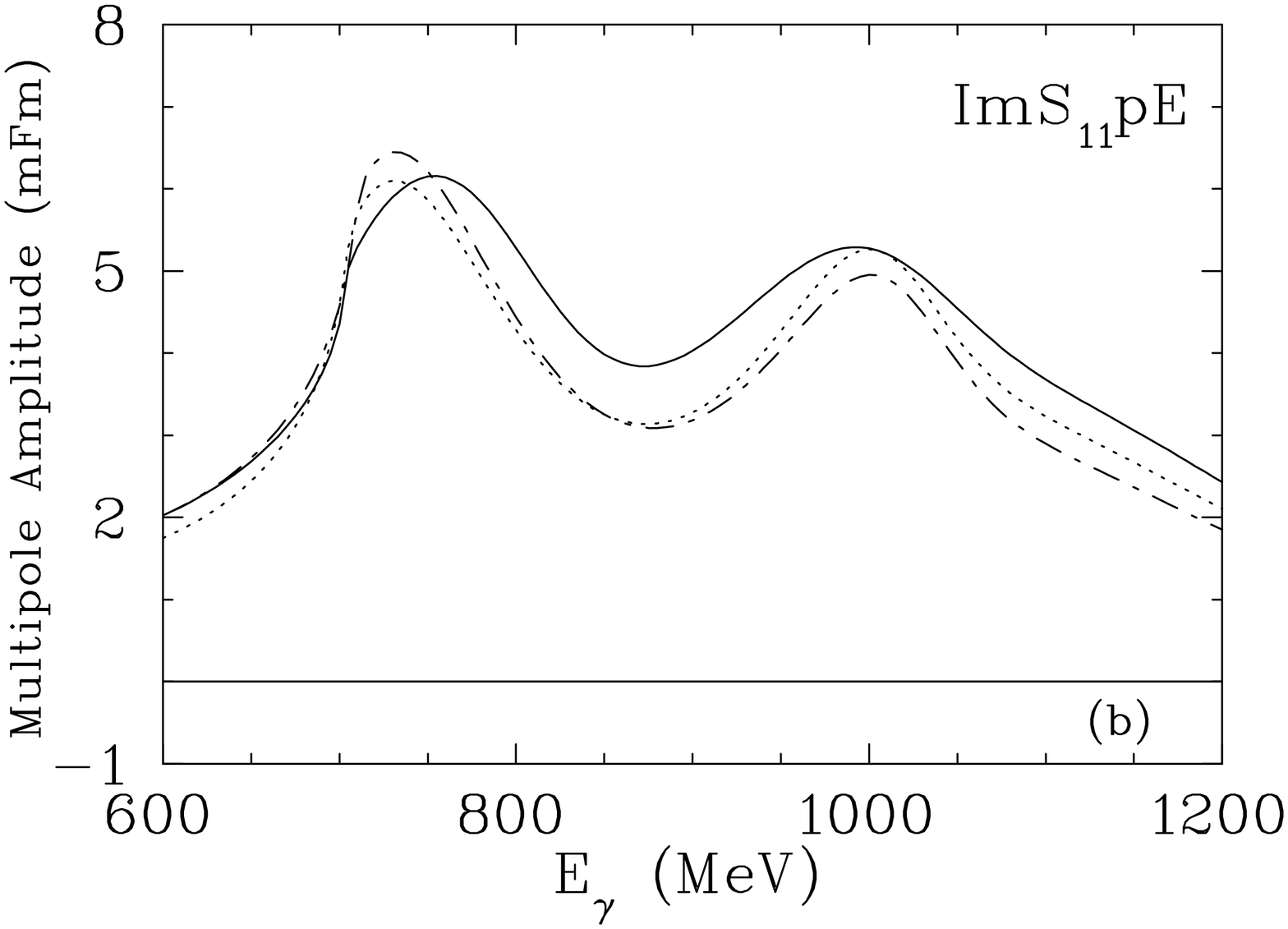,width=3in,clip=,silent=,angle=90}}
\vspace{3mm}
\caption[fig9]{\label{g9}
       $S_{11}pE$ multipole for 600 to 1200~MeV.
       Plotted are (a) real part and (b) imaginary 
       part.  The SM02 (SX99) solution is plotted 
       with a solid (dashed) line and previous~SM95 
       solution~\protect\cite{ar96} with a dash-dotted 
       line.} 
\end{figure}

We find that the $S_{11}pE$ ($E^{1/2}_{0+}$) multipole 
is very sensitive to both the database and 
parametrization in the range associated with the
$N(1535)$.  The range of variation, particularly 
large for the real part of $S_{11}pE$, is displayed 
in Fig.~\ref{g9}.  This sensitivity is not surprising, 
as the quantity (${\rm Im}T_{\pi N} - T_{\pi N}^2$) 
has a very sharp structure at the $\eta N$ threshold.  
This variability in the multipole amplitude is 
reflected in the $N(1535)$ resonance coupling,  
which we feel is presently too uncertain to quote. 

Other couplings significantly altered using 
the revised parameterization scheme include the 
$P_{11}(1710)$, which is still essentially 
undetermined, $P_{13}(1720)$, $S_{31}(1620)$, 
and $F_{35}(1905)$.  The $F_{37}(1950)$ has an easily 
identifiable magnetic but, essentially no electric 
multipole. Multipoles associated with the
$D_{35}(1930)$ show very little resonance 
signature.

We were particularly surprised to see a large
change in the $\Delta (1232)$ photo-decay amplitudes.
Comparison with our SM95 fit shows a significant 
decrease in cross section at the resonance position.
This shift has resulted due to the inclusion of 
recent Mainz $\pi^0p$ measurements~\cite{be97}, which 
are systematically lower (particularly at backward 
angles) than an older set of Bonn measurements. More 
recent Mainz fits for MAID2001~\cite{sabit1} give 
$-133 \pm 4$ and $-252 \pm 6$ (in $10^{-3} GeV^{-1/2}$ 
units) for $A_{1/2}$ and $A_{3/2}$, respectively. 
These results have also shifted lower and are 
consistent with our determination.

\subsection{Sum Rules}
\label{sec:sum}

\begin{figure}[ht]
\centerline{
\psfig{file=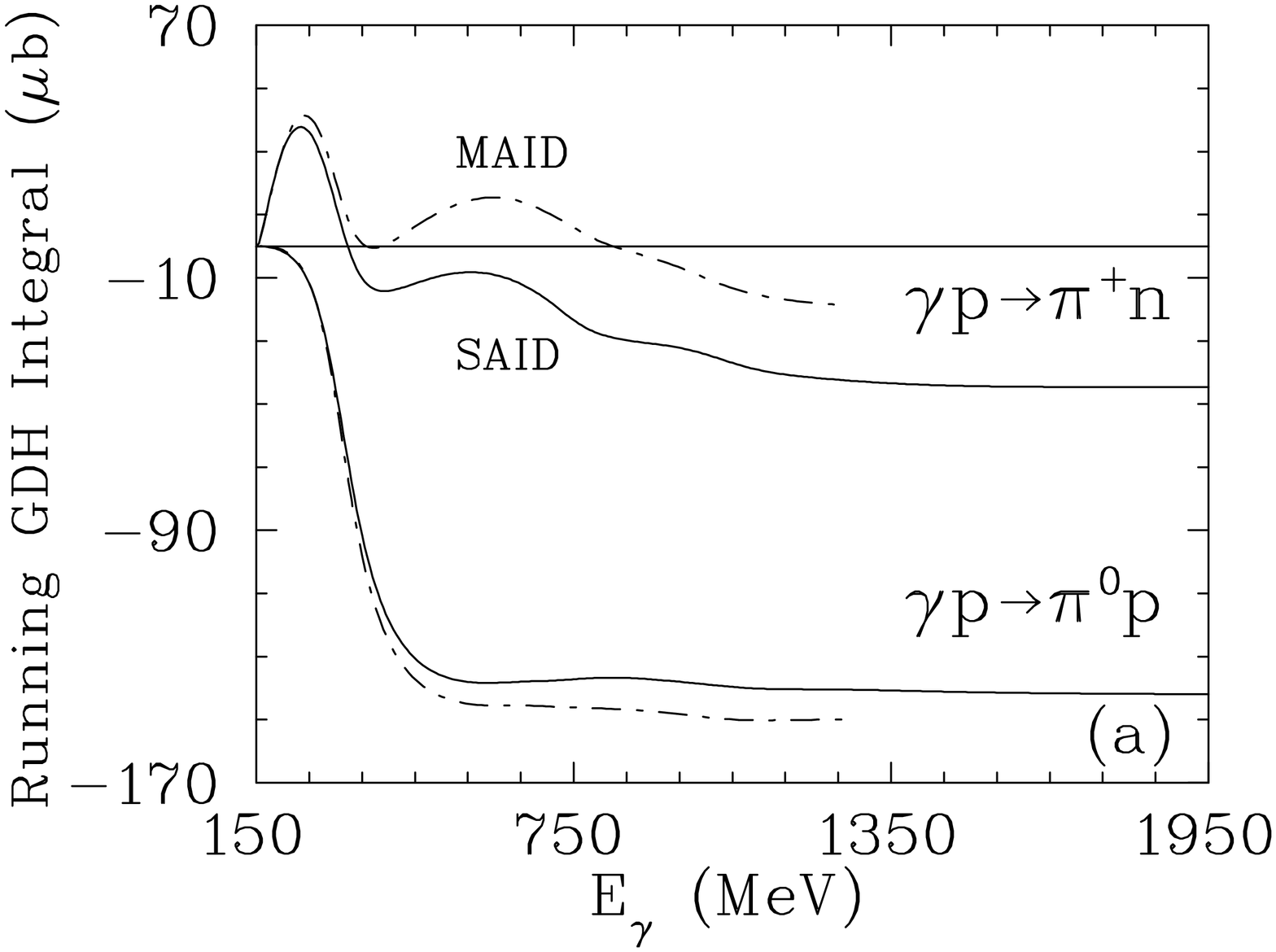,width=3in,clip=,silent=,angle=90}\hfill
\psfig{file=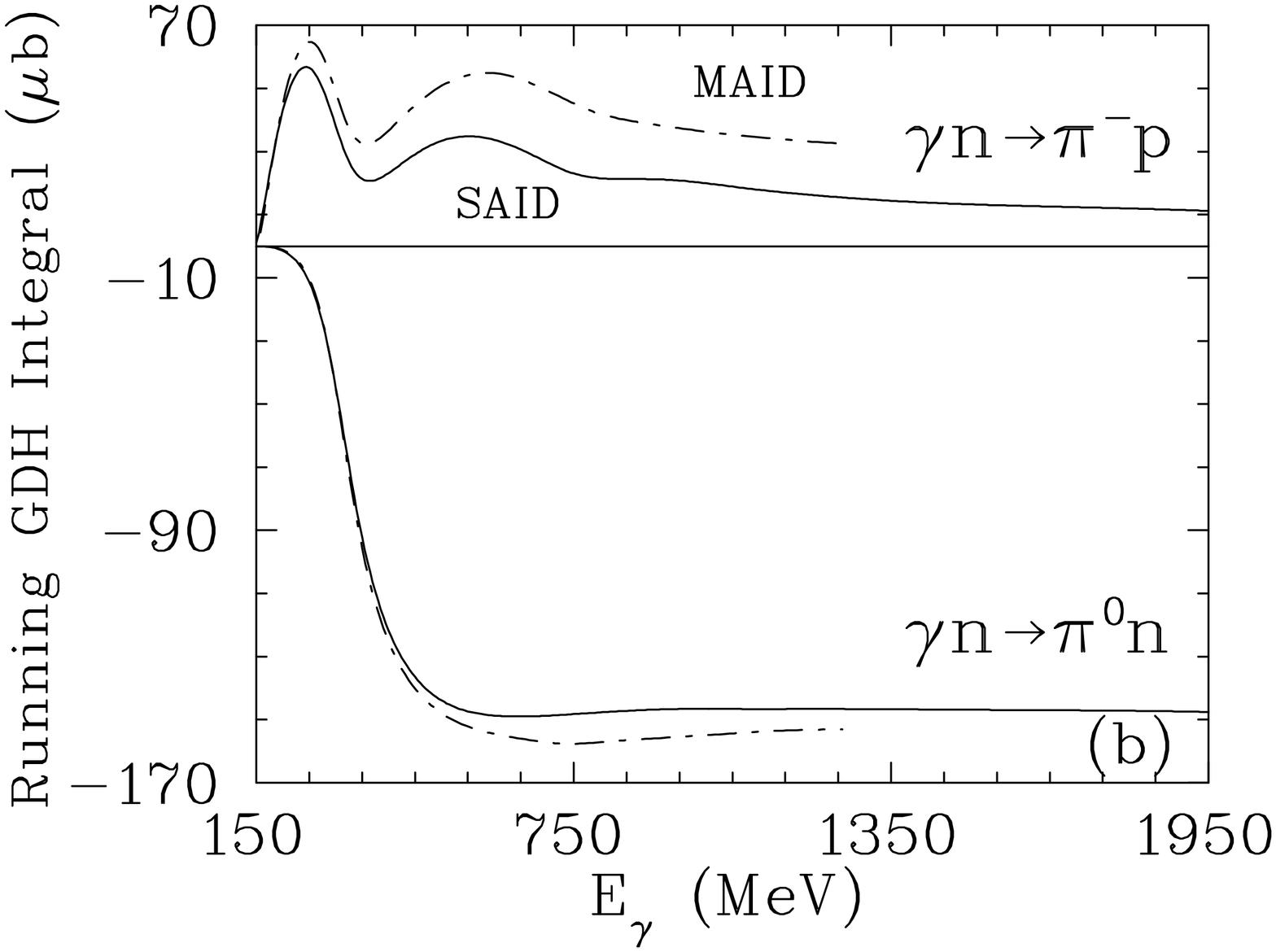,width=3in,clip=,silent=,angle=90}}
\vspace{3mm}
\caption[fig10]{\label{g10}
       Running GDH integral.  (a) for proton and 
       (b) neutron targets.  The solid (dash-dotted) 
       line represents the SM02 (MAID2000
       ~\protect\cite{maid}) solution.}
\end{figure}

The amplitudes obtained in our analyses can be used 
to evaluate the single-pion production component of 
several sum rules.  The GDH integral~\cite{gdh} 
relates the anomalous magnetic moment $\kappa$, the 
charge $e$, and nucleon mass $M$ to the difference 
in the total photoabsorption cross sections for 
circularly polarized photons on longitudinally 
polarized nucleons  
\begin{eqnarray}
I_{GDH} =  \int_{\nu_0}^{\infty}\frac{\sigma_{1/2}
          -\sigma_{3/2}}{\nu}d \nu 
        = -\frac{\pi e^2}{2 M^2}\kappa^2,
\label{3}\end{eqnarray}
where $\sigma_{1/2}$ and $\sigma_{3/2}$ are the 
photoabsorption cross sections for the 
helicity states $1/2$ and $3/2$, respectively, 
with $\nu$ being the photon energy.  For the proton 
(neutron) target, Eq.~(\ref{3}) predicts $-$205~
($-$233)~$\mu b$.  The running GDH integrals for 
the proton and neutron are shown in Fig.~\ref{g10}, 
where a comparison with MAID is also given. 

\begin{figure}[ht]
\centerline{
\psfig{file=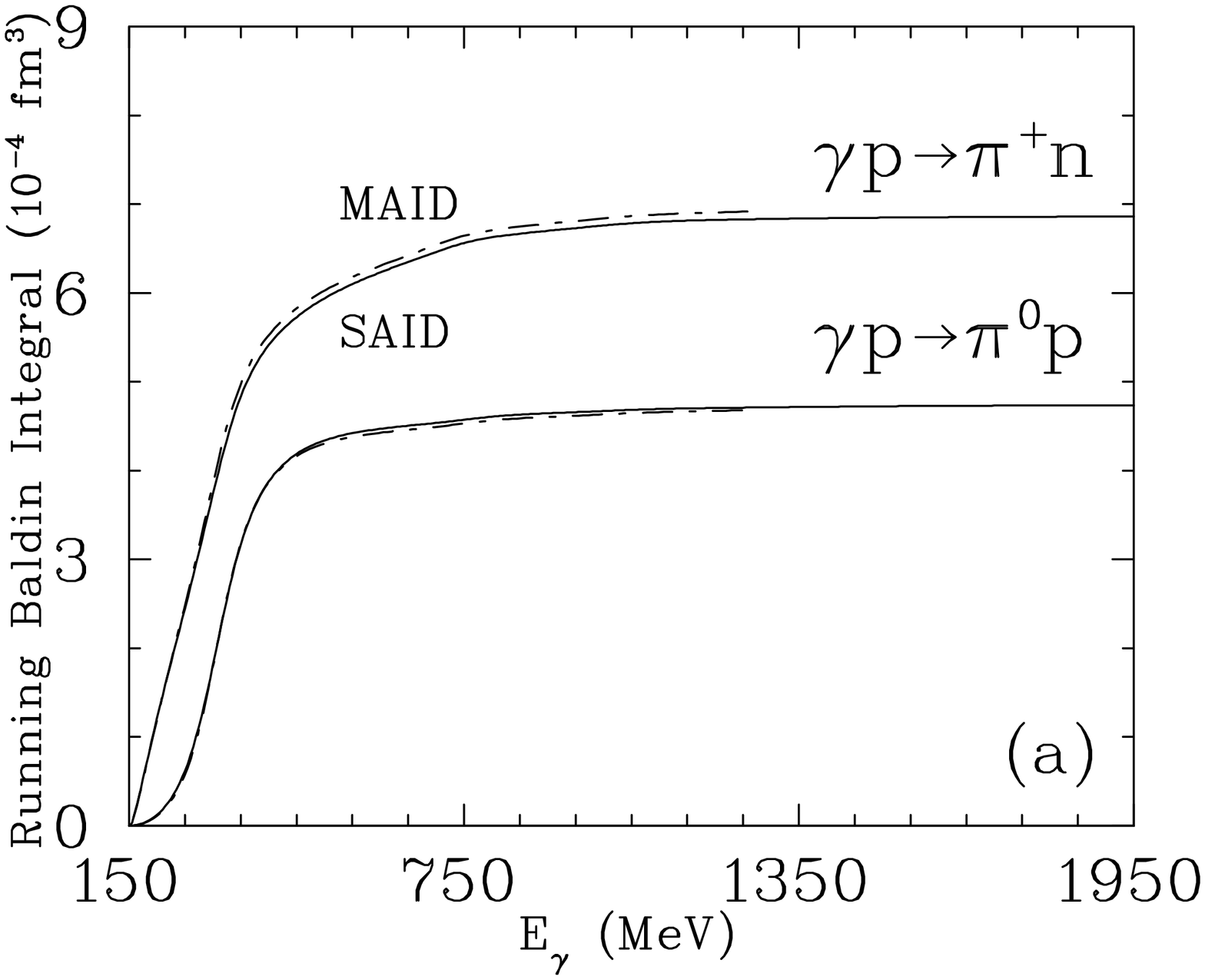,width=3in,clip=,silent=,angle=90}\hfill
\psfig{file=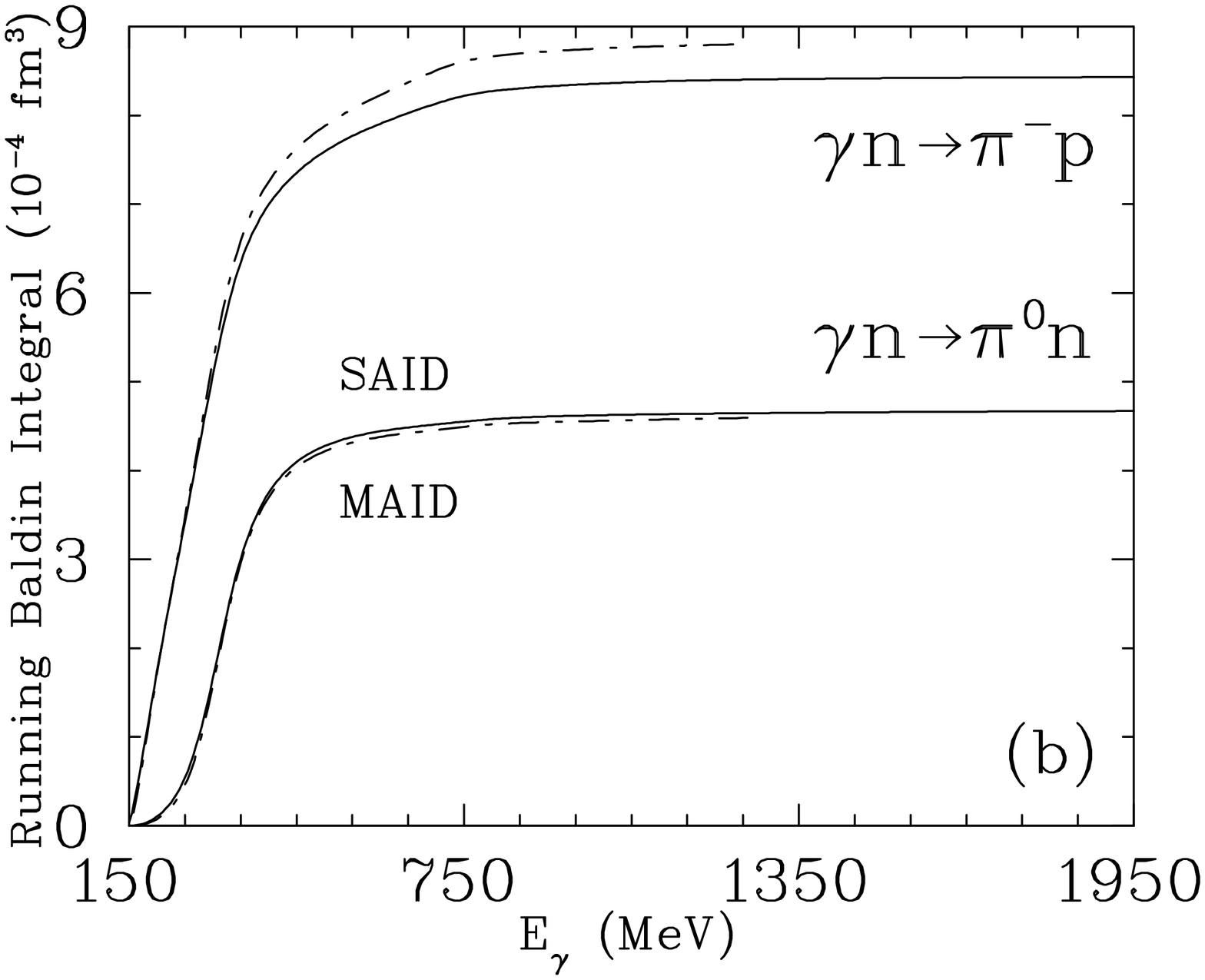,width=3in,clip=,silent=,angle=90}}
\vspace{3mm}
\caption[fig11]{\label{g11}
       Running Baldin integral. (a) for proton and 
       (b) neutron targets.  The solid (dash-dotted) 
       line represents the SM02 (MAID2000
       ~\protect\cite{maid}) solution.}
\end{figure}

The Baldin sum rule~\cite{bl} relates the sum of 
electric and magnetic polarizabilities of the 
nucleon to the total photo-absorption cross section 
$\sigma_{tot}$  
\begin{eqnarray}
I_{Baldin} = \frac{1}{2 \pi^2}\int_{\nu_0}^{\infty}
             \frac{\sigma_{tot}}{\nu^2} d \nu 
           = \frac{1}{2 \pi^2}\int_{\nu_0}^{\infty}
             \frac{\sigma_{1/2}+\sigma_{3/2}}{2 \nu^2}
             d \nu.
\label{4}\end{eqnarray}
For the proton (neutron) target, the recent dispersion
calculations by Levchuk and L'vov give ($14.0\pm 
0.3) 10^{-4} fm^3$ [($15.2\pm 0.5) 10^{-4} fm^3$]~
\cite{lvov}.  For the proton, an independent Mainz 
determination gives $(13.8\pm 0.4) 10^{-4} fm^3$~
\cite{ol01} and the LEGS group quotes $(13.25\pm 0.86^{+0.23}
_{-0.58}) 10^{-4} fm^3$~\cite{bl01}.  The isospin
averaged nucleon polarizabilities determined by
MAX-lab measurements of Compton scattering from the
deuteron is ($16.4\pm 3.6) 10^{-4} fm^3$~\cite{max}.
Running Baldin integrals, with comparisons to MAID, 
are given in Fig.~\ref{g11}.

\begin{figure}[ht]
\centerline{
\psfig{file=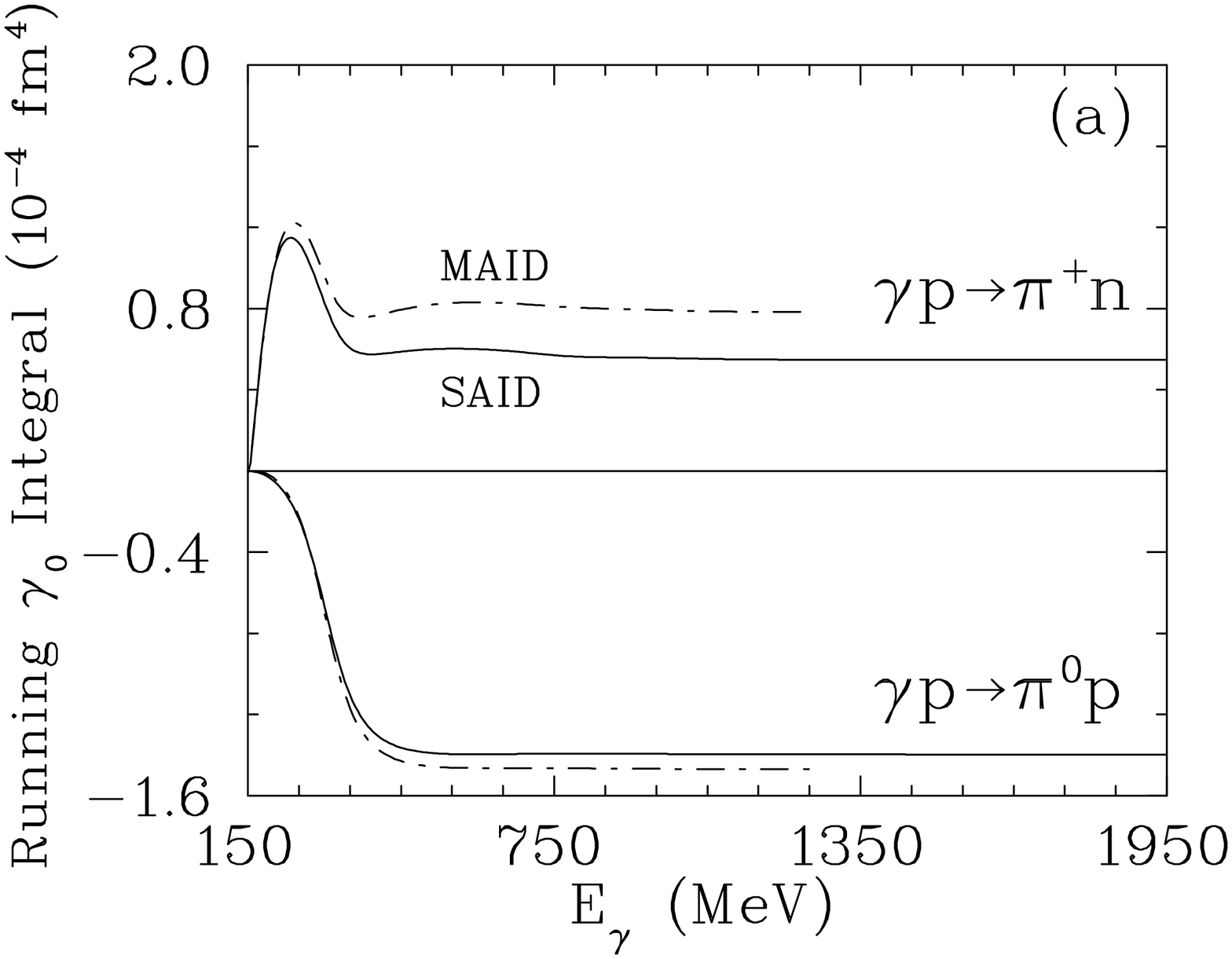,width=3in,clip=,silent=,angle=90}\hfill
\psfig{file=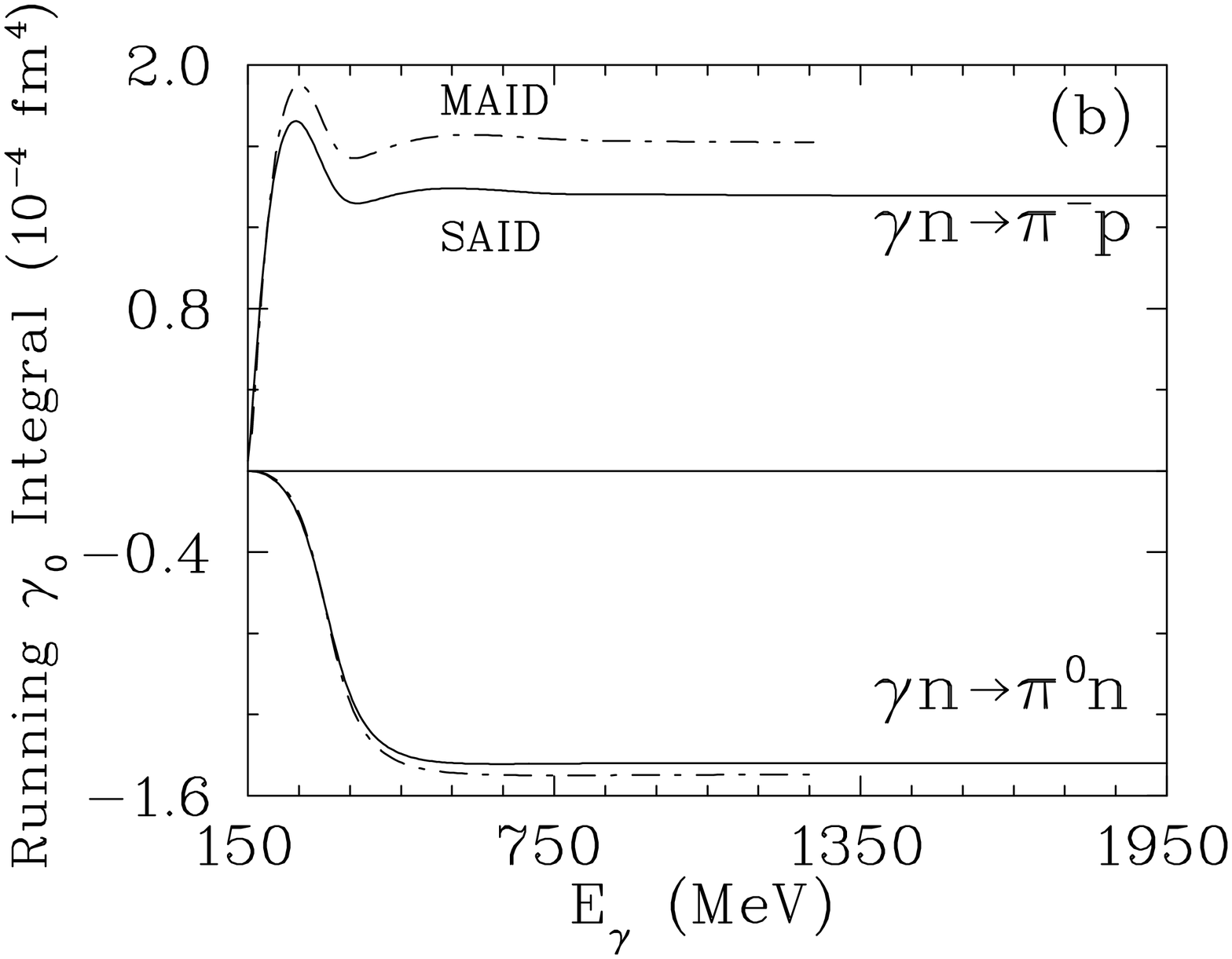,width=3in,clip=,silent=,angle=90}}
\vspace{3mm}
\caption[fig12]{\label{g12}
       Forward spin polarizability $\gamma_0$.
       (a) for proton and (b) neutron targets.
       The solid (dash-dotted) line represents the
       SM02 (MAID2000~\protect\cite{maid}) solution.}
\end{figure}

The forward spin polarizability $\gamma_0$~
\cite{ra93} is
\begin{eqnarray}
\gamma_0 = \frac{1}{4 \pi^2}\int_{\nu_0}^{\infty}
\frac{\sigma_{1/2}-\sigma_{3/2}}{\nu^3} d \nu.
\label{5}\end{eqnarray}
A recent dressed K-matrix model approach for the 
proton gives $\gamma _0 = -0.9~10^{-4} fm^4$~
\cite{kond}.  The LEGS analysis gives, for a proton 
target, $\gamma _0 = (-1.55\pm 0.15\pm 0.003) 
10^{-4} fm^4$~\cite{bl01}.  The running integral 
is shown in Fig.~\ref{g12}.  

For charge states $\pi^+n$ and $\pi^-p$, all 
three quantities are sensitive to the threshold 
energy range ($E_{\gamma} < 200~MeV$), as shown in 
Table~\ref{tbl7}.  From Figs.~\ref{g10}$-$\ref{g12}, 
one can see that each integral of the single-pion 
contribution, based upon the SM02 solution, has 
essentially converged by 2~GeV.

Experimental data for the GDH and $\gamma_0$ 
quantities have been obtained from measurements at 
MAMI, covering ranges from 200 to 450~MeV~\cite{ah00} 
and to 800~MeV~\cite{ah01}.  In Tables~\ref{tbl8} 
and~\ref{tbl9}, we show SM02 and MAID results for 
the abovementioned quantities over energy ranges 
corresponding to measurements.  Clearly, 
calculations above 450~MeV have to take into account 
contributions beyond single-pion photoproduction.  

\section{Summary and Conclusion}
\label{sec:conc}

The single-pion photoproduction database, for proton
targets, has 
increased significantly since the publication of 
our fit SM95.  The inclusion of these precise new 
measurements has resulted in a fit with a lower 
overall $\chi^2$/data. However, some polarization 
quantities have been difficult to fit, and these 
difficulties have prompted an examination of the 
phenomenological forms we use.  By changing the way 
we extrapolate beyond the two-pion threshold, an 
improved fit was obtained.

This new multipole solution was found to differ 
significantly from SM95 in a number of partial 
waves.  The largest changes were associated with 
multipoles connected to $\pi N$ resonances with 
$\Gamma_{\pi} / \Gamma \lesssim 0.3$. Also quite 
different was the $S_{11}$ multipole, for which 
the associated $\pi N$ inelasticity has a sharp 
increase at the $\eta N$ threshold. As might be 
expected, states with large $\pi N$ branching 
fractions remained stable.  This stability held 
for the $N(1520)$ as well, though investigations 
based upon $\eta N$ photoproduction, and 
quantities related to the GDH integral, have 
suggested a shift in its ratio of photo-decay 
amplitudes.  Given the sensitivity of weaker 
resonances to the choice of phenomenology, we 
are now attempting to replace the dependence on
(${\rm Im}T_{\pi N} - T_{\pi N}^2$) with a form 
more directly connected to the opening of 
specific channels, such as $\eta N$ and $\pi \Delta$. 

The evaluation of sum rules (GDH, Baldin, and 
forward spin polarizability) for a single-pion 
contribution exhibits convergence by 2~GeV. 
Agreement with Mainz is now good below 450~MeV, 
with larger deviations at higher energies. 

In both $\pi N$ elastic scattering~\cite{alekseev}
and pion photoproduction~\cite{ku01,gi00}, the 
measurements of precise new single- and 
double-polarization data have highlighted problems 
existing in the ``standard" fits.  Further 
polarization measurements will be required to test 
assumptions implicit in the SAID and MAID programs. 

\acknowledgments

The authors express their gratitude to J.~Ahrens, 
G.~Anton, H.-J.~Arends, R.~Beck, J.~C.~Bergstrom, 
J.-P.~Bocquet, D.~Branford, H.~M.~Fischer, K.~G.~
Fissum, M.~Fuchs, R. Gilman, H.~H.~Hakopian, D.~A.~
Hutcheon, M.~Khandaker, V.~Kouznetsov, B.~Krusche, 
B.~A.~Mecking, A. Omelaenko, P.~Pedroni, I.~
Preobrajenski, D.~Rebreyend, G.~V.~O'Rielly, A.~
M.~Sandorfi, C.~Schaerf, and A.~Shafi for providing 
experimental data prior to publication or for 
clarification of information already published.  
We also thank L.~Tiator and S.~Kamalov for the MAID 
contribution, A.~L'vov for a discussion of the sum 
rules, and A.~E.~Kudryavtsev for comments on the 
$S_{11}(1535)$ problem.  This work was supported in 
part by the U.~S.~Department of Energy under Grant 
DE--FG02--99ER41110.  The authors acknowledge partial 
support from Jefferson Lab, by the Southeastern 
Universities Research Association under DOE contract 
DE--AC05--84ER40150.


\eject
\begin{table}[t]
\caption{Comparison of present (SM02) and previous 
         (SM95~\protect\cite{ar96}, SP93~
         \protect\cite{li93}, and SP89~\protect\cite{ar90}) 
         energy-dependent partial-wave analyses of charged 
         and neutral pion photoproduction data.  The $\chi 
         ^2$ values for the previous solutions correspond 
         to our published results.  $N_{prm}$ is the 
         number parameters varied in the fit.}
\label{tbl1}
\begin{tabular}{ccccccc}
Solution & Range~(MeV) & 
$\chi^2$/$\pi^0p$ & $\chi^2$/$\pi^+n$ & 
$\chi^2$/$\pi^-p$ & $\chi^2$/$\pi^0n$ & $N_{prm}$  \\
\tableline
SM02 & 2000 & 18686/8092 & 12246/7279 & 4123/2080 & 241/120 & 148 \\
SM95 & 2000 & 13087/4711 & 12284/6359 & 6156/2225 & 282/120 & 135 \\
SP93 & 1800 & 14093/4015 & 22426/6019 & 8280/2312 & 275/120 & 134 \\
SP89 & 1000 & 13073/3241 & 11092/3847 & 4947/1728 & 461/120 &  97 \\
\end{tabular}
\end{table}
\begin{table}[t]
\caption{Comparison of $\chi^2$/data for normalized (Norm) 
         and unnormalized (Unnorm) data for SM02 solution.}
\label{tbl2}
\begin{tabular}{ccc}
Data & Norm & Unnorm \\
\tableline
$\pi^0p$ & 2.3 & 3.8 \\
$\pi^+n$ & 1.7 & 2.7 \\
$\pi^-p$ & 2.0 & 2.6 \\
$\pi^0n$ & 2.0 & 2.0 \\
\end{tabular}
\end{table}
\begin{table}[t]
\caption{Comparison of $\chi^2$/data for the SM02 solution 
         over the 600$-$900~MeV range, associated with the
         N(1535), versus full database and more recent 
         measurements.}
\label{tbl3}
\begin{tabular}{ccccc}
Reaction &   Observable      &     All   & 1980$-$Present & 1985$-$Present \\
\tableline
$\pi^0p$ & $d\sigma/d\Omega$ & 3946/1644 & 1808/992       & 1211/903 \\
         &     $\Sigma$      &  479/179  &  345/122       &  277/105 \\
         &         P         &  361/181  &  177/103       &   56/33  \\
         &         T         &  376/72   &    2/2         &    2/2   \\
\tableline
$\pi^+n$ & $d\sigma/d\Omega$ & 1343/1407 &  768/930       &  349/596 \\
         &     $\Sigma$      &  954/251  &  594/134       &  565/112 \\
         &         P         &  158/62   &    5/5         &    $-$   \\
         &         T         &  437/228  &   44/60        &   44/60  \\
\end{tabular}
\end{table}
\begin{table}[t]
\caption{Single-energy (binned) fits of combined 
         charge and neutral pion photoproduction data, 
         with $\chi^2$ values.  $N_{prm}$ is the number 
         parameters varied in the single-energy fits, and 
         $\chi^2_E$ is given by the energy-dependent fit, 
         SM02, over the same energy interval.} 
\label{tbl4}
\begin{tabular}{cccccccccc}
E$_{\gamma}$~(MeV)&Range~(MeV)&$N_{prm}$&$\chi^2$/data&$\chi^2_E$&
E$_{\gamma}$~(MeV)&Range~(MeV)&$N_{prm}$&$\chi^2$/data&$\chi^2_E$ \\
\tableline
 152 & 148 $-$ 155 &  5 & 494/269 & 596 &  885 & 873 $-$ 896 & 28 & 210/161 & 374 \\
 157 & 153 $-$ 160 &  5 & 463/240 & 596 &  905 & 893 $-$ 916 & 28 & 461/329 & 740 \\
 162 & 158 $-$ 165 &  5 & 370/211 & 401 &  925 & 913 $-$ 936 & 28 & 139/149 & 293 \\
 167 & 163 $-$ 170 &  5 & 422/173 & 540 &  945 & 933 $-$ 956 & 29 & 476/263 & 704 \\
 172 & 168 $-$ 175 &  5 & 116/ 86 & 157 &  965 & 953 $-$ 975 & 20 & 145/130 & 280 \\
 177 & 173 $-$ 180 &  5 & 136/105 & 148 &  985 & 973 $-$ 996 & 20 & 130/130 & 227 \\
 182 & 178 $-$ 185 &  5 & 149/105 & 174 & 1005 & 993 $-$1016 & 29 & 508/294 & 766 \\
 187 & 183 $-$ 190 &  5 &  69/ 99 & 124 & 1025 &1013 $-$1037 & 20 & 230/131 & 313 \\
 192 & 188 $-$ 195 & 10 & 124/133 & 153 & 1045 &1033 $-$1057 & 29 & 256/205 & 417 \\
 197 & 193 $-$ 200 &  5 & 131/105 & 154 & 1065 &1053 $-$1076 & 29 & 126/124 & 191 \\
 202 & 199 $-$ 205 &  5 & 128/ 95 & 157 & 1085 &1073 $-$1096 & 29 &  92/ 97 & 155 \\
 207 & 203 $-$ 210 &  5 & 154/111 & 173 & 1105 &1093 $-$1115 & 29 & 359/212 & 479 \\
 212 & 208 $-$ 215 & 10 & 184/148 & 223 & 1125 &1114 $-$1136 & 29 & 139/ 98 & 224 \\
 217 & 213 $-$ 220 &  9 & 182/138 & 297 & 1145 &1133 $-$1155 & 29 & 212/160 & 328 \\
 222 & 218 $-$ 225 & 15 & 239/178 & 367 & 1165 &1153 $-$1176 & 21 & 110/ 98 & 167 \\
 227 & 223 $-$ 230 & 15 & 124/148 & 188 & 1185 &1173 $-$1194 & 21 &  54/ 73 &  98 \\
 232 & 228 $-$ 235 & 15 & 230/176 & 254 & 1205 &1193 $-$1216 & 29 & 218/185 & 314 \\
 237 & 233 $-$ 240 & 15 & 266/215 & 306 & 1225 &1213 $-$1236 & 11 &  98/ 83 & 149 \\
 242 & 238 $-$ 245 & 15 & 319/218 & 398 & 1245 &1233 $-$1255 & 29 & 116/118 & 217 \\
 247 & 243 $-$ 250 & 15 & 352/211 & 472 & 1265 &1253 $-$1276 & 21 &  65/ 67 & 129 \\
 265 & 253 $-$ 276 & 15 &1371/698 &1641 & 1285 &1273 $-$1297 & 11 &  34/ 45 &  68 \\
 285 & 273 $-$ 297 & 16 &1060/491 &1207 & 1305 &1293 $-$1315 & 21 & 275/139 & 435 \\
 305 & 293 $-$ 316 & 16 & 898/492 & 957 & 1325 &1313 $-$1335 & 11 &  42/ 48 &  79 \\
 325 & 313 $-$ 336 & 16 &1124/473 &1344 & 1345 &1334 $-$1355 & 21 & 178/102 & 262 \\
 345 & 333 $-$ 356 & 16 & 823/532 &1043 & 1365 &1354 $-$1376 & 11 &  41/ 40 &  62 \\
 365 & 353 $-$ 376 & 16 & 579/443 & 799 & 1385 &1374 $-$1395 & 11 &  59/ 36 &  81 \\
 385 & 373 $-$ 396 & 17 & 450/406 & 587 & 1405 &1394 $-$1416 & 21 & 324/147 & 509 \\
 405 & 393 $-$ 416 & 18 & 639/448 & 785 & 1425 &1414 $-$1436 & 21 &  53/ 57 & 110 \\
 425 & 413 $-$ 436 & 18 & 497/358 & 654 & 1445 &1434 $-$1456 & 21 &  89/ 91 & 147 \\
 445 & 433 $-$ 456 & 18 & 409/275 & 524 & 1465 &1453 $-$1475 & 11 &  62/ 36 & 100 \\
 465 & 453 $-$ 476 & 18 & 298/230 & 345 & 1485 &1473 $-$1495 & 11 &  64/ 35 &  95 \\
 485 & 473 $-$ 496 & 19 & 284/192 & 405 & 1505 &1493 $-$1515 & 21 & 107/ 98 & 198 \\
 505 & 493 $-$ 516 & 19 & 430/272 & 517 & 1525 &1514 $-$1536 & 11 &  74/ 37 & 103 \\
 525 & 513 $-$ 536 & 20 & 223/202 & 285 & 1545 &1533 $-$1555 & 11 &  46/ 53 &  63 \\
 545 & 533 $-$ 556 & 20 & 348/237 & 336 & 1565 &1554 $-$1575 & 11 &  11/ 19 &  30 \\
 565 & 553 $-$ 577 & 20 & 311/211 & 369 & 1585 &1574 $-$1595 & 21 &  42/ 34 &  73 \\
 585 & 573 $-$ 596 & 20 & 432/275 & 560 & 1605 &1593 $-$1616 & 21 &  90/ 87 & 146 \\
 605 & 593 $-$ 616 & 20 & 318/286 & 406 & 1625 &1613 $-$1635 & 11 &  40/ 24 &  77 \\
 625 & 613 $-$ 636 & 20 & 384/294 & 482 & 1645 &1634 $-$1655 & 11 & 118/ 73 & 149 \\
 645 & 633 $-$ 656 & 20 & 469/332 & 583 & 1665 &1654 $-$1675 & 21 &  24/ 36 &  36 \\
 665 & 653 $-$ 676 & 21 & 528/405 & 637 & 1685 &1673 $-$1695 & 11 &  25/ 31 &  34 \\
 685 & 673 $-$ 696 & 21 & 647/436 & 795 & 1705 &1693 $-$1715 & 21 &  98/ 88 & 139 \\
 705 & 693 $-$ 716 & 21 &1112/625 &1219 & 1725 &1714 $-$1735 & 11 &  10/ 15 &  19 \\
 725 & 713 $-$ 736 & 26 & 425/402 & 620 & 1745 &1734 $-$1755 & 11 & 101/ 42 & 121 \\
 745 & 733 $-$ 757 & 26 & 932/580 &1179 & 1765 &1753 $-$1775 & 11 &  53/ 35 &  57 \\
 765 & 753 $-$ 776 & 26 & 552/400 & 787 & 1785 &1774 $-$1796 & 11 &  25/ 21 &  36 \\
 785 & 773 $-$ 797 & 26 & 470/387 & 699 & 1805 &1794 $-$1815 & 21 &  78/ 70 & 129 \\
 805 & 793 $-$ 816 & 27 & 510/347 & 758 & 1845 &1824 $-$1865 & 11 & 125/ 67 & 155 \\
 825 & 813 $-$ 836 & 27 & 243/183 & 332 & 1900 &1879 $-$1920 & 21 & 158/ 99 & 212 \\
 845 & 833 $-$ 856 & 27 & 542/338 & 761 & 1990 &1969 $-$2005 & 21 & 150/100 & 326 \\
 865 & 853 $-$ 876 & 27 & 164/150 & 280 &      &             &    &         &     \\
\end{tabular}
\end{table}
\begin{table}[t]
\caption{Comparison of $\chi^2$ for the SM95~\protect\cite{ar96},
         SM02, and SX99 solutions to 2~GeV versus the present
         database.  Only a fraction of this database was used 
         in genereting SM95.}
\label{tbl5}
\begin{tabular}{cccccc}
Reaction &  Observable       &  SM95 &  SM02 &  SX99 & Data \\
\tableline
$\pi^0p$ & $d\sigma/d\Omega$ & 32235 & 12681 & 12745 & 5523 \\
         &  $\sigma_{tot}$   &  1577 &  1331 &  1681 &  713 \\
 &$\sigma_{1/2}-\sigma_{3/2}$&     7 &    10 &    10 &   13 \\
         &     $\Sigma$      &  3238 &  1918 &  1975 &  772 \\
         &       P           &  1269 &  1390 &  1393 &  576 \\
         &       T           &  1499 &  1651 &  1678 &  389 \\
\tableline
$\pi^+n$ & $d\sigma/d\Omega$ &  7587 &  6364 &  6952 & 4995 \\
         &  $\sigma_{tot}$   &   132 &    70 &    84 &   76 \\
 &$\sigma_{1/2}-\sigma_{3/2}$&   147 &    29 &    67 &   13 \\
         &     $\Sigma$      &  8672 &  2881 &  3780 & 1047 \\
         &       P           &   512 &   471 &   430 &  250 \\
         &       T           &  1678 &  1512 &  1854 &  694 \\
\tableline
$\pi^-p$ & $d\sigma/d\Omega$ &  3702 &  3098 &  3118 & 1570 \\
         &  $\sigma_{tot}$   &   227 &   157 &   165 &  117 \\
         &     $\Sigma$      &   541 &   600 &   576 &  216 \\
         &       P           &   159 &   144 &   161 &   88 \\
         &       T           &   117 &   157 &   129 &   96 \\
\end{tabular}
\end{table}
\begin{table}[t]
\caption{Resonance couplings from a Breit-Wigner fit 
         to the SM02 solution [GW] and SES [GWSES]
         (to illustrate the difference, we include
         $D_{35}(1930)$ and $D_{37}(1950)$ twice, 
         once with the GWSES database and
         once with the GW (global fit) database),  
         our previous solution SM95 [VPI]~
         \protect\cite{ar96}, the analysis of Crawford 
         and Morton [CM83]~\protect\cite{cm83}, 
         Crawford [CR01]~\protect\cite{cr01}, 
         Drechsel~\textit{et al.} [MAID98]~
         \protect\cite{maid98}, an average from the 
         Particle Data Group [PDG]~\protect\cite{pdg}, 
         and quark model predictions of Capstick [CAP92]~
         \protect\cite{cp92}.}
\label{tbl6}
\begin{tabular}{lccccc}
  &       &\multicolumn{2}{c}{\bf $\gamma p(GeV)^{-1/2} *10^{-3}$}
          &\multicolumn{2}{c}{\bf $\gamma n(GeV)^{-1/2} *10^{-3}$} \\
{\bf Resonance State} & {\bf Reference}  & $A_{1/2}$ & $A_{3/2}$
                                         & $A_{1/2}$ & $A_{3/2}$   \\
\tableline
${\bf S_{11}(1650)}$                        & GWSES
                       & $ 74\pm  1$ &
                       & $-28\pm  4$ &                             \\
\hspace*{0.15in} $W_{R}=1674~MeV     $      & VPI
                       & $ 69\pm  5$ &
                       & $-15\pm  5$ &                             \\
\hspace*{0.15in} $\Gamma _{\pi}/\Gamma=0.77$& CM83
                       & $ 33\pm 15$ &
                       & $-68\pm 40$ &                             \\
\hspace*{0.15in} $\Gamma =191~MeV    $      & CR01
                       & $ 71      $ &
                       & $-        $ &                             \\
                                            & MAID98
                       & $ 39      $ &
                       & $-32      $ &                             \\
                                            & PDG
                       & $ 53\pm 16$ &
                       & $-15\pm 21$ &                             \\
                                            & CAP92
                       & $ 54      $ &
                       & $-35      $ &                             \\
\hspace*{0.15in}       &&&&&                                       \\
${\bf P_{11}(1440)}$                        & GWSES
                       & $-67\pm  2$ &
                       & $ 47\pm  5$ &                             \\
\hspace*{0.15in} $W_{R}=1472~MeV     $      & VPI
                       & $-63\pm  5$ &
                       & $ 45\pm 15$ &                             \\
\hspace*{0.15in} $\Gamma _{\pi}/\Gamma=0.64$& CM83
                       & $-69\pm 18$ &
                       & $ 56\pm 15$ &                             \\
\hspace*{0.15in} $\Gamma =434~MeV    $      & CR01
                       & $-88      $ &
                       & $-        $ &                             \\
                                            & MAID98
                       & $-71      $ &
                       & $ 60      $ &                             \\
                                            & PDG
                       & $-65\pm  4$ & 
                       & $ 40\pm 10$ &                             \\
                                            & CAP92
                       & $  4      $ & 
                       & $ -6      $ &                             \\
\hspace*{0.15in}       &&&&&                                       \\
${\bf D_{13}(1520)}$                        & GWSES  
                       & $-24\pm  2$ & $ 135\pm  2$ 
                       & $-67\pm  4$ & $-112\pm  3$                \\
\hspace*{0.15in} $W_{R}=1517~MeV     $      & VPI
                       & $-20\pm  7$ & $ 167\pm  5$
                       & $-48\pm  8$ & $-140\pm 10$                \\
\hspace*{0.15in} $\Gamma _{\pi}/\Gamma=0.63$& CM83
                       & $-28\pm 14$ & $ 156\pm 22$
                       & $-56\pm 11$ & $-144\pm 15$                \\
\hspace*{0.15in} $\Gamma = 109~MeV   $      & CR01
                       & $-15      $ & $ 162      $
                       & $-        $ & $-         $                \\
                                            & MAID98
                       & $-17      $ & $ 164      $
                       & $-40      $ & $-135      $                \\
                                            & PDG
                       & $-24\pm  9$ & $ 166\pm  5$
                       & $-59\pm  9$ & $-139\pm 11$                \\
                                            & CAP92
                       & $-15      $ & $ 134      $
                       & $-38      $ & $-114      $                \\
\hspace*{0.15in}       &&&&&                                       \\
${\bf D_{15}(1675)}$                        & GWSES  
                       & $ 33\pm  4$ & $   9\pm  3$
                       & $-50\pm  4$ & $ -71\pm  5$                \\
\hspace*{0.15in} $W_{R}=1678~MeV     $      & VPI
                       & $ 15\pm 10$ & $  10\pm  7$
                       & $-49\pm 10$ & $ -51\pm 10$                \\
\hspace*{0.15in} $\Gamma _{\pi}/\Gamma=0.39$& CM83
                       & $ 21\pm 11$ & $  15\pm  9$
                       & $-59\pm 15$ & $ -59\pm 20$                \\
\hspace*{0.15in} $\Gamma =144~MeV    $      & CR01
                       & $ 13      $ & $  38      $
                       & $-        $ & $-         $                \\
                                            & MAID98
                       & $-        $ & $-         $
                       & $-        $ & $-         $                \\
                                            & PDG
                       & $ 19\pm  8$ & $  15\pm  9$
                       & $-43\pm 12$ & $ -58\pm 13$                \\ 
                                            & CAP92
                       & $  2      $ & $   3      $ 
                       & $-35      $ & $ -51      $                \\
\hspace*{0.15in}       &&&&&                                       \\
${\bf F_{15}(1680)} $                       & GWSES  
                       & $-13\pm  2$ & $ 129\pm  2$
                       & $ 29\pm  6$ & $ -58\pm  9$                \\
\hspace*{0.15in} $W_{R}=1682~MeV     $      & VPI
                       & $-10\pm  4$ & $ 145\pm  5$
                       & $ 30\pm  5$ & $ -40\pm 15$                \\
\hspace*{0.15in} $\Gamma _{\pi}/\Gamma=0.68$& CM83
                       & $-17\pm 18$ & $ 132\pm 10$
                       & $ 44\pm 12$ & $ -33\pm 15$                \\
\hspace*{0.15in} $\Gamma =120~MeV    $      & CR01
                       & $-14      $ & $ 135      $
                       & $-        $ & $-         $                \\
                                            & MAID98
                       & $-10      $ & $ 138      $
                       & $ 35      $ & $ -41      $                \\
                                            & PDG
                       & $-15\pm  6$ & $ 133\pm 12$
                       & $ 29\pm 10$ & $ -33\pm  9$                \\
                                            & CAP92
                       & $-38      $ & $  56      $
                       & $ 19      $ & $ -23      $                \\
\hspace*{0.15in}       &&&&&                                       \\
${\bf S_{31}(1620)} $                       & GWSES  
                       & $-13\pm  3$ & $          $
                       & $         $ & $          $                \\
\hspace*{0.15in} $W_{R}=1612~MeV     $      & VPI
                       & $ 35\pm 20$ & $          $
                       & $         $ & $          $                \\
\hspace*{0.15in} $\Gamma _{\pi}/\Gamma=0.34$& CM83
                       & $ 35\pm 10$ & $          $
                       & $         $ & $          $                \\
\hspace*{0.15in} $\Gamma = 117~MeV   $      & CR01
                       & $ 17      $ & $          $
                       &             &                             \\
                                            & MAID98
                       & $-        $ &
                       & $         $ &                             \\
                                            & PDG
                       & $ 27\pm 11$ & $          $
                       & $         $ & $          $                \\
                                            & CAP92
                       & $ 81      $ & $          $
                       & $         $ & $          $                \\
\hspace*{0.15in}       &&&&&                                       \\
${\bf P_{33}(1232)}$                        & GWSES  
                       &$-129\pm  1$ & $-243\pm  1$
                       & $         $ & $          $                \\
\hspace*{0.15in} $W_{R}=1235~MeV     $      & VPI
                       &$-141\pm  5$ & $-261\pm  5$
                       & $         $ & $          $                \\
\hspace*{0.15in} $\Gamma _{\pi}/\Gamma=1.00$& CM83
                       &$-145\pm 15$ & $-263\pm 26$
                       & $         $ & $          $                \\
\hspace*{0.15in} $\Gamma =119~MeV    $      & CR01
                       &$-149      $ & $-259      $
                       &             &                             \\
                                            & MAID98
                       & $-138     $ & $-256      $
                       & $         $ & $          $                \\
                                            & PDG
                       &$-135\pm  6$ & $-255\pm  8$
                       & $         $ & $          $                \\
                                            & CAP92
                       &$-108      $ & $-186      $
                       & $         $ & $          $                \\
\hspace*{0.15in}       &&&&&                                       \\
${\bf D_{33}(1700)}$                       & GWSES
                       & $ 89\pm 10$ & $  92\pm  7$
                       & $         $ & $          $                \\
\hspace*{0.15in} $W_{R}=1668~MeV     $     & VPI
                       & $ 90\pm 25$ & $  97\pm 20$
                       & $         $ & $          $                \\
\hspace*{0.15in} $\Gamma _{\pi}/\Gamma=0.16$& CM83
                       & $111\pm 17$ & $ 107\pm 15$
                       & $         $ & $          $                \\
\hspace*{0.15in} $\Gamma =300~MeV$      & CR01
                       & $ 79      $ & $  90      $
                       &             &                             \\
                                            & MAID98
                       & $ 86      $ & $  85      $
                       & $         $ & $          $                \\
                                            & PDG
                       & $104\pm 15$ & $  85\pm 22$
                       & $         $ & $          $                \\
                                            & CAP92
                       & $ 82      $ & $  68      $
                       & $         $ & $          $                \\
\hspace*{0.15in}       &&&&&                                       \\
${\bf D_{35}(1930)}$                        & GWSES  
                       & $  4\pm  6$ & $  -3\pm  6$
                       & $         $ & $          $                \\
\hspace*{0.15in} $W_{R}=2113~MeV     $      & GW
                       & $  6\pm  2$ & $  -4\pm  2$
                       & $         $ & $          $                \\
\hspace*{0.15in} $\Gamma _{\pi}/\Gamma=0.14$& VPI
                       & $ -7\pm 10$ & $   5\pm 10$
                       & $         $ & $          $                \\
\hspace*{0.15in} $\Gamma =524~MeV    $      & CM83
                       & $-38\pm 47$ & $ -23\pm 80$
                       & $         $ & $          $                \\
                                            & CR01
                       & $  8      $ & $   8      $
                       &             &                             \\
                                            & MAID98
                       & $-        $ & $-         $
                       & $         $ & $          $                \\
                                            & PDG
                       & $ -9\pm 28$ & $ -18\pm 28$
                       & $         $ & $          $                \\
                                            & CAP92
                       & $-        $ & $-         $
                       & $         $ & $          $                \\
\hspace*{0.15in}       &&&&&                                       \\
${\bf F_{35}(1905)}$                        & GWSES  
                       & $  2\pm  5$ & $ -56\pm  5$
                       & $         $ & $          $                \\
\hspace*{0.15in} $W_{R}=1845~MeV     $      & VPI
                       & $ 22\pm  5$ & $ -45\pm  5$
                       & $         $ & $          $                \\
\hspace*{0.15in} $\Gamma _{\pi}/\Gamma=0.13$& CM83
                       & $ 21\pm 10$ & $ -56\pm 28$
                       & $         $ & $          $                \\
\hspace*{0.15in} $\Gamma =300~MeV    $      & CR01
                       & $ 17      $ & $ -18      $
                       &             &                             \\
                                            & MAID98
                       & $-        $ & $-         $
                       & $         $ & $          $                \\
                                            & PDG
                       & $ 26\pm 11$ & $ -45\pm 20$
                       & $         $ & $          $                \\
                                            & CAP92
                       & $ 26      $ & $  -1      $
                       & $         $ & $          $                \\
\hspace*{0.15in}       &&&&&                                       \\
${\bf F_{37}(1950)} $                       & GWSES   
                       & $-62\pm  4$ & $ -80\pm  3$
                       & $         $ & $          $                \\
\hspace*{0.15in} $W_{R}=1929~MeV     $      & GW
                       & $-64\pm  4$ & $ -83\pm  4$
                       & $         $ & $          $                \\
\hspace*{0.15in} $\Gamma _{\pi}/\Gamma=0.49$& VPI
                       & $-79\pm  6$ & $-103\pm  6$
                       & $         $ & $          $                \\
\hspace*{0.15in} $\Gamma =274~MeV    $      & CM83
                       & $-67\pm 14$ & $ -82\pm 17$
                       & $         $ & $          $                \\
                                            & CR01
                       & $-59      $ & $ -62      $
                       &             &                             \\
                                            & MAID98
                       & $-        $ & $-         $
                       & $         $ & $          $                \\
                                            & PDG
                       & $-76\pm 12$ & $ -97\pm 10$
                       & $         $ & $          $                \\
                                            & CAP92
                       & $-33      $ & $ -42      $
                       & $         $ & $          $                \\
\end{tabular}
\end{table}
\begin{table}[t]
\caption{\label{tbl7}
         Comparison of the SM02 and recent MAID2000~
         \protect\cite{maid} calculations for the GDH 
         and Baldin integrals and the forward spin 
         polarizability from threshold to 2~GeV (for 
         MAID to 1.25~GeV) [upper set] and from 
         threshold to 200~MeV [lower set] displayed 
         as SAID/MAID.}
\begin{tabular}{cccc}
Reaction & GDH           & Baldin  & $\gamma_0$      \\
         & ($\mu b$)     & ($10^{-4} fm^3$) & ($10^{-4} fm^4$) \\
\tableline  
$\pi^0p$ & $-$142/$-$150 & 4.7/4.7 & $-$1.40/$-$1.47 \\
$\pi^+n$ &  $-$45/$-$18  & 6.8/6.9 &    0.55/   0.79 \\
$\pi^0n$ & $-$148/$-$153 & 4.6/4.6 & $-$1.44/$-$1.50 \\
$\pi^-p$ &     11/    33 & 8.3/8.8 &    1.36/   1.64 \\
\tableline
$\pi^0p$ &   $-$2/$-$1   & 0.1/0.1 & $-$0.05/$-$0.04 \\
$\pi^+n$ &     30/    32 & 1.2/1.2 &    0.99/   1.02 \\
$\pi^0n$ &   $-$1/$-$1   & 0.1/0.1 & $-$0.04/$-$0.04 \\
$\pi^-p$ &     42/    47 & 1.7/1.8 &    1.39/   1.53 \\
\end{tabular}
\end{table}
\begin{table}[t]
\caption{\label{tbl8}
         Comparison of the SM02 and recent MAID2000
         ~\protect\cite{maid} calculations, and recent 
         Mainz data~\protect\cite{ah00} for the GDH 
         and Baldin integrals and the forward spin 
         polarizability for proton and neutron 
         targets from 200 to 450~MeV.  Units are $\mu b$,
         $10^{-4} fm^3$, and $10^{-4} fm^4$ for
         the GDH and Baldin integrals and 
         polarizability, respectively.}
\begin{tabular}{ccccc}
Integral  &Reaction  & SAID & MAID & Mainz \\
\tableline
GDH       & $\pi^0p$ &$-$129&$-$136&$-$144$\pm$7$\pm$9\\
          & $\pi^+n$ & $-$42& $-$25& $-$32$\pm$3$\pm$2\\
          & proton   &$-$171&$-$161&$-$176$\pm$8$\pm$11\\
\tableline
          & $\pi^-p$ & $-$14&  $-$1&       \\
          & $\pi^0n$ &$-$135&$-$139&       \\
          & neutron  &$-$149&$-$140&       \\
\tableline
Baldin    & $\pi^0p$ & 4.0  & 4.0  &       \\
          & $\pi^+n$ & 4.5  & 4.6  &       \\
          & proton   & 8.5  & 8.6  &       \\
\tableline
          & $\pi^-p$ & 5.5  & 5.9  &       \\
          & $\pi^0n$ & 3.9  & 4.0  &       \\
          & neutron  & 9.5  & 9.8  &       \\
\tableline
$\gamma_0$& $\pi^0p$ &$-$1.31&$-$1.39&$-$1.45$\pm$0.09$\pm$0.09\\
          & $\pi^+n$ &$-$0.39&$-$0.21&$-$0.23$\pm$0.04$\pm$0.01\\
          & proton   &$-$1.71&$-$1.61&$-$1.68$\pm$0.10$\pm$0.10\\
\tableline
          & $\pi^-p$ &$-$0.05&   0.11&       \\
          & $\pi^0n$ &$-$1.35&$-$1.40&       \\
          & neutron  &$-$1.41&$-$1.29&       \\
\end{tabular}
\end{table}
\begin{table}[t]
\caption{\label{tbl9}
         Comparison of the SM02 and recent MAID2000
         ~\protect\cite{maid} calculations, and recent
         Mainz data~\protect\cite{ah01} for the GDH
         and Baldin integrals and the forward spin
         polarizability for proton and neutron 
         targets from 200 to 800~MeV.  Units are $\mu b$,
         $10^{-4} fm^3$, and $10^{-4} fm^4$ for
         the GDH and Baldin integrals and polarizability,
         respectively.}
\begin{tabular}{ccccc}
Integral  &Reaction  & SAID & MAID & Mainz \\
\tableline
GDH       & proton   &$-$193&$-$175&$-$226$\pm$5$\pm$12\\
          & neutron  &$-$167&$-$160&       \\
\tableline
Baldin    & proton   & 9.8  & 9.9  &       \\
          & neutron  &10.9  &11.3  &       \\
\tableline
$\gamma_0$& proton   &$-$1.76&$-$1.63&$-$1.87$\pm$0.08$\pm$0.10\\
          & neutron  &$-$1.46&$-$1.34&       \\
\end{tabular}
\end{table}
\eject
{\Large\bf Figure captions} \\
\newcounter{fig}
\begin{list}
{Figure \arabic{fig}.}
{\usecounter{fig}\setlength{\rightmargin}{\leftmargin}}
\item {Energy-angle distribution of recent (post-1995) 
      data: (a) unpolarized $\pi^0p$, (b) polarized 
      $\pi^0p$, (c) unpolarized $\pi^+n$, (d) 
      polarized $\pi^+n$.  (a,c) total cross sections 
      and (b,d) ($\sigma_{1/2} - \sigma_{3/2}$) are
      plotted at zero degrees.}
\item {Partial-wave amplitudes (L$_{2I, 2J}$) from 
      threshold to $E_{\gamma}$ = 2~GeV.  Solid (dashed) 
      curves give the real (imaginary) parts of amplitudes 
      corresponding to the SM02 solution.  The real 
      (imaginary) parts of single-energy solutions are 
      plotted as filled (open) circles.  The previous 
      SM95 solution \protect\cite{ar96} is plotted 
      with long dash-dotted (real part) and short 
      dash-dotted (imaginary part) lines.  Plotted 
      are the multipole amplitudes 
      (a) $\rm _pE_{0+}^{1/2}$, (b) $\rm _nE_{0+}^{1/2}$, 
      (c) $\rm _pE_{0+}^{3/2}$, (d) $\rm _pM_{1-}^{1/2}$,
      (e) $\rm _nM_{1-}^{1/2}$, (f) $\rm _pE_{1+}^{1/2}$, 
      (g) $\rm _pM_{1+}^{1/2}$, (h) $\rm _nE_{1+}^{1/2}$, 
      (i) $\rm _nM_{1+}^{1/2}$, (j) $\rm _pM_{1-}^{3/2}$,
      (k) $\rm _pE_{1+}^{3/2}$, (l) $\rm _pM_{1+}^{3/2}$, 
      (m) $\rm _pE_{2-}^{1/2}$, (n) $\rm _pM_{2-}^{1/2}$, 
      (o) $\rm _nE_{2-}^{1/2}$, (p) $\rm _nM_{2-}^{1/2}$,
      (q) $\rm _pE_{2-}^{3/2}$, (r) $\rm _pE_{2+}^{3/2}$, 
      (s) $\rm _pE_{3-}^{1/2}$, (t) $\rm _pM_{3-}^{1/2}$, 
      (u) $\rm _nE_{3-}^{1/2}$, (v) $\rm _nM_{3-}^{1/2}$,
      (w) $\rm _pE_{3-}^{3/2}$, and (x) $\rm _pM_{3+}^{3/2}$.  
      The subscript p (n) denotes a proton (neutron) 
      target.}
\item {Differential cross section ($d \sigma /d \Omega
       _{1/2}-d \sigma /d \Omega _{3/2}$) for $\vec{\gamma}
       \vec{p}\to\pi^0p$ at $\theta = 85\pm 4^{\circ}$.  
       The solid (dash-dotted) line plots the SM02 
       (MAID2001~\protect\cite{maid}) solution.  
       Experimental data are from Mainz~
       \protect\cite{dx13}.}
\item {Photon asymmetry for $\pi^0$ photoproduction on
       the proton at 159.5~MeV.  Data are from Mainz
       (solid circles)~\protect\cite{sc01}.  Plotted
       are the SM02 (solid line), the 162~MeV-SES (158
       $-$ 165~MeV) fit associated with SM02
       (dotted lines represent uncertainties of the SES
       fit) and the MAID2000 results (dash-dotted)~
       \protect\cite{maid}.}
\item {$\Sigma$ beam asymmetry for $\pi^+n$ at 1100~MeV.
       Black circles show GRAAL results~
       \protect\cite{ku01}, open circles
       indicate the results of the Daresbury group~ 
       \protect\cite{bs79}, open triangles indicate
       the results from Saclay~\protect\cite{as72}.  The
       solid (dash-dotted) line represents the SM02
       (MAID2001~\protect\cite{maid}) solution.}
\item {Forward ($5^{\circ}$) differential cross section
       for $\gamma p\to\pi^+n$ as a function of energy.
       Experimental data for the range of $5\pm 
       2^{\circ}$ are from Orsay~\protect\cite{bt68} 
       (black circles), SLAC~\protect\cite{ec67} (open
       circles), \protect\cite{by61} (open triangles),
       \protect\cite{ki62} (black square), and
       DESY~\protect\cite{bu67} (black diamonds.)
       The solid (dash-dotted) line represents the SM02
       (MAID2001~\protect\cite{maid}) solution.}
\item {Difference of the total cross sections for the
       helicity states 1/2 and 3/2.
       (a) $\vec{\gamma}\vec{p}\to\pi^0p$ and
       (b) $\vec{\gamma}\vec{p}\to\pi^+n$.  The solid
       (dash-dotted) line represents the SM02 (MAID2001~
       \protect\cite{maid}) solution.  Experimental
       data are from Mainz~\protect\cite{ah00}.}
\item {Selected partial-wave amplitudes to $E_{\gamma}$
      = 1250~MeV.  Solid (dashed) curves give the real 
      (imaginary) parts of amplitudes corresponding to 
      the SM02 solution.  The recent MAID2001 solution~
      \protect\cite{maid} is plotted with long 
      dash-dotted (real part) and short dash-dotted
      (imaginary part) lines.  Plotted are the 
      multipole amplitudes (a) $S_{11}pE$
      [$\rm _pE_{0+}^{1/2}$], (b) $P_{13}pE$
      [$\rm _pE_{1+}^{1/2}$], (c) $P_{31}pM$
      [$\rm _pM_{1-}^{3/2}$], and (d) $D_{13}pE$
      [$\rm _pE_{2-}^{1/2}$].  The subscript p (n) 
      denotes a proton (neutron) target.}
\item {$S_{11}pE$ multipole for 600 to 1200~MeV.  
       Plotted are (a) real part and (b) imaginary
       part.  The SM02 (SX99) solution is plotted 
       with a solid (dashed) line and previous~SM95 
       solution~\protect\cite{ar96} with a dash-dotted 
       line.}     
\item {Running GDH integral.  (a) for proton and 
       (b) neutron targets.  The solid (dash-dotted) 
       line represents the SM02 (MAID2000
       ~\protect\cite{maid}) solution.}
\item {Running Baldin integral.  (a) for proton 
       and (b) neutron targets.  The solid 
       (dash-dotted) line represents the SM02 (MAID2000
       ~\protect\cite{maid}) solution.}
\item {Forward spin polarizability $\gamma_0$.
       (a) for proton and (b) neutron targets.
       The solid (dash-dotted) line represents the
       SM02 (MAID2000~\protect\cite{maid}) solution.}
\end{list}
\eject
\eject
\eject
\eject
\eject
\eject
\eject
\end{document}